\documentclass[a4,11pt]{article}

\usepackage[latin1]{inputenc}

\usepackage{dingbat}
\usepackage{color}
\usepackage{graphicx}  

\usepackage{psfrag}

\usepackage{tocbibind} 

\usepackage{epsf}
\usepackage{epic}
\usepackage{eepic}

\usepackage{epsfig}
\usepackage{wrapfig}
\usepackage{amsmath,amssymb,amsfonts}
\usepackage[mathscr]{eucal}
\usepackage{citesort}
\usepackage[active]{srcltx}
\usepackage{deleq}
\usepackage{url}
\usepackage{fancybox}
\usepackage{ntheorem}

\usepackage{mhequ}

\newcommand{\m}{\mathfrak{g}}
\newcommand{\f}{\mathfrak{h}}

\newcommand{\otimess}{\mathring\otimes}%
\newcommand{\stau}{y}%
\newcommand{\gbeta}{{\gamma}}%

\newcommand{\myOmega}{\mcU}

\newcommand{\no}{\nonumber\\}
\newcommand{\mcG}{{\mycal G}}

\newcommand{\si}{\sigma}
\newcommand{\de}{\delta}

\newcommand{\n}{\nabla}

\newcommand{\Om}{{\Omega}}%

\newcommand{\newF}{\lambda}
\newcommand{\divv}{\mathrm{div}}

\newcommand{\tmT}{t+T}

\newcommand{\pT}{-T}
\newcommand{\ppT}{+T}

\newcommand{\NLc}{\emph{NL}-condition}

\newcommand{\barf}{\bar f}

\newcommand{\beijing}[1]{}

\newcommand{\nmoh}{\frac{n-1}2}
\newcommand{\nmfh}{\frac{n-5}2}

\newcommand{\npth}{\frac{n+3}2}

\newcommand{\npsh}{\frac{n+7}2}

\newcommand{\al}{\alpha}
\newcommand{\bet}{\beta}
\newcommand{\p}{\partial}
\newcommand{\et}{\eta}
\newcommand{\la}{\lambda}

\newcommand{\SetOmega}{\mcU}

\newcommand{\zpsi}{\mathring{\psi}}

\newcommand{\mcH}{{\mycal H}}
\newcommand{\fCy}[2]{\mcC_{\{y=0\},{#2}}^{#1}}
\newcommand{\fCx}[2]{\mcC_{\{x=0\},{#2}}^{#1}}
\newcommand{\fCtx}[2]{\mcC_{\{\tilde{x}=0\},{#2}}^{#1}}
\newcommand{\fCxy}[2]{\mcC_{\{\zlxy \},{#2}}^{#1}}

\newcommand{\cDask}{{\mycal C}_{\{\zlxy \},k}^{\alpha,\sigma}}
\newcommand{\cDak}{{\mycal C}_{\{\zlxy \},k}^\alpha}
\newcommand{\cDmimn}%
{{\mycal C}_{\{\zlxy
\},\infty}^{\min(\alpha+\delta-\epsilon,(mq-p)\delta-\epsilon,-\epsilon)}}
\newcommand{\cDmimns}%
{{\mycal C}_{\{\zlxy
\},\infty}^{\min(\sigma+\delta-\epsilon,(mq-p)\delta-\epsilon,-\epsilon)}}

\newcommand{\argw}{w}
\newcommand{\varz}{z}

\newcommand{\zlxy}{0\le x \le y}

\newcommand{\Ayd}{{\mycal A}^\delta_{\{y=0\}}}
\newcommand{\Axd}{{\mycal A}^\delta_{\{x=0\}}}
\newcommand{\Axyd}{{\mycal A}^\delta_{\{\zlxy \}}}

\newcommand{\Ax}{{\mycal A}_{\{x=0\}}}
\newcommand{\Axy}{{\mycal A}_{\{\zlxy \}}}

\newcommand{\tx}{{\tilde x}}%

\newcommand{\beq}{\begin{equation}}

\newcommand{\ep}{\epsilon}
\newcommand{\vp}{\varphi}
\newcommand{\wt}{\widetilde}
\newcommand{\wh}{\widehat}

\newcommand{\tauz}{{\tau_{0}}}
\newcommand{\tauone}{{\tau_{1}}}

\newcommand{\mya} {a}
\newcommand{\myab} {b}
\newcommand{\myc} {c}

\newcommand{\xUp}{{}^{(2)}\Upsilon}%
\newcommand{\yUp}{{}^{(1)}\Upsilon}%
\newcommand{\yUpsilon}{\yUp}%
\newcommand{\FS}       
                  {F}

\newcommand{\HS} 
       {H_{\mbox{\scriptsize volume}}}

{\ptc{this should be removed in the oberwolfach version}}%

\newcommand{\eeal}[1]{\label{#1}\end{eqnarray}}
\newcommand{\bed}{\begin{deqarr}}
\newcommand{\eed}{\end{deqarr}}
\newcommand{\bedl}[1]{\begin{deqarr}\label{#1}}
\newcommand{\eedl}[2]{\arrlabel{#1}\label{#2}\end{deqarr}}

\newcommand{\loc}{\textrm{\scriptsize\upshape loc}}

\newcommand{\mcO}{{\mycal O}}
\newcommand{\mcU}{{\mycal U}}

\newcommand{\mcN}{{\mycal N}}

\newcommand{\bel}[1]{\begin{equation}\label{#1}}
\newcommand{\bea}{\begin{eqnarray}}
\newcommand{\bean}{\begin{eqnarray}\nonumber}
\newcommand{\beal}[1]{\begin{eqnarray}\label{#1}}
\newcommand{\eea}{\end{eqnarray}}

\newcommand{\Eq}[1]{Equation~\eq{#1}}

\newcommand{\qed}{\hfill $\Box$ \medskip}
\newcommand{\proof}{\noindent {\sc Proof:\ }}
\newcommand{\be}{\begin{equation}}
\newcommand{\eeq}{\end{equation}}
\newcommand{\ee}{\end{equation}}
\newcommand{\beqa}{\begin{eqnarray}}
\newcommand{\eeqa}{\end{eqnarray}}
\newcommand{\beqan}{\begin{eqnarray*}}
\newcommand{\eeqan}{\end{eqnarray*}}
\newcommand{\ba}{\begin{array}}
\newcommand{\ea}{\end{array}}

\newcommand{\const}{\mbox{\rm const}} 

\newcommand{\hyp}{\mycal S}

\newcommand{\mcM}{{\mycal M}}
\newcommand{\mcD}{{\mycal D}}

\newcommand{\mcV}{{\mycal V}}

\newtheorem{Theorem} {\sc  Theorem\rm} [section]
\newtheorem{Corollary} [Theorem] {\sc  Corollary\rm}
\newtheorem{Lemma} [Theorem] {\sc  Lemma\rm}
\newtheorem{Proposition} [Theorem] {\sc  Proposition\rm}

\theorembodyfont{\upshape}
\newtheorem{Remark}[Theorem]{\sc Remark\rm}

\newtheorem{example}[Theorem]{\sc Example\rm}

\theoremsymbol{\ensuremath{\diamondsuit}}
\newcommand{\fcoco}{\small}
\theorembodyfont{\fcoco} \theoremseparator{} \theoremindent0.5cm

\theoremstyle{nonumberplain} \theorembodyfont{\fcoco}
\theoremseparator{} \theoremindent0.5cm

\DeclareFontFamily{OT1}{rsfs}{} \DeclareFontShape{OT1}{rsfs}{m}{n}{
<-7> rsfs5 <7-10> rsfs7 <10-> rsfs10}{}
\DeclareMathAlphabet{\mycal}{OT1}{rsfs}{m}{n}
\def\scri{{\mycal I}}%

{\catcode `\@=11 \global\let\AddToReset=\@addtoreset}
\AddToReset{equation}{section}
\renewcommand{\theequation}{\thesection.\arabic{equation}}

{\catcode `\@=11 \global\let\AddToReset=\@addtoreset}
\AddToReset{table}{section}

{\catcode `\@=11 \global\let\AddToReset=\@addtoreset}
\AddToReset{figure}{section}

\newcounter{mnotecount}[section]

\renewcommand{\themnotecount}{\thesection.\arabic{mnotecount}}

\newcommand{\mnote}[1]
{\protect{\stepcounter{mnotecount}}$^{\mbox{\footnotesize
$
\bullet$\themnotecount}}$ \marginpar{
\raggedright\tiny\em $\!\!\!\!\!\!\,\bullet$\themnotecount: #1} }

\newcommand{\warn}[1]
{\protect{\stepcounter{mnotecount}}$^{\mbox{\footnotesize
$
\bullet$\themnotecount}}$ \marginpar{
\raggedright\tiny\em $\!\!\!\!\!\!\,\bullet$\themnotecount: {\bf
Warning:} #1} }

\newcommand{\zvarphi}{{\mathring{\varphi}}}

\newcommand{\R}{\mathbb R}

\newcommand{\N}{\mathbb N}
\newcommand{\Z}{\mathbb Z}

\newcommand{\eq}[1]{(\ref{#1})}

\newcommand{\ptc}[1]{\mnote{{\bf ptc:}#1}}

\newcommand{\Hess}{\mathrm{Hess}\,}

\newcommand{\Ric}{\mbox{\rm Ric}}

\newcommand{\mcC}{{\mycal C}}
\newcommand{\mcS}{{\mycal S}}

\newcommand{\beqar}{\begin{deqarr}}
\newcommand{\eeqar}{\end{deqarr}}

\newcommand{\beaa}{\begin{eqnarray*}}
\newcommand{\eeaa}{\end{eqnarray*}}

\newcommand{\tr}{\mbox{tr}}

\newcommand{\mcB}{{\mycal B}}

\let\a=\alpha\let\b=\beta

\title{Solutions of quasi-linear wave equations polyhomogeneous at null infinity in high dimensions}

\author{Piotr T. Chru\'sciel\\
Gravitational Physics\\
University of Vienna
\\
\\
Roger Tagne Wafo\\
D\'epartement de Math\'ematiques et Informatique\\ Facult\'e des
Sciences, Universit\'e de Douala}

\begin{document}
\maketitle

\begin{abstract}
We prove propagation of weighted Sobolev regularity for
solutions of the hyperboloidal Cauchy problem for a class of
quasi-linear symmetric hyperbolic systems, under structure
conditions compatible with the Einstein-Maxwell equations in
space-time dimensions $n+1\ge 7$. Similarly we prove
propagation of  polyhomogeneity in dimensions $n+1\ge 9$. As a
byproduct we obtain, in those last dimensions, polyhomogeneity
at null infinity of small data solutions of vacuum Einstein, or
Einstein-Maxwell equations evolving out of initial data which
are stationary outside of a ball.
\end{abstract}

\tableofcontents

\section{Introduction}
\label{Sintro}

 A problem of current interest is the asymptotic behavior
of solutions of hyperbolic equations in the radiation zone. For
large (however, not for all) sets of initial data, this
question can be reduced to one where the initial data are given
on a Cauchy surface that resembles a hyperboloid in Minkowski
space-time. In recent works~\cite{ChLeski,ChLengardnwe},
polyhomogeneity of solutions of such Cauchy problems, with
polyhomogeneous initial data, has been proved for a large class
of semi-linear symmetric hyperbolic systems. The object of this
work is to extend those results to quasi-linear equations
satisfying certain structure conditions which are compatible
with the vacuum Einstein equations, or with the
Einstein-Maxwell equations, in space-time dimensions $n+1 \ge
9$.

A special case of our results is Theorem~\ref{TEMphg2} below,
where polyhomogeneity at null infinity of small data global
solutions of the Einstein-Maxwell equations, evolving out of
initial data which are stationary outside of a compact set, is
established; this is perhaps the most significant result in
this work. For clarity we repeat the relevant part of that
theorem here:

\begin{Theorem}
 \label{TEMphg3}
In dimensions $n+1\ge 9$ the  global solutions of
Einstein-Maxwell equations constructed  in
\cite{Loizelet:these,LoizeletCRAS} out from small initial data
stationary outside of a compact set are polyhomogeneous at null
infinity.
\end{Theorem}

The polyhomogeneous expansions above  are in terms of powers of
$\log r$ and negative integer powers of $r$ in odd space
dimension, while one has powers of $\log r$ and negative
half-integer powers of $r$ in even space dimension.

Theorem~\ref{TEMphg3} should be compared with~\cite{CCL}, where
even space-time dimension $n+1\ge 6$ is assumed, where initial
data \emph{Schwarzschildian} outside of a compact set are
considered, and where solutions which are smooth at null
infinity are obtained. The methods of that last reference
completely fail in odd space-time dimensions. Furthermore, in
odd space dimensions, generic initial data which are only
\emph{stationary}, as opposed to \emph{Schwarzschildian}, are
likely to be polyhomogeneous, \emph{but not smooth}, at null
infinity, and generic such initial data are expected to be too
singular to be covered by the approach in~\cite{CCL}. We also
note the analysis in~\cite{ChBeig3}, which implies smoothness
at null infinity of \emph{exactly} stationary vacuum or
electro-vacuum space-times, in even space-dimension, in
space-time harmonic gauge. But the dimensions covered
in~\cite{ChBeig3} are precisely those not covered by the
evolution theorems in~\cite{CCL,AndersonChruscielConformal}.

\section{Polyhomogeneity of solutions}
\label{SPos}

\subsection{Notation}
\label{sSnot}

The notation of~\cite{ChLeski} is used unless explicitly stated
otherwise. However, to avoid a clash of notation with the
symbol which is customarily used for the conformal factor
arising in the rescaling of the metric, we will use the symbol
$\SetOmega$ for the sets $\Omega$ of~\cite{ChLeski}:
%
\bel{Uset} \SetOmega =\left\{(x,v^{A},y):\ 0 < x < y\;,\ v=(v^A) \in
\mcO\;,\ 0 < y < 2T \right\} \;, \ee
where $\mcO$ is a compact manifold without boundary. We will write
$\varz$ for the joint set of variables $(x,y,v^A)$.

Let $W^\alpha$ be a family of spaces, where $\alpha$ is a decay
index, e.g.  $W^\alpha=\fCx{\alpha}{k}(\SetOmega )$, or
$W^\alpha=\fCxy{\alpha}{\infty}(\SetOmega )$, etc. We define
$$
W^{<\alpha}=\cap_{\sigma < \alpha} W^\sigma \;.
$$
This notation is very useful to accommodate  $\ln^n x$ factors that
arise in the problem at hand: for example, in this notation we have
$$
x^\alpha \ln^N x \in  \fCx{<\alpha}{\infty}(\SetOmega ) \;.
$$

We use  a slight generalization of a definition
of~\cite{ChLengardnwe}: We shall say that a function $H(\varz
,\argw)$ is {\em $\Axyd$-polyhomogeneous in $\varz $ with a uniform
zero of order $l$ in $\argw$} if the following hold: First, $H$ is
smooth in $\argw\in\R^N$ at fixed $\varz \in \SetOmega $. Next, it
is required that for all $B\in \R$ and $k\in \N$ there exists a
constant $\hat{C}(B)$ such that, for all $|\argw|\le B$ and $0\leq
i\leq \min(k,l)$,
\be\label{S2.71} \left\| \partial^i_\argw H(\cdot,\argw)
\right\|_{{\mcC}_{\{0\le x \le y\},k-i}^0(\SetOmega )} \leq
\hat{C}(B) |\argw|^{l-i}\;.\ee Further,
\begin{equation}
  \label{eq:m1}
\forall i\in\N \quad \partial^i_\argw H(\cdot, \argw)\in\Axyd
\end{equation}
at fixed constant $\argw$. Finally we demand the uniform estimate
for constant $\argw$'s: $  \forall \epsilon >0,  M \geq 0 , i,k\in
\N \ \; \exists\; C(\epsilon,M,i,k) \ \forall |\argw|\leq M$ such
that
\begin{equation}
  \label{m1}
\|
\partial^i_\argw H(\cdot, \argw)\|_{\mcC^{-\epsilon}_{\{0\le x \le y\},k}(\SetOmega )} \leq
C(\epsilon,M,i,k)\;.
\end{equation}
The qualification ``in $\argw$'' in ``uniform zero of order $l$ in
$\argw$'' will often be omitted. Similarly to~\cite{ChLengardnwe},
the small parameter $\epsilon$ has been introduced above to take
into account the possible logarithmic blow-up of functions in
$\Axyd$ at $x=0$; for the applications to the nonlinear scalar wave
equation or to the wave map equation on Minkowski space-time, the
alternative simpler requirement would actually suffice:  $\forall \;
M \geq 0 , i,k\in \N \ \; \exists\; C(M,i,k) \ \forall |\argw|\leq M
$
\begin{equation}
  \label{m2}
\| \partial^i_\argw H(\cdot, \argw)\|_{\mcC^{0}_{\{0\le x \le
y\},k}(\SetOmega )} \leq C(M,i,k)\;,
\end{equation}
again for constant $\argw$'s. Functions which are smooth in
$(\argw, \varz )$, and have a zero  of  order $l$ in $\argw$ at
$\argw=0$, satisfy the above conditions.

\subsection{The theorem}
\label{ssTt}

Let $\psi=(\psi_1,\psi_2)$ and set
\bel{fdef} f:=( {\psi},\varphi)\;, \quad \barf:=(
{\psi_1,x\psi_2},x\varphi) \;. \ee
We shall say that a function $G$ \emph{satisfies the \NLc}\ if there
exist $N,p_i,q_i,m_i\in \N$ and functions $H_i$ with a uniform zero
of order $m_i$ in the variable
\beaa w_i&:=& x^{q_i \delta}( \barf, x^2\partial_x f, x^2\partial_y
f, x\partial_A f)
\\
& \equiv&
  x^{q_i \delta}( {\psi_1,x\psi_2},x\varphi , x^2\partial_x f, x^2\partial_y  f, x\partial_A f)
\eeaa
such that
\bel{gdef} G = \sum_{i=1}^N x^{-p_i\delta} H_i(z,w_i) \;, \ee
with, for $i=1,\ldots,N$,
\begin{equation}
m_i  > \frac{p_i- \frac{1}{\delta}}{q_i} \;. \label{cocoi}
\end{equation}

Our first main result  is the following:

\begin{Theorem}
  \label{Tmain}
Let $\mcU$ be defined in \eq{Uset}, suppose that $p\in \Z$, $ q,
1/\delta\in \N^*$, $k\in \N\cup\{\infty\}$, and let
$$
\psi=(\psi_1,\psi_2)
$$
and $\varphi$, with
\bel{23I.2} \psi_1 \in \mcC^{<-1}_{\{0\le x \le y\},\infty}\cap
\mcC^{<0}_{\{0\le x \le y\},0}\;, \qquad \psi_2, \varphi \in
\mcC^{<-1}_{\{0\le x \le y\},\infty}
 \;,
\ee
be a solution on $\mcU$ of  the following system of equations:
\begin{equation}
\label{S1} \left \{
\begin{array}{c}
\partial_{y}\varphi+B_{\varphi\varphi}\varphi+ B_{\varphi\psi}\psi = L_{\varphi\varphi}\varphi + L_{\varphi\psi} \psi + a + G_\varphi \\
\partial_{x}\psi+B_{\psi\varphi}\varphi+B_{\psi\psi}\psi=L_{\psi\varphi}\varphi + L_{\psi\psi}\psi+ b + G_\psi
\end{array}
\right. \;,
  \end{equation}
with  the operators
\bel{HLdelta0} L_{ij}= L_{ij}^{A}\partial_A + xL_{ij}^y
\partial_y + xL_{ij}^x\partial_x
\ee
satisfying
  \be
  L_{\varphi\varphi}^{\mu}\in x^\delta\Axyd\;, \quad L_{\psi\varphi}^{\mu}\;,
  L_{\varphi\psi}^{\mu}\;, L_{\psi\psi}^{\mu}\in \Axyd
\label{HLdelta} \ee (no symmetry hypotheses  are made on the
matrices $L^\mu_{ij}$), while
{\rm
  \begin{deqarr}&
    B_{\varphi\varphi}\in C_\infty(\overline \Omega) +x^\delta\Axyd
\;,\qquad B_{\varphi\psi}, B_{\psi\psi}, B_{\psi\varphi}\in \Axyd
\;,& \label{H3first}\arrlabel{H3}
\\ &a,b \in x^{-1+\delta}\Axyd\;,&
\label{H3SEC}
\\ &\varphi|_{x=y} =\zvarphi\in x^{-1+\delta}\Axd\;,\
\psi|_{x=y} =\zpsi\in x^{-1+\delta}\Axd\;.&
\heqno\label{H4first}\arrlabel{H4}
  \end{deqarr}}
If the non-linear terms $G_\varphi$, $G_\psi$ satisfy the \NLc, then
$$(\psi,\varphi)\in \Axyd\times
x^{\delta-1}\Axyd\;;$$ more precisely
\minilab{stsn}\begin{equs}\label{stst1n}\psi &\in x^{\delta} \Axyd+ \Ayd\;,\\
   \varphi&\in x^{\delta-1} \Axd + x^{\delta-1} y
   \Axyd\;.\label{stst2n} \end{equs} In particular for any $\tau>0$ we
have $$(\psi,\varphi)|_{\{y\ge \tau\}} \in \Axd\times x^{\delta-1}
\Axd\;,$$ which shows that the solution is polyhomogeneous with
respect to $\{x=0\}$ on $\{y\ge\tau\}$.
\end{Theorem}

\proof This theorem is a   generalization of the semi-linear case,
Theorem 3.7 of~\cite{ChLeski}, and can be proved by following step
by step the proof given there. A detailed exposition can be found
in~\cite{WafoPhD}.
\qed

\section{Propagation of polyhomogeneity for the Einstein-Maxwell equations}
\label{SaEMe}

Let us show that Theorem~\ref{Tmain} applies to the source-free
Einstein-Maxwell equations; we will make extensive appeal
to~\cite{CCL}. More generally, consider a system of second
order wave equations of the form
\begin{equation}
\label{3.6}
\eta^{\alpha\beta}\frac{\partial^{2}f}{\partial x^{\alpha}\partial x^{\beta}%
}=-H^{\alpha\beta}(x^\mu,f,\partial f,\partial
\partial f)\frac{\partial^{2}f}{\partial x^{\alpha}\partial
x^{\beta}}+F(f,\partial f,x^\mu)\;,
\end{equation}
for a map $f$ with values in $\R^N$ for some $N$, where $\eta$
is the $(n+1)$--dimensional Minkowski metric. (The map $f$ in
this section should \emph{not} be confused with the map $f$
appearing in \eq{fdef}, compare \eq{coresp} below.)  The
Einstein-Maxwell equations in the harmonic-Lorenz gauge can be
written in this form, with
$f:=(g_{\mu\nu}-\eta_{\mu\nu},A_\mu)$, then
$$
H^{\mu\nu}:=g^{\mu\nu}-\eta^{\mu\nu}= \eta^{\mu \alpha}
\eta^{\nu\beta}(g_{\alpha\beta}-\eta_{\alpha\beta}) +  \mbox{
quadratic terms}
$$
depends only upon $g_{\mu\nu}-\eta_{\mu\nu}$, while $F$ is a
quadratic form in $\partial f$ with coefficients depending upon
$g_{\mu\nu}-\eta_{\mu\nu}$. Thus, in the Einstein-Maxwell case
the source function $F$ has a uniform zero of order two, while
the functions $H^{\mu\nu}$ all have a uniform zero of order
one.

As in~\cite{CCL}, and similarly%
\footnote{In~\cite{CCL} one works within $I_{\eta,x}^{+}(0)$, while
in~\cite{ChLengardnwe} the complement of $I_{\eta,x}^{+}(0)$ is
considered. However, the methods of ~\cite{ChLengardnwe} apply to
both situations.}
%
to~\cite{ChLengardnwe}, we use a mapping $\phi:x\mapsto y$ from the
of the future timelike cone with vertex $0$,
$I_{\eta,x}^{+}(0)$, of a Minkowski space-time, which we denote
$(\R_{x}^{n+1},\eta_{x})$, into
the past
timelike cone with vertex $0$
of another Minkowski space-time, $(\R_{y}^{n+1}%
,\eta_{y})$, defined by
\begin{equation}
\phi: I_{\eta,x}^{+}(0)\rightarrow \R_{y}^{n+1}\text{ \ \ by
\ }x^{\alpha}\mapsto y^{\alpha}:=\frac{x^{\alpha}}{\eta_{\lambda\mu}%
x^{\lambda}x^{\mu}}\text{ }. \label{ytox}
\end{equation}

It is easy to check that $\phi$ is a bijection from
$
I_{\eta,x}^{+}(0)$ onto
$
I_{y,\eta}^{-}(0)$, with inverse
\begin{equation}
\phi^{-1}:y^{\alpha}\mapsto x^{\alpha}\text{ \ by \ \ \ }x^{\alpha}:=\frac{y^{\alpha}}%
{\eta_{\lambda\mu}y^{\lambda}y^{\mu}} \;.
\end{equation}
Moreover $\phi$ is a conformal mapping between Minkowski metrics:
\begin{equation}
\eta_{\alpha\beta}dx^{\alpha}dx^{\beta}=\Omega^{-2}\eta_{\alpha\beta
}dy^{\alpha}dy^{\beta}
\;, %
\end{equation}
where $\Omega$ is a function defined on all $\R_{y}^{n+1}$, given by
\begin{equation}
\Omega:=-\eta_{\alpha\beta}y^{\alpha}y^{\beta}\;.
\end{equation}

We work within $I^-_{y,\eta}(0)$ and to the future of a
hypersurface
$$
\hyp_{\tau_0}:=\{y^0=\tau_0\}\;, \quad  \tau_0 <0 \;,
$$
where we set
$$
\rho \equiv |\vec y|:= \sqrt{\sum_{i=1}^n (y^i)^2}\;, \quad x:=
-|\vec y|-y^0\ge 0\;,\quad y:= y^0-|\vec y|+1 \ge  0 \;,
$$
so that
$$
\Omega=-x(1-y)\;,\quad \partial_x = -\frac 12
\Big(\sum_{i=1}^n\frac{y^i}{|\vec y|} \frac{\partial}{\partial y^i}
+ \frac{\partial}{\partial y^0}\Big) \;,\quad \partial_y = -\frac 12
\Big(\sum_{i=1}^n\frac{y^i}{|\vec y|} \frac{\partial}{\partial y^i}
- \frac{\partial}{\partial y^0}\Big) \;,
$$
and
$$
y^\alpha\frac{\partial}{\partial y^\alpha} = (y-1)\partial_y + x
\partial_x \;.
$$
Furthermore the flat d'Alembertian $ \Box_{\eta,y}$ associated with
the coordinates $y^\mu$ equals
$$
\Box_{\eta,y} = 4 \partial_x \partial_y   - \frac{2(n-1)}
{1-x-y}\Big( {\partial_x}+{\partial _y}\Big) +
\frac{4\Delta_h}{(1-x-y)^2} \;,
$$
where $\Delta_h$ is the canonical Laplacian on $S^{n-1}$.

It should be kept in mind that we are interested in $x$ small and
$y$ bounded away from one.

The general relation between the wave operator on scalar functions
in two conformal metrics transforms the left-hand-side of \eq{3.6}
into the following partial differential operator
\begin{equation}
\label{3.7}
\eta^{\alpha\beta}\frac{\partial^{2}(\Omega^{-\frac{n-1}{2}}f\circ\phi^{-1}%
)}{\partial y^{\alpha}\partial y^{\beta}}\equiv\Omega^{-\frac{n+3}{2}}%
(\eta^{\alpha\beta}\frac{\partial^{2}f}{\partial x^{\alpha}\partial x^{\beta}%
})\circ\phi^{-1}\;. \end{equation}
We introduce the following new set of scalar functions on $\R_{y}^{n+1}$%
\begin{equation}
 \label{23I.1}
\hat{f}:=\Omega^{-\frac{n-1}{2}}f\circ\phi^{-1}\;,
\end{equation}
so that the system \eq{3.6} reads
\begin{equation}
\label{3.8} \eta^{\alpha\beta}\frac{\partial^{2}\hat{f}}{\partial
y^{\alpha}\partial
y^{\beta}}=-\Omega^{-\frac{n+3}{2}}\Big\{H^{\alpha\beta}(x,f,\partial
f,\partial\partial f)\frac{\partial^{2}f}{\partial
x^{\alpha}\partial x^{\beta}}-F(x,f,\partial f)
\Big\}\circ\phi^{-1}\;,
\end{equation}
and we need to analyse the  structure of the right-hand side.
As calculated in detail in~\cite{CCL}, if we set
\begin{equation}
\label{3.9} A_{\mu}^{\alpha}:=\frac{\partial y^{\alpha}}{\partial
x^{\mu}}\circ\phi ^{-1} \equiv
 -\Omega\delta_{\mu}^{\alpha}-2y^{\alpha}\eta_{\mu\beta}y^\beta\;,
\end{equation} which is bounded on any
bounded set of $\R_{y}^{n+1}$, we can write
\begin{equation}
\label{3.10}
\frac{\partial f}{\partial x^{\mu}}\circ\phi^{-1}=A_{\mu}^{\alpha}%
\frac{\partial(f\circ\phi^{-1})}{\partial y^{\alpha}}= \Big(-x(1-y)
\frac{\partial}{\partial y^\mu} - 2 \eta_{\mu\alpha}y^\alpha(
(y-1)\frac{\partial}{\partial y} +x\frac{\partial}{\partial
x})\Big)f\circ\phi^{-1}\;.
\end{equation}
(We emphasise the occurrence of a factor of $x$ in front of
each derivative except $\partial_y$; that will play an
important role in what follows.) Similarly
\bean \frac{\partial^{2}f}{\partial x^{\lambda}\partial
x^{\mu}}\circ\phi ^{-1} &= & \Big\{x^2(1-y)^2 \frac{\partial^{2}
}{\partial y^{\lambda}\partial y^{\mu}} +4x(1-y)
\eta_{\alpha(\lambda}y^\alpha\frac{\partial}{\partial
y^{\mu)}}\left(x\partial_x+(y-1)\partial_y \right)
\\
\nonumber && \quad +4\eta_{\lambda \alpha}\eta_{\mu \beta} y^\alpha
y^\beta \big((y-1)\partial_y+x\partial_x\big)^2
\\
&& +\Big[
4\eta_{\lambda \alpha}\eta_{\mu \beta} y^\alpha y^\beta
+2x(1-y)\eta_{\lambda\mu}
\Big] \big((y-1)\partial_y+x\partial_x\big) \Big\}f\circ\phi^{-1}
\;. \phantom{xxxxxxx} \label{secderyy} \eea
%
%
If we now set $f\circ\phi^{-1}=\Omega^{{\nmoh }}\hat{f}$, we find:
\begin{equation}
\frac{\partial(f\circ\phi^{-1})}{\partial y^{\alpha}}\equiv\frac
{\partial(\Omega^{{\nmoh }}\hat{f})}{\partial
y^{\alpha}}=\Omega^{{\nmoh }}\frac {\partial\hat{f}}{\partial
y^{\alpha}} - {(n-1)}\Omega^{{(n-3)/2}}y_{\alpha}\hat{f} \;,
\end{equation}
and
\bean
\frac{\partial^{2}(f\circ\phi^{-1})}{\partial y^{\alpha}\partial y^{\beta}%
} &\equiv& \big(x(1-y)\big)^{{\nmoh
}}\frac{\partial^{2}\hat{f}}{\partial y^{\alpha}\partial y^{\beta}}
-
{(n-1)}\big(x(1-y)\big)^{{(n-3)/2}}\left(\eta_{\beta\sigma}y^\sigma\frac{\partial\hat{f}}{\partial y^{\alpha}%
}+\eta_{\alpha\sigma}y^\sigma\frac{\partial\hat{f}}{\partial
y^{\beta}}\right)
\\
&&+{(n-1) \over 2}\big(x(1-y)\big)^{{\nmfh }}%
D_{\alpha\beta}\hat{f} \;, \eea
with
\begin{equation}
D_{\alpha\beta}:=2{(n-3) } \eta_{\lambda \alpha}\eta_{\mu \beta}
y^\lambda y^\mu+2\big(x(y-1)\big)\eta_{\alpha\beta} \;.
\end{equation}
Collecting all this, we conclude that
\bean \frac{\partial^{2}f}{\partial x^{\lambda}\partial
x^{\mu}}\circ\phi ^{-1} &=& \big(x(1-y)\big)^\nmoh \Bigg\{x^2(1-y)^2
\frac{\partial^{2} }{\partial y^{\lambda}\partial y^{\mu}} +4x(1-y)
\eta_{\alpha(\lambda}y^\alpha\frac{\partial}{\partial
y^{\mu)}}\left(x\partial_x+(y-1)\partial_y \right)
\\
\nonumber & & +4\eta_{\lambda \alpha}\eta_{\mu \beta} y^\alpha
y^\beta ((y-1) \partial_y+x\partial_x)^2
+2(n-1)x(1-y)y^\alpha\eta_{\alpha (\lambda}\frac{\partial}{\partial
y^{\mu)}}
\\
&&  +\Big[ 4n\eta_{\lambda \alpha}\eta_{\mu \beta} y^\alpha y^\beta
+2x(1-y)\eta_{\lambda\mu} \Big] ((y-1)\partial_y+x\partial_x)
\nonumber
\\
&& +(n-1)\big[(n+1)\eta_{\alpha\mu}\eta_{\lambda\beta}y^\alpha
y^\beta+x(1-y)\eta_{\lambda\mu}\big]  \Bigg\}\hat f \;.
\label{secderyyh} \eea

The second term on the right-hand side of \eq{3.8} is
\begin{equation}
\Omega^{-\frac{n+3}{2}} F\Big(f,\frac{\partial f}{\partial x}
\Big)\circ\phi^{-1} = \Omega^{-\frac{n+3}{2}}
  F\Big(
\Omega^{\nmoh }
\hat{f},\Omega^{\nmoh }A_{\mu}^{\alpha}
(\frac{\partial\hat{f}}{\partial y^{\alpha}} -
{(n-1)}\Omega^{-1}y_{\alpha}\hat{f} )\Big) \label{firtdernon} \;.
\end{equation}
Now, $A^\alpha_\mu y_\alpha= \Omega y_\mu$, and   it follows
from \eq{3.10} that   the right-hand side of the last equation
can be rewritten as
\begin{eqnarray}
\lefteqn{
\big(x(1-y)\big)^{-\frac{n+3}{2}}
\times}
&&
 \nonumber
\\
 &&
  \!\!\!\!\!\!\!\!
  F\Big(\big(x(1-y)\big)^{\frac
{n-1}{2}}\hat{f},\big(x(1-y)\big)^{\frac {n-1}{2}}\Big(x(y-1)
\frac{\partial \hat f}{\partial y^\mu} - 2
\eta_{\mu\alpha}y^\alpha((y-1) {\partial_y}\hat f +x\ {\partial
_x}\hat f)-(n-1)y_{\mu}\hat{f} \Big)\Big) \;.
\nonumber
\\
&&\label{3.6x}
\end{eqnarray}
As shown in~\cite{ChLengardnwe,ChLeski}, the left-hand-side of
\eq{3.6} can be brought to the form needed in Theorem~\ref{Tmain} by
setting
\bel{coresp} \psi_1= \hat f\;,\quad \psi_2 = (\partial_y \hat f,
\partial_A \hat f)\;, \quad \varphi = \partial_x \hat f\;. \ee
Here $\partial_A f= \partial_{v^A}f$, where the $v^A$'s are local
coordinates on the sphere. To bring \eq{3.6x} to the desired form
\eq{gdef}, the choice
$$
p_2\delta= \frac{n+3}2\;, \quad q_2\delta= \frac {n-3}2 \;, \quad
$$
provides the supplementary power of $x$ needed in the arguments of
$F$  to satisfy the structure conditions of Theorem~\ref{Tmain},
provided that we choose $1/(2\delta) \in \N^*$ in even
space-dimensions; any $1/\delta\in \N^*$ is admissible in odd ones.
If we assume that $F$ has a uniform zero of order $m_2$, condition
\eq{cocoi} will now be satisfied for
\bel{m2cond} m_2 > \frac{n+1}{n-3}=1+\frac 4 {n-3} \quad
\Longleftrightarrow \quad n \ge 4 \ \mbox{ and } m_2 \ge \left\{
       \begin{array}{ll}
         6, & \hbox{$n=4$;} \\
         4, & \hbox{$n=5$;} \\
         3, & \hbox{$n=6,7 $;} \\
         2, & \hbox{$n\ge 8$.}
       \end{array}
     \right.
\ee
(In the Einstein-Maxwell case we have $m_2=2$, which enforces
$n\ge8$.)

Let us turn our attention to the first term at the right-hand
side of \eq{3.6}. In what follows we will consider the
following restricted class of non-linearities: we   assume
that, after replacing $f$ by $ \Omega^{\nmoh }\hat f$ and
changing variables $x^\mu \to y^\mu$ as above, the terms
$H^{\alpha\beta}$ takes the form
\bel{Hform} H^{\alpha\beta}  = G^{\alpha\beta}( \Omega^{\nmoh } \hat
f,\Omega^{\nmoh +1} \partial_{y^\mu} \hat f,\Omega^{\nmoh +2}
\partial_{y^\nu} \partial_{y^\rho}\hat f) \;, \ee
%
%
%
%
%
%
%
%
%
%
%
%
%
%
%
%
with a uniform zero of order $m_0$.  Such a structure will clearly
be obtained from a function in \eq{3.6} which depends only upon $f$,
in particular this will be the case for the Einstein or the
Einstein-Maxwell equations, with $m_0=1$.

Using \eq{secderyyh} we can write
\beaa 
{ \Omega^{-\npth }H^{\mu\nu}\partial_{x^\mu}\partial_{x^\nu} f =
x^{-\npth+\nmoh }F_1(H^{\alpha\beta}, \hat
f,\partial_{y^\alpha}\partial_{y^\beta} \hat f, \partial_{y^\gamma}
\hat f)}&& \eeaa
where $F_1$ is  linear in the second, third, and fourth argument.
Assuming \eq{Hform}, this can be rewritten as
\beaa
\Omega^{-\npth }H^{\mu\nu}\partial_{x^\mu}\partial_{x^\nu} f &=&
x^{-\npsh }F_1(H,x^{\nmoh+2}\hat
f,x^{\nmoh+2}\partial_{y^\alpha}\partial_{y^\beta} \hat
f,x^{\nmoh+2}\partial_{y^\gamma} \hat f)
\\ &=&
x^{-\npsh  }F_2(x^{\nmoh}\hat
f,x^{\nmoh+2}\partial_{y^\alpha}\partial_{y^\beta} \hat
f,x^{\nmoh+1}\partial_{y^\gamma} \hat f)
  \;,
\eeaa
where $F_2$ has a uniform zero of order $m_1=m_0+1$. With the
restrictions on $\delta $ as before, we will obtain the right
structure by setting
$$
p_1\delta= \frac{n+7}2\;, \quad q_1\delta= \frac {n-1}2 \;,
$$
and the \NLc\  will hold provided that $m_1:=m_0+1$ satisfies
\bel{m1cond} m_1 > \frac{n+5}{n-1}=1+\frac{6} {n-1}\;, \quad
\Longleftrightarrow \quad
m_0 \ge \left\{
       \begin{array}{ll}
         7, & \hbox{$n=2$;} \\
         4, & \hbox{$n=3$;} \\
        3, & \hbox{$n=4$;} \\
        2, & \hbox{$n=5,6,7$;} \\
         1, & \hbox{$n\ge 8$.}
       \end{array}
     \right.
\ee
In particular the structure conditions will be satisfied by the
Einstein-Maxwell equations in space-dimensions larger than or equal
to eight.

The hypothesis \eq{Hform} will not be satisfied in general if
$H^{\mu\nu}$ in \eq{3.6} is a non-linear function of $f$ and
$\partial_{x^\mu}  f$, for then $H$ will belong instead to the
following class of functions  (compare \eq{firtdernon})
\bel{Hform2} H  = G( \Omega^{\nmoh } \hat f,\Omega^{\nmoh }
\partial_{y^\mu} \hat f,\Omega^{\nmoh+1 }  \partial_{y^\nu}
\partial_{y^\rho}\hat f) \;, \ee
An analysis similar to the one above shows that, for $H^{\mu\nu}$'s
which are a finite sum of terms of the form \eq{Hform2}, we will
obtain the right structure by setting
$$
p_1\delta= \frac{n+5}2\;, \quad q_1\delta= \frac {n-3}2 \;, \quad
$$
and the \NLc\  will hold provided that $m_1=m_0+1$ satisfies
\bel{m1condx} m_1 > \frac{n+3}{n-3}=1+ \frac{6}{n-3} \quad
\Longleftrightarrow \quad n \ge 4 \ \mbox{ and } m_0 \ge \left\{
       \begin{array}{ll}
         7, & \hbox{$n=4$;} \\
         4, & \hbox{$n=5,$;} \\
        3, & \hbox{$n=6$;} \\
        2, & \hbox{$n=7,8,9$;} \\
         1, & \hbox{$n\ge 10$.}
       \end{array}
     \right.
\ee

The reader should have no troubles similarly working out the
conditions on the nonlinearity for general $H$'s which depend
on $f$, $\partial_{x^\mu} f$ and $\partial_{x^\mu}
\partial_{x^\nu} f$.

Summarizing, we have proved:

\begin{Theorem}
 \label{Twavephg}
Let $f$ be a solution of equation \eq{3.6}, define $\psi_1$,
$\psi_2$, and $\varphi$  by \eq{coresp}, where $\hat f$ is
given by \eq{23I.1}. Suppose that \eq{m2cond} holds, and assume
that \emph{either} \eq{Hform} with \eq{m1cond} hold, \emph{or}
\eq{Hform2} with \eq{m1condx} hold. If \eq{23I.2} and
\eq{H4first} hold, then the conclusions of Theorem~\ref{Tmain}
apply. In particular Theorem~\ref{Tmain} applies to the
Einstein-Maxwell equations in space-time dimensions $n+1\ge 9$.
\end{Theorem}

\section{Towards solutions with a polyhomogeneous Scri}
\label{Sgsps}

In order to establish existence of solutions of the vacuum
Einstein equations, in sufficiently high dimensions, with a
polyhomogeneous Scri, it remains to construct appropriate
initial data, and show that the corresponding solutions are in
the right function spaces.

Recall, now, that  large classes of polyhomogeneous
hyperboloidal initial data have been constructed
in~\cite{AndChDiss} (the emphasis in that reference is on $n=3$
at several places, but the general results there show that the
conformal method, starting from smooth or polyhomogeneous seed
fields, provides polyhomogeneous solutions of the general
relativistic vacuum constraint equations in any dimension $n\ge
3$). There is little doubt that large collections of  initial
data so constructed provide polyhomogeneous data for the
harmonically reduced equations of the last section, but we have
not checked this in  detail.
Instead, we will follow the standard-by-now strategy of using
initial data which are stationary outside of a compact set. So,
in Section~\ref{sSId}, we provide large classes of
Corvino-Schoen type initial data with polyhomogeneous
asymptotics on hyperboloids. One of the reasons for proceeding
this way is that small such initial data lead to global,
geodesically complete
solutions~\cite{Loizelet:these,Loizelet:AFT}.

One then needs to verify that the associated solutions satisfy
the space-time weighted regularity conditions needed in
Theorem~\ref{Tmain}. One could hope that the
Lindblad-Rodnianski type estimates of
Loizelet~\cite{Loizelet:these,Loizelet:AFT} would provide that
information.  It turns out that the available estimates, for
space-times obtained by evolving small initial data of
Section~\ref{sSId}, are not sufficient for our polyhomogeneity
result; this is analyzed in Section~\ref{sLLR}. This means that
the desired estimates have to be derived from scratch, which
will be done in the remainder of this paper.

\subsection{Stationary vacuum metrics in higher dimensions}
\label{sSst}

The only way, so far, of obtaining space-times with controlled
asymptotic behavior near $i^0$ is to use initial data sets
which are stationary at large distances. We will outline the
construction of such data in Section~\ref{sSId}, but before
doing this it is convenient to start with a short discussion of
stationary metrics in higher dimensions; our presentation
follows~\cite{ChBeig3}.

Consider a vacuum Lorentzian metric ${}^{n+1}g$ in any
space-time-dimension $n \geq 3$, with Killing vector
$X=\partial/\partial t$. In the region where $X$ is timelike there
exist adapted coordinates in which ${}^{n+1}g$ takes the form
\beal{gme1} &^{n+1}g = -V^2(dt+\underbrace{\theta_i
dx^i\!}_{=\theta}\;)^2 + \underbrace{g_{ij}dx^i dx^j}_{=g}\;, & \\ &
\partial_t V = \partial_t \theta = \partial_t g=0
\;. \eeal{gme2}
The vacuum Einstein equations (with vanishing cosmological constant)
read (see, e.g.,~\cite{Coquereaux:1988ne})
\begin{equation}\label{mainequation}
\left\{\begin{array}{l} V\nabla^*\nabla V=\frac 1{4} |\newF|_g^2\;,\\
     \Ric(g)-V^{-1}\Hess_gV=\frac{1}{2V^{2}}\newF\circ \newF\;,
\\
\divv  (V \newF)=0\;,
\end{array}\right.
\end{equation}
where
$$
\newF_{ij}=-V^2(\partial_i \theta_j - \partial_j
\theta_i)\;,\;\;\;(\newF\circ \newF)_{ij}=\newF_i{^k}\newF_{kj}\;.
$$
We assume that there exists $\alpha>0$ such that
\bel{foff}
  g_{ij}-\delta_{ij}=O(r^{-\alpha})\;, \ \partial_k
g_{ij}=O(r^{-\alpha-1})\;,
\ee
similarly for $V-1$ and $\theta_i$. A redefinition  $t\to t+ \psi$,
introduces a gauge transformation
$$
\theta \to \theta + d\psi \;,
$$
and one can exploit this freedom to impose restrictions on $\theta$.
For our purposes it is convenient to impose the \emph{harmonic
gauge}, $\Box t=0$, which reads
\bel{harmonic}
\partial_i(\sqrt{\det g} V g^{ij} \theta_j)=0
\;. \ee
Equation~\eq{harmonic} can always be achieved by replacing $\theta$
by $\theta+d\psi$, and  solving the resulting  linear equation for
$\psi$, \emph{cf.,
e.g.},\/~\cite{choquet-bruhat:christodoulou:elliptic,Bartnik} for
the relevant isomorphism theorems.) One can then introduce new
coordinates~\cite{Bartnik}  which are harmonic for $g$.

In space-harmonic coordinates, and in the gauge \eq{harmonic}, the
system \eq{mainequation} is elliptic, and  standard  considerations
show that the functions $g_{ij}$, $V$ and $\theta_i$ have a
polyhomogeneous expansion in terms of $\log r$ and inverse powers of
$r$. Furthermore, $^{n+1}g_{\mu\nu}$ is Schwarzschild in the leading
order, and there exist constants $\alpha_{ij}$ such that
$$
\theta_i = \frac {\alpha_{ij}x^j}{r^n} + O(r^{-n}) \;.
$$

It is of interest to enquire whether or not the logarithmic
powers are essential in the polyhomogeneous expansion. It has
long been know in space-dimension three that, for metrics which
are stationary and vacuum in the asymptotic region, coordinate
systems \emph{exist} where no $\log r$ terms arise whenever the
ADM mass is non-zero~\cite{SimonBeig}. The same property is
true for \emph{static} solutions with non-zero ADM mass in
space-dimension four~\cite{ChBeig3}. Now, in the evolution
theorems used below we need all coordinates to satisfy the wave
equation,
\bel{harmcond} \Box x^\mu = 0 \;, \ee
and the transition from the coordinates used in~\cite{ChBeig3} to
the coordinates satisfying \eq{harmcond} might introduce log terms:
This is exactly what happens for the Schwarzschild metric in $n=4$,
which  \emph{does have} a logarithmic term in its asymptotic
expansion in a natural choice of wave coordinates~\cite{CCL}, but
this is the only dimension where this happens for Schwarzschild.

In general, \eq{harmcond} is achieved by changing
space-coordinates $x^i \to x^i + \psi^i(x^j)$ (recall that $t$
is already harmonic), thus solving a linear equation for
$\psi^i$;  by standard results (see, e.g.,~\cite{ChAFT}) the
$\psi^i$'s will have a full asymptotic expansion in terms of
powers of $\ln r$ and inverse powers of $r$, and so will the
space-time metric in the new coordinate system, when
transformed from the space-harmonic ones. In view of the
calculations in \cite{CCL}, this implies the existence of
polyhomogeneous asymptotics of the initial data on hyperboloids
at $\scri$, as needed in Theorem~\ref{Tmain}.

Rather surprisingly, {in even space-dimensions} larger than or
equal to six the space-coordinates used in~\cite{ChBeig3}
satisfy \eq{harmcond}, and so does the time coordinate. It
follows that the analysis of stationary solutions
in~\cite{ChBeig3} directly provides wave coordinates in which
no log terms occur in those dimensions.

\subsection{Corvino-Schoen data in higher dimensions}
\label{sSId}

So far we have considered metrics which are exactly stationary.
Now, there exists a construction due to Corvino and
Schoen~\cite{CorvinoSchoen2,Corvino} (see
also~\cite{ChDelay,ChDelay2}, and also the more recent
Reference~\cite{CCI2}, where the construction is carried out
under considerably weaker asymptotic conditions) which allows
one to glue   exactly stationary ends to  asymptotically
Euclidean initial data sets. Some details of this construction
have been presented in those references in dimension three
only, but the construction generalises to any dimension, as
follows: Recall that the construction requires a family of
stationary reference metrics which cover the whole range of
asymptotic charges. In dimension $3+1$ this is provided by the
family of metrics obtained by boosting and translating the Kerr
metrics. In higher dimensions one such  family can be obtained
by boosting and translating the Myers-Perry
metrics~\cite{myersperry}. Note that the question, whether or
not the reference solutions have naked singularities is
irrelevant for the problem at hand because here one only needs
the solutions at large distances. (Similarly to the Kerr
family, all the metrics in the family so obtained have a
timelike ADM momentum, and therefore can only be glued to
asymptotically flat initial data which also have this property;
this is no restriction for well behaved initial data sets which
are spin, or for space-dimensions up to seven, and is expected
not to be a restriction for well behaved initial data sets in
general, but this has not been proved at the moment of writing
of this work.)

So let $R_x$, $\epsilon_k$ be positive constants and consider
the collection, say ${\bf \large C}_{R_x,\epsilon_k}$ of
general relativistic electro-vacuum initial data sets
$(\R^n,g,K)$ which are stationary outside a coordinate ball
$B(R_x)$ and with weighted Sobolev norm controlling
$k$-derivatives of the metric smaller than $\epsilon_k$. Here
$k$ should be sufficiently large as in~\cite{Loizelet:AFT,CCL},
and the norm should be the one described in those references.
From what has been said this collection is non-empty, and
contains an open set (in the topology associated to the norm)
around Minkowski space-time.

Now, for the Schwarzschild metric in dimension $n+1$ with $n\ge
4$, and in harmonic coordinates, the boundary of the domain of
influence of a ball is sandwiched between two hypersurfaces
$t-r=\const$   \cite[Section~5.3]{CCL}. This remains true for
stationary electro-vacuum metrics because the leading order
behaviour of the metric coincides with the Schwarzschild one
(compare~\cite[Appendix~A]{poorman2}). This implies that the
maximal globally hyperbolic development of  all initial data in
${\bf \large C}_{R_x,\epsilon_k}$ contains hyperboloidal
hypersurfaces, the asymptotic region of which is contained in
that part of the space-time where the metric is stationary. So
our considerations of the previous section apply to this
region, leading to polyhomogeneous initial data on such
hypersurfaces. Since the leading order deviation of the metric
from the flat one is Schwarzschildian, the tensor field
$\hat{h}:= \Omega^{-\frac{n-1}2}(g-\eta)$, that plays a key
role in our analysis, is $O(x^{(n-2) - (n-1)/2}) =
O(x^{(n-3)/2})$, and in fact
\bel{1IX0.8}
 \hat{h}\in x^{(n-3)/2}(\Axd\cap L^\infty)
  \;,
\quad
 \partial_{y^0} \hat{h}\in x^{(n-5)/2}(\Axd\cap L^\infty)
 \;,
\ee
with $\delta = 1$ on any hyperboloid whose asymptotic part is
contained in the stationary region.

\subsection{Lindblad-Rodnianski-Loizelet metrics near $\scri$}
\label{sLLR}

In this section we analyze how the asymptotic behavior of  the
small-data space-times constructed in~\cite{Loizelet:these}
(compare~\cite{LindbladRodnianski,LindbladRodnianski2}) relates
to the differentiability conditions needed in
Theorem~\ref{Tmain}. We find that sharper decay rates along
outgoing null geodesics would be needed for a direct proof of
polyhomogeneity using our approach. The estimates established
here are then combined with the results of our analysis in
subsequent sections to provide a rather more involved proof of
polyhomogeneity.

We start by recalling some notation
of~\cite{Loizelet:these,LindbladRodnianski,LindbladRodnianski2}. Let
$\mathcal{Z}$ denote the following set of vectors on Minkowski
space-time:
$$
\p_\al\equiv \frac \p {\p x^\alpha}, \; \al = 0,1,\ldots,n;
$$
$$
Z_{\al\bet} = x_\al\p_\bet-x_\bet\p_\al,\; \al,\bet = 0,1,\ldots,n;
$$
$$ Z_0 = \sum_{\al = 0}^n x^\al\p_\al= t\p_t + \sum_{i = 1}^n x_i\p_i=t\p_t+r\p_r\; .$$
Here, as usual, $ x_0 = -x^0 = -t$, $   x_i = x^i $ for $ i = 1
\ldots,n$.
Let the spherical coordinates $(r,\theta^A)$ be defined as
\bel{20VIII0.7}
  \left\{ \begin{array}{l}  t= x^0  \;,
    \\ r=  \big( \sum\limits_{i=1}^n (x^i)^2\big)^{1/2}\;,
    \\ x^i =r\omega^i(\theta^A), \; i=1,\ldots,n ,\end{array}
 \right.
\ee
where $\theta^A$ denotes any local coordinates on the sphere
$S^{n-1}$. The vector fields
$$ L = \p_t+\p_r= \p_t +  \omega^i \partial_i\;, \qquad \underline{L} = \p_t-\p_r= \p_t -\omega^i \partial_i\;.
$$
are tangent, respectively transverse, to the light cones $t-r =
\const$. We note
$$
Z_0 = t\p_t + r\p_r\;.
$$
Furthermore, the  $Z_{ij}$'s, $i,j = 1 \ldots,n$ are tangent to the
spheres $S^{n-1}\subset \mathbb{R}^n$, and can be purely expressed
in terms of the $\theta^A$'s.

Let $T\ge 0$, set $  T^\mu =(T,0, \ldots,0) $, in this section it is
more convenient to consider instead the following variation of
\eq{ytox}:
\bel{ytox2} y^\mu = \frac{x^\mu\ppT^\mu}{(x^\al\ppT^\al)(
x_\al\ppT_\al)} \quad \Longleftrightarrow \quad x^\mu\ppT^\mu =
\frac{y^\mu}{y^\al y_\al} \;. \ee
This provides a conformal transformation from the future causal cone
centred at $T^\mu$ in the Minkowski space-time  with coordinates
$x^\mu$ to the past causal cone of the origin in the Minkowski
space-times with coordinates $y^\mu$, and with conformal factor $\Om
= y^\al y_\al = \frac{1}{-(\tmT)^2+r^2}.$

To make contact with Section~\ref{SPos} we set
$$
x = -y^0 - \rho, \quad y =  y^0 - \rho+1 \quad \text{where} \quad
\rho = \big(\sum_{i=1}^n (y^i)^2\big)^{1/2}
  \;,
$$
so that
\bel{20VIII0.6}
 \left\{ \begin{array}{l}  y^0 =\frac 12 (y-x-1)
    \\ \rho= \frac 12 (-y-x+1)
    \\ y^i =\frac 12 (-y-x+1)\omega^i(v^A), \; i = 1,
        \ldots,n\end{array}\right.
        \;.
\ee
Here  $\omega^i$ is a unit vector, and the $v^A$'s denote local
coordinates on $S^{n-1}$ in the $y$--coordinates. One can take $
\omega^i(\theta^A) = \omega^i(v^A)$, $ i = 1, \ldots,n$; we will
make this choice, and simply write $\omega^i$ in both $x^\mu$ and
$y^\mu$ coordinates.

Letting $\mcH_s$ be the following family of hyperboloids,
$$ \mcH_s =\left\{ x^0-s= \sqrt{s^2+r^2}\right\}
\;, s>0\;,
$$
we will have
$$\phi(\mcH_s)= \{y^0 = -\frac{1}{2s}\}$$
in particular $\phi(\mcH_1)= \{y^0 = -\frac{1}{2}\}$.

The methods of Section~\ref{SPos} involve the vector fields
$$ x\p_x, \; y\p_y, \; \p_{A}= \frac{\p}{\p v^A}, \; A = 1,\ldots n-1  .$$
By straightforward calculations one finds, keeping in mind that $
\rho = \frac{r}{(\tmT )^2-r^2}$ for $\tmT\ge r$,
$$ x = \frac{1}{\tmT +r}, \quad 1-y =\frac{1}{\tmT -r} \quad \Longleftrightarrow \quad r= \frac 1{2x}-\frac 1{2(1-y)} ,
\quad t =\frac 1{2x}+ \frac 1{2(1-y)}\pT $$
\bel{xpxypy} \left\{ \begin{array}{l}  x\p_x =
-\frac 12(\tmT +r) (\partial_t + \p_r)
\;,
\\
(1-y)\p_y 
=\frac 12 (\tmT -r)
  (\partial_t - \p_r)
  \;,
  \\
   \p_A = \text{linear combinations of $Z_{ij}, \; i,j = 1, \ldots,n$
   \;.}\end{array}\right.
\ee
The coefficients in the   equation for $\partial_A$ above depend
only upon the angular variables, and a finite number of coordinate
patches $v^A$ can be chosen so that in each of those patches the
coefficients are uniformly bounded together with derivatives of any
order.

This leads us to

\begin{Proposition}
\label{P1} Let $T,T_0>0$, $t\ge0$
and suppose that
\bel{ranges}
1-T\le t-r \le T_0
  \quad
  \Longleftrightarrow
  \quad
0 \le y \le 1 - \frac 1 {T+T_0}
  \;.
\ee For all $ k\in \;\mathbb{N}, \; \forall \;(i,j,\gamma) \;
\in \; \mathbb{N}\times\mathbb{N}\times\mathbb{N}^{n-1}$
satisfying $i+j+|\gamma| \le k,$ and for any function $f\in
C^k$ we have
\be [x\p_x]^i\p_y^j\p_v^\gamma f =   \sum_{|I|\le k,\; Z \in
\mathcal{Z}}H^{ij\gamma}_I( \theta,y)Z^I f
\ee
with $ |H^{ij\gamma}_I(\theta,y)| \le  C(i,j,I,T,T_0) $ .
\end{Proposition}

\proof Using \eq{xpxypy} one can rewrite $x\partial_x$ and $
\partial_y$ as
\beal{xpx} x\p_x = -\frac 12  (Z_0 -\omega^iZ_{0i}\ppT (\partial_t +
\omega^i
\partial_i))  \;,
\\
  \p_y = \underbrace{\frac  1{2(1-y)}}_{=:\varphi_1(y)} \underbrace{(Z_0 +\omega^iZ_{0i}\ppT
(\partial_t - \omega^i \partial_i))}_{=:\tilde Z}
  \;.
\eeal{ypy}
It is thus clear that $x\p_x$, and any of its powers, have the right
structure. Next, the factor $ \varphi_1(y) $ appearing in \eq{ypy}
is  bounded on any compact subinterval of $[0,1)$ (note that $y=1$
corresponds to the tip of the past causal cone centred at the origin
of the $y^\mu$-coordinates). One easily finds by induction that
$$
\p_y^j = \sum_{i=1}^j \varphi_i(y) \tilde Z^i \;,
$$
where the functions $\varphi_i$ are  bounded on compact subsets of
$[0,1)$, whence the result.
\qed

\medskip

We wish to obtain the asymptotic behavior of the fields
occurring in Theorem~\ref{Tmain} for the global solutions
$$
f:= (h_{\mu\nu},A_{\mu})
$$
of the Einstein-Maxwell equations constructed in
\cite{Loizelet:these}. In order to apply  Theorem~\ref{Tmain}
we need
$$\psi_1= \hat f \in \mcC^{<0}_{\{0\le x\le y\},0}; \; \big(\psi_2=
(\p_y\hat f, \p_A \hat f), \; \vp = \p_x \hat f \big)\; \in
\mcC^{<-1}_{\{0\le x\le y\},\infty} \;, $$
where
$$\hat f = \Om^{-\frac{n-1}{2}} f\circ \phi^{-1}.$$

Now,
$$
\Omega= -x(1-y)
$$
which implies that for any $\alpha \in \R$ we have
\bel{xomc} (x\partial_x)^i(\Omega^{\alpha} f)=
\Omega^\alpha\sum_{j=0}^i C(\alpha,i,j)(x \partial_x)^j f\;. \ee
Similarly,
\bel{yomc} (y\partial_y)^i(\Omega^{\alpha} f)=
\Omega^\alpha\sum_{j=0}^i C'(\alpha,i,j,x,y)(y \partial_y)^j
f\;,
  \qquad
  \partial_y ^i(\Omega^{\alpha} f)= \Omega^\alpha\sum_{j=0}^i C''(\alpha,i,j,x,y)  \partial_y ^j
f\;, \ee
where the functions $C'$ and $C''$ are bounded for $x$ in, say,
$[0,x_0]$,
  and for $y$ bounded away from $1$.

The solutions constructed in~\cite{Loizelet:these} satisfy the
following: there exists  $0<\delta<1/4 $ such that for $t\ge 0$ and
$|t-r|\le C_1$, and for all $I$
there exists a constant $C$, depending upon $I$ and $C_1$, such that
\beal{estder} |Z^I f(t,x^i)| &\le & C  (1+t+r)^{\frac{1-n}2 +\delta}
\;,
  \\
  |
  \bar \p Z^I f(t,x^i)| &\le &
C  (1+t+r)^{\frac{-1-n}2 +\delta} \;, \eeal{estder2}
where
\bel{dbardef} \bar \p\in\Big\{\p_t+\p_r, \ r^{-1}\partial_A\Big\} =
\Big\{ -2x^2\partial_x, \frac{2x(1-y)}{1-x-y} \partial_A\Big\}
  \;.
\ee
Now,
$$ \frac{1+t+r}{\tmT +r} = 1+ \frac{1+T}{\tmT +r}\in\left[ 1, \frac {1+T}T \right]\quad \text{for} \quad T>0,\ t\ge0\;, $$
so \eq{estder}-\eq{estder2}  imply
\beal{estder3} |Z^I f(t,x^i)| &\le & C  x^{\frac{n-1}2 -\delta }
  \;,
  \quad
|\bar \p Z^I f(t,x^i)|  \le C  x^{\frac{n-1}2 +1-\delta } \;. \eea
{}From \eq{xomc}-\eq{yomc} and Proposition~\ref{P1}  we obtain
\beaa
 \lefteqn{
  [x\p_x]^i\p_y^j\p_v^\gamma  \hat f(x,y,v)}
 &&
\\  &=&  [x\p_x]^i\p_y^j\p_v^\gamma  \Om^{\frac{1-n}{2}}
 f(t,x^i)
\\
 &=& \Om^{\frac{1-n}{2}}
 \sum_{0\le m\le i}\sum_ {0\le \ell \le j}c(i,j,m,\ell,n,x,y)[x\p_x]^m\p_y^\ell \p_v^\gamma f(t,x^i)\\
&=&\Om^{\frac{1-n}{2}}\sum_{0\le m\le i}\sum_ {0\le \ell \le
j}c(i,j,m,\ell,n,x,y) \sum_{|I|\le k,\; Z \in \mathcal{Z}}
{H}^{m\ell\gamma}_I( \theta,y) Z^If(t,x^i) \;. \eeaa
Using the first inequality in \eq{estder3} we conclude that for
any $0<\epsilon\le 1$ and for $0\le y \le 1-\epsilon$ we have
$$
  \Big|[x\p_x]^i[y\p_y]^j\p_v^\gamma
  \hat f(x,y,v)\Big|\le   \Big|[x\p_x]^i\p_y^j\p_v^\gamma
  \hat f(x,y,v)\Big|\le C x^{-\delta}\;,
$$
while it should be clear from \eq{dbardef} that the second
inequality in  \eq{estder3} does not provide any new
information in the coordinate ranges assumed above. In any case
the property
\bel{10I.1} \Big(\psi_1= \hat f,\psi_2= (\p_y\hat f, \p_A \hat
f)\Big) \in \mcC^{-\delta}_{\{0\le x\le y\},0}\;,\quad  \vp = \p_x
\hat f   \in \mcC^{-1-\delta}_{\{0\le x\le y\},\infty} \ee
immediately follows. Unfortunately, to apply
Theorem~\ref{Tmain} one would need $\delta$ to be an arbitrary
positive number, while in \eq{10I.1} $\delta$ is a small number
determined by the initial data. So, as already pointed out, we
need to derive  the necessary estimates by different methods.
This is the purpose of the sections that follow.

\section{Weighted energy estimates near a null boundary}
\label{sWee}

Let $(\mcM,\m)$ be an $(n+1)$-dimensional space-time. We
consider systems of quasi-linear  of nonlinear  wave equations,
with diagonal principal part of the form
\bel{E0}
  \Box_\m u = F(.\;,u,\p u) \;,
\ee
on a neighborhood  of a null hypersurface  of $\mcM$. We
suppose that the background metric $\m$ is a smooth function of
the coordinates, of the unknown vector valued function $u$, as
well as its first order derivatives.

All calculations below will be done  for a real valued function
$u$, the result for a vector valued function is obtained by
summing over the components.

\subsection{The hypotheses, and the geometry of the problem}
\subsubsection{The hypotheses}
\label{SH}
We will consider the Cauchy problem associated to  equation
(\ref{E0}), the initial data will be given on a hypersurface
$\mcS_0$. We will evolve these initial data to obtain a
solution of our problem in a \emph{past} one-sided neighborhood
of a null hypersurface
$$
\mcN =
\{ x =0 \}
$$
forming the boundary, or a subset thereof, of the domain of
dependence of $\mcS_0$. Here, and throughout, $x$ stands for a
positive function such that $dx$ has no zeros on $\{ x =0 \}$.
We will be working in   a neighborhood of $\{ x =0 \}$, chosen
so that $x$ is a coordinate there, of the form
$$
{\mcV} \equiv
\left[\tauz ,  \right. \tauone \left[\right.\times\left.
\right]0, x_0 \left[ \right.\times \mcO
  \; ,
$$
where $ \left[\tauz ,  \right. \tauone \left[\right. $
corresponds to the time interval, $\left. \right]0, x_0 \left[
\right.$ the range of the variable $x$, and $\mcO$ is an
$(n-1)$-dimensional compact submanifold of $\mcM$ without
boundary. The coordinates will be denoted by $(\tau, x, v)$,
with $v=(v^A)_{A=1}^{n-1}$ the coordinates on $\mcO$. We assume
that  $\partial_\tau$ is timelike, and we choose the
time-orientation on $\mcM$ such that the vector $\p_\tau$ is
everywhere future directed.

One can think of the set $\mcU$ of \eq{Uset} as a subset of the
coordinate patch above,  compare Figure~\ref{F1}, page
\pageref{F1}.

On the components of the metric $\m$  with respect to the
coordinates $(\tau,x,v)$, we assume the following:
\begin{enumerate}
\item We suppose that
\bel{G0} \exists \ep_0
>0\;, \quad \text{such that} \quad -\m^{\tau\tau} \ge \ep_0 \; \ee
everywhere on ${\mcV}.$
\item The components $\m^{\tau\tau}$  and $\m^{\tau x}$ can
    be written as \bel{H1} \m^{\tau \tau}= -1 + x\f^0(\tau,
    x,v^A) \quad \text{and} \quad \m^{\tau \tau}+\m^{\tau
    x}=   x\f^1(\tau, x,v^A)\ee where the functions $\f^0$
    and $\f^1$ are bounded on bounded sets.
\item On the components  $\m^{xA}$ and $\m^{xx}$ we assume
    that \bel{H2}  \m^{xA} \;  = \text{O}(x)\quad
    \text{and} \quad \m^{\tau\tau}+2\m^{\tau x}+
    \m^{xx}\;=\; 1+ \text{O}(x) \ee
and we set $\m^{xA} = x\f^A $  and $\m^{\tau\tau}+2\m^{\tau
x}+ \m^{xx}=1+ x\f$, where $\f$ and $\f^A$  are bounded
functions  on bounded sets. We further suppose that
$$\m^{\tau\tau}+2\m^{\tau
x}+ \m^{xx}>0
 \;.
$$
\item The vector field
\bel{Ydef}
 Y^\nu\p_\nu:= \p_\tau-\p_x
\ee
 is assumed to be everywhere timelike on ${\mcV}$  and
    future directed. This vector will be used to contract
    the energy momentum tensor.
\end{enumerate}
The set of functions $( \f\;,\f^\mu)$ will be denoted by
$\f^\sharp\; $ and $\m^\sharp$ will denote the inverse matrix
of the matrix $(\m_{\mu\nu})$.
\begin{Remark}%
It follows from the above that the vector $\nabla x$
    (where $\nabla$ is the covariant derivative compatible
    with the metric $\m$) can be decomposed as
\bel{BC}\nabla x = \omega^{(1)} + \bet(x)\omega^{(2)}\ee
where $\omega^{(1)}  $ is  causal future directed,    and that
there exists a constant $C_0$ such that
\bel{betineq}
 |\bet(x)|
 \le C_0 x\;, \ |\omega^{(2)}| \le C_0 |\f^\sharp|
 \;.
\ee
%
%
\end{Remark}

\begin{example}
As an example, consider a conformally rescaled asymptotically
flat solution of asymptotically vacuum Einstein equations in
\emph{Bondi coordinates} near Scri~\cite{Tamburino:Winicour},
with the metric taking the form
\bel{cangnnB} \tilde \m_B=e^{2\beta} dx \otimess d\stau+ \chi
d\stau \otimess d\stau + 2 \gbeta \otimess d\stau + \mu \;, \ee
for some functions  $\beta$  and  $\chi$, and a one-form field
$\gbeta$. (Here $y$ corresponds to the Bondi retarded time $u$,
and $x=1/2r$ is half the inverse of the luminosity distance
$r$. E.g., for the Minkowski metric in any dimensions,
$\beta=\chi=0=\gbeta$.) In $3+1$ dimensions, for smoothly
compactifiable metrics, the Einstein equations imply, for
matter fields decaying sufficiently fast, that $\beta=O(x^2)$
as well as \bel{Bonddec} \chi = O(x^2)\;,\quad \gbeta_A =
O(x^2)\;,\ee with derivatives behaving in the obvious way.
\Eq{Bonddec} remains valid for asymptotically vacuum metrics
which, after conformal rescaling, are polyhomogeneous and $C^1$
(see~\cite[Section~6]{ChMS} or~\cite[Appendix~C.1.2]{CJK}),
while for general $\Axd\cap L^\infty$--polyhomogeneous
asymptotically vacuum metrics one
has~\cite[Equations~(2.15)-(2.19) with $H=X^a=0$]{ChMS} the
asymptotic behaviors $\beta=O(x^2\ln^Nx)$ and \bel{Bonddec2}
\chi = O(x^2)\;,\quad \gbeta_A = O(x^2\ln^Nx)\;,\ee for some
$N$. Here ``asymptotically vacuum'' requires, for
polyhomogeneous metrics, that the components of the
energy-momentum tensor in asymptotically Minkowskian
coordinates satisfy (see~\cite[end of Section~2]{ChMS})
\bel{asvac} T_{\mu\nu}=o(r^{-2})\;.\ee

We have
$$\det \m = - \frac 14 \det \mu
\;,
$$
which, for a Lorentzian metric, shows that $\mu$ must be a
non-degenerate $(n-1)\times (n-1)$ tensor field. It is simple
to check that the inverse metric
$\m^\sharp=\m^{\alpha\gbeta}\partial_\alpha\otimess
\partial_\gbeta$ is given by the formula
\bean  \m^\sharp&=&
{4(-\chi+ |\gbeta|_\mu^2)}
\partial_x \otimess
\partial_x + 4\partial_x  \otimess \partial_\stau
  - 4\gbeta^\sharp \otimess \partial_x + \mu^\sharp
  \\
  &=&
4\partial_x  \otimess \Big(\partial_\stau + {(-\chi+
|\gbeta|_\mu^2)}
\partial_x -\gbeta^\sharp \Big)+ \mu^\sharp
\;, \eeal{cangnni}
with $\mu^\sharp= \mu^{AB}\partial_A \otimess \partial_B$,
where $\mu^{AB}$ is the matrix inverse to $\mu_{AB}$,
$\gbeta^\sharp = \mu^{AB}\gbeta_A\partial_B$,
$|\gbeta|_\mu^2=\mu^\sharp(\gbeta,\gbeta)=\mu^{AB}\gbeta_A\gbeta_B$,
and $\otimess$ denotes the symmetric tensor product. We note
   $$\m(\nabla \stau, \nabla \stau)=\m^{\stau \stau}=0\;,  $$
which makes clear the null character of the level sets of
$\stau$, and implies, by a well-known argument, that the
integral curves of
$$\nabla \stau=\m^{\alpha\gbeta}\partial_\alpha \stau
\partial_\gbeta=\m^{\stau
\gbeta}\partial_\gbeta=2\partial_x $$ are null geodesics.

Consider a new coordinate system $(x, v^A,\tau)$, where
\bel{chofvn} (x,y)\longrightarrow (x,\tau=\frac{y-x}2)\;, \ee
so that \bel{chofvn2}
\partial_x\longrightarrow \partial_x -\frac{1}2\partial_\tau\;,
\quad \partial_y=\frac 12 \partial_\tau \;. \ee
Thus
\bean  \m^\sharp&=&
{4(-\chi+ |\gbeta|_\mu^2)} (\partial_x -\frac{1}2\partial_\tau)
\otimess (\partial_x -\frac{1}2\partial_\tau) + 4(\partial_x
-\frac{1}2\partial_\tau) \otimess (\frac 12 \partial_\tau)
  - 4\gbeta^\sharp \otimess (\partial_x -\frac{1}2\partial_\tau) +
  \mu^\sharp
\;, \eeal{21X.1}
giving
\beal{21X.3} &  \m^{xx}= {4(-\chi+ |\gbeta|_\mu^2)} \;, \quad
\m^{x\tau}= 1 {-2(-\chi+ |\gbeta|_\mu^2)} \;,
   \quad \m^{xA}= -2\mu^{AB}\gamma_B
\;, \phantom{xX} &
\\
&
   \quad \m^{\tau A}=    \mu^{AB}\gamma_B
\;,
   \quad \m^{\tau\tau}= -1+{ (-\chi+ |\gbeta|_\mu^2)}
\;,
   \quad \m^{AB}= \mu^{AB}
   \;.
  &
\eeal{21X.2}
This, together with \eq{Bonddec2}, shows that \eq{H1}-\eq{H2}
hold  for such metrics.
\end{example}
\subsubsection{The slices}

\begin{figure}[t]
\begin{center} {
\psfrag{mcN}{\large $\mcN=\{x=0\}$} \psfrag{xzero}{\large $x_0$}
\psfrag{Surf}{\large $S$} \psfrag{Slam}{\large $\;\;S_\lambda$}
\psfrag{Htau}{\large $H_T$} \psfrag{Hlamt}{\large $H_{\lambda,t}$}
\psfrag{Htprim}{\large $H_{t'}$} \psfrag{Hzero}{\large $H_0\equiv H_{\tauz}$}
\resizebox{3in}{!}{\includegraphics{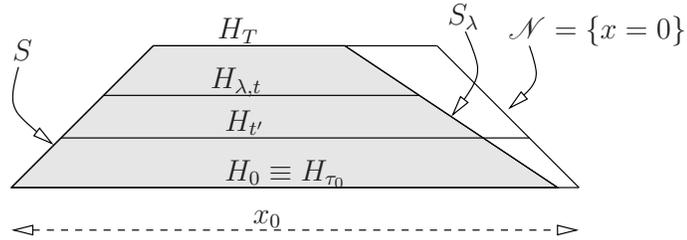}} } \caption{The
sets $\myOmega_{\lambda,T}$ (shaded) and $\myOmega_T$ (the outermost trapezium). In this picture
(but \emph{not} in our hypotheses) the light-cones have forty-five
degrees slopes, as in Minkowski space-time. \label{U-set}}
\end{center}
\end{figure}

In this section we describe the sets within which we  obtain
our estimates, see Figure~\ref{U-set}.
Let $t \in  [\tauz ,  0 [$ run over the range of the time
coordinate $\tau$ of the previous section.
\begin{itemize}
\item Let $ \la \in [0,1]$ parameterize a family of
    spacelike hypersurfaces  $S_{\la }$, which approach
    $\{x=0\}$ when $\lambda$ approaches zero, of the form
$$S_{\la }= \{(\tau, x, v^A): \; x
=\sigma_\la(\tau)\} \;,
$$
where $\sigma_\la$ is a $C^1$function such that:
   \begin{itemize}
    \item $\sigma_0(\tau) \equiv 0 \quad $ i.e. $\quad
        S_0 = \{x=0\}$
    \item $S_\la$ is everywhere spacelike.
       \end{itemize}

One can legitimately raise concerns about existence of the
family $S_\lambda$ with global behaviour as above when the
space-time under consideration is being constructed as a
solution of a Cauchy problem. While the aim of this work is
to prove  that the resulting space-time will have
properties as in Figure~\ref{U-set}, this is not known a
priori. Now, one way to  proceed is to construct the
solution as the limit of solutions of linear equations on a
sequence of metrics, each of those metrics satisfying
controlled weighted energy estimates as proved below. In
particular each space-time in this sequence is globally
hyperbolic, with the set $\{x=0\}$ being part of the
boundary of the domain of dependence of the initial
surface. For each metric in the sequence a relevant family
$S_\lambda$ can be constructed using e.g. Cauchy time
functions; no details will be given as no significant
difficulties are involved. This can then be used to justify
our estimates for each metric in the sequence, and for the
solution.

\item By $S$ we denote a smooth spacelike hypersurface
    transverse to $\{\tau = \tauz \}$ defined by
    \begin{equation}
     \label{7X0.1}
      S= \{(\tau, x, v^A): \; x= \sigma(\tau) \}
      \;,
    \end{equation}
    where
    $\sigma$ is a smooth function of $\tau$ such that $$0 <
\sigma(\tauone ) \le \sigma(\tau) \le \sigma(\tauz ) =
x_0\;. $$
\item $ {\bf{\Large{H}}}_{\la,t} = \{(\tau,x,v^A);\;
    \tau=t,  \;\sigma_\la(x)\le x\le \sigma(\tau) \} , \;
    \myOmega_{\la,T} = \underset{\tauz \le t\le
    T}\cup{\bf{\Large{H}}}_{\la,t}$.
\item $   {\bf{\Large{H}}}_{t} = \{(\tau,x,v^A);\; \tau=t,
    \;0\le x\le \sigma(\tau) \} , \quad \myOmega_{T} =
    \underset{\tauz \le t\le T}\cup{\bf{\Large{H}}}_{t}$.
\end{itemize}
%
%
%
Note that the boundary $\p\myOmega_{\la,t}$ of the region
$\myOmega_{\la,t}$
%
is made of four pieces, $S_{\la},\;S,\; {\bf{\Large{H}}}_{\la,
\tauz } $ and ${\bf{\Large{H}}}_{\la,t}$.  We recall that, for
$\theta \in \R, \; j\in \N$ the spaces
$\mcC^\theta_j({\bf{\Large{H_{\la,\tau}}}}), \;
\mcB^\theta_j({\bf{\Large{H_{\la,\tau}}}}), \;
\mcH^\theta_j({\bf{\Large{H_{\la,\tau}}}})\; $ and
$\mcG^\theta_j({\bf{\Large{H_{\la,\tau}}}})$ are defined in the
appendix of~\cite{ChLengardnwe}.
\subsubsection{The causality properties of the boundary}
We want to show that under the assumptions we made on certain
components of the metric, all the hypersurfaces defined above
have the  nature which  will be needed when  applying the
Stokes' theorem or when we will like to use the positivity of
the stress energy momentum tensor.

The vector $\nabla \tau =  \nabla^\mu(\tau)\p_\mu =\m^{\mu\nu}
\delta_\nu^\tau\p_\mu = \m^{\tau\tau}\p_\tau+ \m^{x\tau}\p_x+
\m^{A\tau}\p_A $ is normal to the hypersurfaces
${\bf{\Large{H}}}_t$ and ${\bf{\Large{H}}}_{\la, t}$, and the
square of its norm is $\m(\nabla \tau, \nabla \tau) =
\m^{\tau\tau} <0$. Therefore $\nabla \tau$ is time-like and
thus these hypersurfaces are space-like. Their \emph{past
directed} unit normal is
\bel{L1} \eta = \eta^\mu\p_\mu =
 \frac{1}{\sqrt{|\m^{\tau\tau}|}}(\m^{\tau\tau}\partial_\tau+
\m^{x\tau}\partial_x+ \m^{A\tau}\partial_A) \;. \ee
We also note de following
$$ \eta_\mu = \m_{\mu\nu}\eta^\nu=
\frac{1}{\sqrt{|\m^{\tau\tau}|}}\m_{\mu\nu}\m^{\nu \tau} =
\frac{1}{\sqrt{|\m^{\tau\tau}|}}\delta_\mu^\tau$$
that is
\bel{L3}
 \eta_\mu
 dx^\mu= \frac{1}{\sqrt{|\m^{\tau\tau}|}}d\tau
 \;.
\ee
As far as the hypersurfaces $S_{\la}$ are concerned, the
functions $\sigma_\la$ are assumed to be such that the normal
$N = \nabla \{-x+\sigma_\la(\tau )\}$ is timelike and the
outward unit normal to this hypersurface is such that the
integral of the contracted energy momentum tensor is negative
(see  \eq{L4}).  The same remark holds for the hypersurface
$S$.

\subsection{Estimates on the space derivatives of the solution}%
\label{ssEsds}

We  want to  derive  weighted energy inequalities for solutions
of (\ref{E0}). These inequalities will be used to prove
existence of a solution satisfying the hypothesis of the
theorem of polyhomogeneous solution of quasi-linear wave
equation near scri.

\subsubsection{The stress energy momentum tensor and its properties}
\label{sssem}

The stress-energy tensor of the system (\ref{E0})  is given by
$$T_{\mu\nu}:= \nabla_\mu u\nabla_\nu u - \frac 12 \m_{\mu\nu}
\nabla^\al u\nabla_\al u \;.
$$
The explicit form of $T_0^{\phantom{0}{0}}$, (the component
of the tensor $T$ which in general determines the energy
density of the system) in   local coordinates system is given
by:
\bea T_0^{\phantom{0}{0}}&=& \nabla^0 u\nabla_0 u - \frac 12
\nabla^\al u\nabla_\al u \nonumber\\
&=& \m^{0\bet}\nabla_\bet u\nabla_0 u -\frac 12 \m^{\al\bet}\nabla_\al u\nabla_\bet u\nonumber\\
&=& \big\{\m^{00}\nabla_0 u\nabla_0 u+\m^{0i}\nabla_iu\nabla_0
u\big\}
  -\frac 12\big\{ \m^{00}\nabla_0 u \nabla_0 u + 2\m^{0i}\nabla_0 u \nabla_i u+ \m^{ij}\nabla_i u \nabla_j u
  \big\}\nonumber\\
\label{E2} &=& \frac 12\big\{\m^{00}(\nabla_0u)^2-
\m^{ij}\nabla_i u \nabla_j u\}\; = \; -\frac
12\big\{-\m^{00}(\nabla_0u)^2+ |Du|^2\} \eea
with $|Du|^2:= \m^{ij}\nabla_i u \nabla_j u$.

The tensor $T$ is symmetric and its divergence is given by
\bea \nabla_\mu T_\nu^{\phantom {\nu}{\mu}}&=&  \Box_\m u
\nabla_\nu u \nonumber
\\
\label{E1} &=& F \nabla_\nu u \quad \quad \text{when $u$ solves
(\ref{E0}) \;.} \eea
Further, one of the useful properties of the tensor $T$ is its
positivity: For any vectors fields $v^\al$ and $w^\al$ both
causal future-pointing  we have:
\be \label{PT} T_\nu^{\phantom {\nu}{\mu}}v^\nu w_\mu \ge
0\;.\ee

\begin{Remark}
In the particular frame $(\tau, x, v^A)$  we will be interested
with, let us calculate the quantity  $T^Y:= T(\p_\tau -\p_x,
d\tau)= T_\tau^{\phantom {\tau}{\tau}}- T_x^{\phantom
{x}{\tau}}$ which we will use as  energy density. From
(\ref{E2}) we have:
$$ T_\tau^{\phantom {\tau}{\tau}}= \frac 12 \left\{\m^{\tau\tau}\left(\p_\tau u\right)^2
-\m^{xx}\left(\p_x u \right)^2 -2\m^{xA}\p_xu\p_A u -  \m^{AB}\p_A u
\p_Bu  \right\}\;. $$
This expression shows that in the case we are concerned with,
$T_\tau^{\phantom {\tau}{\tau}}$ cannot be used to control the
energy of the system near $\{x=0\}$ since  the metric component
$\m^{xx}$  can degenerate there. On the other hand we have
$$T_x^{\phantom {x}{\tau}} = \m^{\tau\tau}\p_\tau u\p_x u + \m^{\tau
x}\left(\p_x u\right)^2 + \m^{\tau A}\p_x u\p_A u\;,  $$
therefore we deduce the following expression of $T^Y:$
\bea \label{E3} T^Y = \frac 12 \Big\{\m^{\tau\tau}\left(\p_\tau
u\right)^2 &-& 2\m^{\tau\tau}\p_\tau u\p_x u
-\left(\m^{xx}+2\m^{\tau x}\right)\left(\p_x u
\right)^2\nonumber\\&& -2\left(\m^{xA}+\m^{\tau
A}\right)\p_xu\p_A u - \m^{AB}\p_A u \p_Bu\Big\}\;.
\phantom{xxx} \eea
Now, if we set
$$\left\{ \begin{array}{l}\la = \m^{\tau\tau} +
\m^{xx}+2\m^{x\tau} = 1+ \text{O}(x)>0\quad \text{(by hypothesis)}\\
\xi^A = \m^{xA}+\m^{A\tau}
\\ \kappa^{AB}= \frac{\;\xi^A \xi^B}{\la} \end{array}\right.\;,  $$
then we obtain the following decomposition of $T^Y$
\be \label{E47}T^Y= -\frac12 \left\{
-\m^{\tau\tau}\left(\p_\tau u-\p_xu\right)^2 + \la\left(\p_x u+
\frac{\left(\m^{xA}+\m^{A\tau}\right)}{\la}\p_Au\right)^2 +
\left(\m^{AB}-\kappa^{AB}\right) \p_A u\p_B u\right\}. \ee
The above decomposition shows that the quantity $T^Y$ controls
uniformly the energy of the system if and only if there exists
$\epsilon_0>0$ (which can be made to coincide with the one
occurring in \eq{G0}) such that
\bea
 &\mbox{$\lambda>\epsilon_0$, and
 $ \left(\m^{AB}-\kappa^{AB}\right) \zeta_A\zeta_B \ge \epsilon_0\sum
 _A (\zeta_A )^2$} \;;
 &
\eeal{hyphab}
the existence of such a constant follows already from our
previous hypotheses. It turns out that if we have  a priori
bounds on the $L^\infty$ norms of $\m^\sharp$ from above and
below,   this expression can be used to control all the
components of the stress energy tensor. In fact we have
\bel{E43} |T_\nu^{\phantom {\nu}{\mu}}| =
|\m^{\mu\sigma}\p_\sigma u \p_\nu u - \frac
12\delta_\nu^{\phantom {\nu}{\mu}} \m^{\al\beta}\p_\al
u\p_\beta u|\le C|\m^\sharp||\p  u |^2 \le
C|\m^\sharp||T_\tau^{\phantom {\tau}{\tau}}- T_x^{\phantom
{x}{\tau}}|\;;\ee
here the constant $C$ depends upon $\epsilon_0$, and is allowed
to change after each inequality symbol in general.
\end{Remark}
\begin{Remark}
For further purposes we note that, using the vector field
$\p_\tau-\p_x,$ the principal part of the d'Alembertian has the
following form:
\beal{E45} \m^{\al\beta}\p_{\al\beta} &=&
\m^{\tau\tau}(\p_\tau-\p_x)^2 + 2\left(\m^{\tau\tau}+\m^{\tau
x}\right)\left(\p_\tau-\p_x\right)\p_x + 2\m^{\tau
 A}\left(\p_\tau-\p_x\right)\p_A\nonumber
\\&&
 +
 \left(\m^{\tau\tau}+2\m^{\tau x}+ \m^{xx}\right)\p^2_x
+2\left(\m^{xA}+\m^{\tau A}\right)\p_x\p_A +\m^{AB}\p_A\p_B\;.
 \nonumber
\\
 &&
\eea

\end{Remark}

\subsubsection{Estimates on the first derivatives of the solution}
We want to derive some energy inequalities for the  solution
$u$ of the system (\ref{E0}). For this purpose, we consider the
weighted energy  at an instant $t$ of the evolution of the
system defined using the vector field $\partial_\tau -
\partial_x$; recall $T^Y= T_\tau^{\phantom {\tau}{\tau}}-
T_x^{\phantom {x}{\tau}}$:
\bel{EE2}E[u(t)] = -\int_{{\bf{\Large{H}}}_{t} }x^{-2\al}T^Y\;
\frac{dx}{x}d^{n-1}\nu_{t,x} \ee
where $d^{n-1}\nu_{t,x}$ is the measure defined on
$\{t\}\times\{x\}\times\mcO$ by the metric $\m$ (as will be
made precise shortly), and $\al\le 0 $ a real parameter the
range of which will be given later.
We set
\bel{EEEq3}E_\la [u(t)] = -\int_{{\bf{\Large{H}}}_{\la,t}
}x^{-2\al}T^Y \; \frac{dx}{x}d^{n-1}\nu_{t,x}\;.\ee
Our strategy will be to obtain a bound of $E[u(t)]$ from an
uniform bound (with respect to $\la$) of $E_\la[u(t)]$. We will
apply  the divergence theorem to the energy-momentum tensor;
this holds e.g.\ for  $C^{1,1}_\loc $ functions $u$ (first
derivatives locally Lipschitz continuous).
We want to establish the following (recall that $\ep_0$ is the
constant arising in \eq{G0} and in \eq{hyphab}, while $C_0$ is
defined in \eq{betineq}):
\begin{Proposition}
Let   $\al \le -\frac 12 $. Under hypotheses
(\ref{G0})-(\ref{H2}) and (\ref{hyphab}), there exists a
constant $C_1$ depending upon $\ep_0, C_0, \al$ such that for
all
$$ \tau\in  \left[\tauz , \tauone \right] \quad \text{and}\quad u \in
 C^{1,1}_\loc
$$
satisfying (\ref{E0}), we have

\bea \label{I1}
E_\la[u(\tau)]  &\le&  C_1\Bigg\{ E_\la[u(\tauz )] +\int_{\tauz
}^\tau \Big\{ \|F(s)\|^2_{\mcH^\al_0({\bf{\Large{H}}}_{\la,s})}
+
\left(1+\|\f^\sharp\|_{L^\infty}+\|\m^\sharp\|_{L^\infty}\right)\nonumber\\
&& \times \left(1+
\|\m\|^2_{L^\infty({\bf{\Large{H_{\la,s}}}})}
+\|\m^\sharp\|^2_{L^\infty({\bf{\Large{H_{\la,s}}}})} +
\|\left(\p_\tau-\p_x\right)\m^\sharp\|^2_{L^\infty({\bf{\Large{H_{\la,s}}}})}\right)E_\la[u(s)]\Big\}
ds\Bigg\}\nonumber\\\eea
\end{Proposition}

\proof Stokes' theorem for the vector field  $\Lambda^\mu =
x^{-2\al-1}T_\nu^{\phantom {\nu}{\mu}}Y^\nu $ on
$\myOmega_{\la,\tau}$ (compare Fig.~\ref{U-set}) gives
\be \label{ST}
\int_{\p\myOmega_{\la,\tau}}x^{-2\al-1}T_\nu^{\phantom
{\nu}{\mu}}Y^\nu \eta_\mu dS = \int_{\myOmega_{\la,\tau}}
\nabla_\mu \big\{x^{-2\al-1}T_\nu^{\phantom {\nu}{\mu}}Y^\nu
\big\}dV\ee
for an arbitrary differentiable vector field $Y$. Here
\bel{volel} dV = \sqrt{|\det\m|}d\tau\wedge dx\wedge
d^{n-1}v\;, \ee
where $\det\m$ is the determinant of the metric $\m$. Further,
on non-characteristic parts of the boundary, $\eta_\mu$ is the
unit outwards pointing conormal, and
\bel{surfel} dS = \sqrt{|\det\gamma|}d^n y\;, \ee
with $y^i, \; i=1, \ldots, n$, a system of  coordinates on the
corresponding boundary, and $\gamma$ the metric induced on it
by the metric $\m$; i.e. $\gamma = j^*\m$ , $j$ being the
canonical injection of the boundary into the manifold. (On
characteristic parts of the boundary, a convenient choice of
$\eta_\mu$ and $dS$ will be made as need arises). In the case
under consideration, $\p\myOmega_{\la,\tau}$ is made of four
pieces
${\bf{\Large{H}}}_{\la,\tauz }$, $
{\bf{\Large{H}}}_{\la,\tau}$, together with
$$
 S_{\la,\tau}:= S_{\lambda} \cap \{0\le t\le \tau\}
 \ \mbox{and} \  S^\tau:= S \cap \{0\le t\le \tau\}
 \;.
$$
Therefore the identity (\ref{ST}) reads:
\beqa \label{ST1}
\int_{{\bf{\Large{H}}}_{\la,\tau}}x^{-2\al-1}T_\nu^{\phantom
{\nu}{\mu}}Y^\nu \eta_\mu dS
&+&\int_{{\bf{\Large{H}}}_{\la,\tauz
}}x^{-2\al-1}T_\nu^{\phantom {\nu}{\mu}}Y^\nu \eta_\mu dS
+\int_{S_{\la,\tau}}x^{-2\al-1}T_\nu^{\phantom {\nu}{\mu}}Y^\nu
\eta_\mu dS \nonumber\\& +&\int_{S^{\tau}}
x^{-2\al-1}T_\nu^{\phantom {\nu}{\mu}}Y^\nu \eta_\mu dS
  =  \int_{\myOmega_{\la,\tau}}
\nabla_\mu\big\{x^{-2\al-1}T_\nu^{\phantom {\nu}{\mu}}Y^\nu
\big\}dV\;. \nonumber
\\
&& \eeqa
The left-hand-side of equation (\ref{ST1}) is made of four
terms which will be labeled in their order of appearance
$L_1,\; L_2,\; L_3$ and $L_4$.
As mentioned before,  we choose the vector field $Y=Y^\mu\p_\mu
$ to be equal to $\p_\tau-\p_x$. Once this choice is made, let
us look at each of the terms $L_i, i=1,2,3, 4$.
Recall that (see equation (\ref{L3})) on
${\bf{\Large{H_{\la,\tau}}}}$ we have:
$$ \eta_\mu dx^\mu =  \frac{1}{\sqrt{|\m^{\tau\tau}|}}d\tau \quad \text{which implies that} \quad T_\nu^{\phantom {\nu}{\mu}}Y^\nu
\eta_\mu =  \frac{1}{\sqrt{|\m^{\tau\tau}|}}\left\{T_\tau^{\phantom
{\tau}{\tau}}-T_x^{\phantom {x}{\tau}}\right\}$$
and
$dS = \sqrt{|\det\gamma|}dx\wedge d^{n-1}v$  is the surface
element denoted in equations (\ref{EE2}) and (\ref{E3}) by
$dx\, d^{n-1}v_{t,x}$.
Since $\eta_0\sqrt{\det \m}= \sqrt{\det \gamma}$ on
${\bf{\Large{H_{\la,\tau}}}}$,  we obtain that (remember that
$\eta^\mu\p_\mu$ is past directed)
\be \label{E4} L_1 = -E_\la[u(\tau)]\;. \ee
From this, the sign coming from the Stokes' identity shows that
\be \label{EE4}\quad L_2 = E_\la[u(\tauz )]
 \;.
\ee

On the hypersurfaces $S_{\la}$ and  $S$, since  the unit
outward normal is also past directed  and the vector field
$Y^\nu\p_\nu = \p_\tau-\p_x$  future directed, we deduce from
the positivity of the stress energy tensor  that:
\bel{L4} L_3\le 0 \quad \text{and} \quad L_4 \le 0\;.\ee
We can now rewrite  (\ref{ST1}) as:
\bel{E44} -E_\la[u(\tau)]+E_\la[u(\tauz )] + L_3 +L_4=
\int_{\myOmega_{\la,\tau}}
\nabla_\mu\big\{x^{-2\al-1}T_\nu^{\phantom {\nu}{\mu}}Y^\nu
\big\}dV\;. \ee
Now, let us consider the right-hand side of the above equation.
We have:
\beqa
\lefteqn{ \nabla_\mu\big\{x^{-2\al-1}T_\nu^{\phantom
{\nu}{\mu}}Y^\nu \big\}
 }
 &&
\no
 &=& x^{-2\al-1}\Big\{(\nabla_\mu T_\nu^{\phantom
{\nu}{\mu}})Y^\nu + T_\nu^{\phantom {\nu}{\mu}}(\nabla_\mu
Y^\nu) -(2\al+1) x^{-1} T_\nu^{\phantom {\nu}{\mu}}Y^\nu
\nabla_\mu(x)\Big\}\nonumber\\
&=&  x^{-2\al-1}\Big\{(\nabla_\mu T_\nu^{\phantom
{\nu}{\mu}})Y^\nu + T_\nu^{\phantom
{\nu}{\mu}}\left\{\Gamma_{\mu\tau}^\nu- \Gamma_{\mu x}^\nu
\right\} \nonumber\\&& -(2\al+1)
x^{-1}\nabla_\mu x\,\left\{T_\tau^{\phantom{\tau}{\mu}} - T_x^{\phantom{x}{\mu}}\right\}\Big\}\nonumber\\
\label{E'5} &=:& R_1+R_2+R_3, \eeqa
where $$ \Gamma_{\mu\nu}^\rho = \frac 12 \m^{\sigma\rho}(\p_\mu
\m_{\sigma\nu}+ \p_\nu \m_{\mu\sigma} -\p_\sigma
\m_{\mu\nu}),$$ are the Christoffel's symbols of the metric
$\m$.
From (\ref{E1}), we have:
\bea x^{2\al+1}|R_1|= |F||\nabla_\nu u Y^\nu|=
|F||\left(\p_\tau u-\p_x u\right)| &\le&
\frac 12 \left\{F^2+\left(\p_\tau u-\p_x u\right)^2\right\})\nonumber\\
         &\le& c(\ep_0)\left(F^2+ \big|T_\tau^{\phantom{\tau}{\tau}}-T_x^{\phantom{x}{\tau}}\big|\right)
\; . \nonumber
\\
&&\label{E7} \eea
As far as the second term is concerned, we have:
\beaa T_\nu^{\phantom {\nu}{\mu}}\Gamma_{\mu\theta}^\nu &=&
\frac 12 T^{\mu\sigma}\p_\theta \m_{\mu\sigma}\quad = \quad
-\frac 12 T_{\mu \sigma}\p_\theta \m^{\mu\sigma}.
  \eeaa
Thus, replacing successively in the above expression $\theta$
with $\tau$ and $x$ and subtracting the two expressions we find
that
$$
x^{2\al+1}R_2 =-\frac 12 T_\mu^{\phantom
{\mu}{\nu}}\m_{\nu\sigma}\left(\p_\tau -\p_x\right)\m^{\mu\sigma} \;.%
$$
From (\ref{E43})  we obtain:
\bel{E48} x^{2\al+1}|R_2|= |T_{\nu\sigma}\left(\p_\tau
-\p_x\right) \m^{\mu\sigma}|
\le (n+1)C|\m^\sharp|\left( |\m |^2+ |(\p_\tau -\p_x)\m^\sharp
|^2\right)
\big|T_\tau^{\phantom{\tau}{\tau}}-T_x^{\phantom{x}{\tau}}\big|\;.
\ee
For the third term we have, keeping in mind \eq{BC}:
\bea
x^{2\al+1}R_3 &=& -(2\al+1)x^{-1}T_\nu^{\phantom{\nu}{\mu}}\nabla_\mu x\,Y^\nu\nonumber\\
              &=& -(2\al+1)x^{-1}\m^{\mu\sigma}T_{\nu\sigma}\nabla_\mu x\,Y^\nu\nonumber\\
              &=& -(2\al+1) x^{-1}T_{\mu\nu}\nabla^\mu x\, Y^\nu\nonumber\\
              &=&\underbrace{-(2\al+1)x^{-1}T_{\mu\nu}Y^\nu \omega^{(1)\mu}}_{\ge 0  }
              -(2\al+1)\frac{\beta(x)}{x}T_{\mu\nu}Y^\nu\omega^{(2)\mu} \quad \text{for} \quad \al \le\;  - 1/2\nonumber\\
              &\ge & -(2\al+1)\frac{\beta(x)}{x}T_{\mu\nu}Y^\nu\omega^{(2)\mu} =
              -(2\al+1)\frac{\beta(x)}{x}\left(T_{\mu\tau}-T_{\mu x}\right)\omega^{(2)\mu}\nonumber\\
              &\ge & - C(\;\al, C_0,n)|\f^\sharp|\left(1+ |\m|^2+
              |\m^\sharp|^2\right)|T_\tau^{\phantom{\tau}{\tau}}-
T_x^{\phantom{x}{\tau}}|\;.
              \label{E9}
\eea
Let us  justify the last inequality. In other words let us show
that the  expression $T_{\mu\tau}-T_{\mu x}$ is controlled by
$|T_\tau^{\phantom{\tau}{\tau}}- T_x^{\phantom{x}{\tau}}|$. We
have:
\beaa |T_{\mu\tau}-T_{\mu x}| &=&|\p_\mu u
\left(\p_\tau-\p_x\right)u- \frac 12 \left(\m_{\mu\tau}-\m_{\mu
x}\right)\m^{\al\beta}\p_\al u\p_\beta u|\\
&\le&\left(\p_\mu u\right)^2+  \left[\left(\p_\tau
-\p_x\right)u\right]^2 + \left( |\m|^2+
|\m^\sharp|^2\right)\left(\delta^{\al\beta}\p_\al
u\p_\beta u\right) \\
&\le & C(\ep_0)\left(1+ |\m|^2+
|\m^\sharp|^2\right)|T_\tau^{\phantom{\tau}{\tau}}-
T_x^{\phantom{x}{\tau}}|.\quad \quad \text{See  (\ref{E47})}
\eeaa
Inequalities (\ref{E7}), (\ref{E48}) and (\ref{E9}) show that
the right-hand side of (\ref{E'5}) can be estimated as:
\be \label{EE13} R_1+R_2+R_3 \ge
-C_1x^{-(2\al+1)}\left\{\left(1+|\f^\sharp|+|\m^\sharp|\right)\left(1+
|\m |^2+|\m^\sharp |^2+ |(\p_\tau-\p_x)\m^\sharp |^2
\right)|T^Y|+ F^2 \right\} \; ,\ee
where $C_1=C( \al,\ep_0, C_0 , n )\; .$    Now from (\ref{E44})
we have
$$-E_\la[u(t)]+E_\la[u(\tauz )] + L_3 + L_4=  R_1 + R_2+ R_3,$$
thus,
using (\ref{L4}), we obtain the following:%
%
\beaa E_\la[u(t)]\le E_\la[u(\tauz )] &+&    C_1 \int_{\tauz
}^t\int_{{\bf{\Large{H}}}_{\la,s}}
x^{-2\al}\left\{\left(1+|\f^\sharp|+|\m^\sharp|\right)\left(1+
|\m |^2+|\m^\sharp|^2\right.\right.\\&+& \left.\left.
|(\p_\tau-\p_x)\m^\sharp |^2 \right)|T^Y|+ F^2(s) \right\}ds
\frac{dx}{x}d^{n-1}\nu \;. \eeaa
Therefore, there exists a constant $C_1>0$ depending upon $n,
\; \ep_0\;, \; \al\; $ and $C_0$ such that
\bea \label{E6}
E_\la[u(\tau)]  &\le&  C_1\Bigg\{ E_\la[u(\tauz )] +\int_{\tauz
}^\tau \Big\{ \|F(s)\|^2_{\mcH^\al_0({\bf{\Large{H}}}_{\la,s})}
+
\left(1+\|\f^\sharp\|_{L^\infty}+\|\m^\sharp\|_{L^\infty}\right)\nonumber\\
&& \times \left(1+
\|\m\|^2_{L^\infty({\bf{\Large{H_{\la,s}}}})}
+\|\m^\sharp\|^2_{L^\infty({\bf{\Large{H_{\la,s}}}})} +
\|\left(\p_\tau-\p_x\right)\m^\sharp\|^2_{L^\infty({\bf{\Large{H_{\la,\tau}}}})}\right)E_\la[u(s)]\Big\}
ds\Bigg\}\nonumber\\\eea
and the proof is completed. \qed

%
%
%


\subsubsection{Estimates on the  higher space derivatives of the solution}
\label{Energie} To proceed further, we would like to have an
estimate similar to (\ref{I1}) on  space derivatives of the
unknown function in equation (\ref{E0}). For this purpose, for
$k\in \N,\;\beta = (\beta_1, \beta_2, \ldots,\beta_r)\in  \N^r,
\;\text{with}\; |\beta| \le k ;$ we set:
$$
\overset{(\beta)}{T_\nu^{\phantom {\nu}{\mu}}} =
x^{-2\al-1+2\beta_1}\left\{\nabla^\mu \mcD^\beta u\nabla_\nu
\mcD^\beta u - \frac 12 \delta_\nu^{\phantom {\nu}{\mu}}
\nabla^\al\mcD^\beta  u\nabla_\al\mcD^\beta  u\right\},$$
where  $\al\le -1/2$ is the real parameter of the previous
section,
  $\mcD^\beta = X_1^{\beta_1}X_2^{\beta_2} \ldots
  X_r^{\beta_r}
  $, with
the  $X_i$'s being the vector fields defined in \cite[page
51]{ChLengardnwe}: for $i= 2, \ldots, r$, $ X_i=
\sum\limits_{A=2}^r X_i^A(v)\p_A$,
  where the $X_i^A$'s are smooth functions bounded on bounded set with all their
  derivatives, and $X_1=\partial_x$.
Since the operator $\n$ is linear, as in (\ref{E1}), we have
$$\nabla_\mu \overset{(\beta)}{T_\nu^{\phantom {\nu}{\mu}}}= x^{2\al-1+2\beta_1}\Box_\m(\mcD^\beta  u)\nabla_\nu(\mcD^\beta  u )
+(-2\al -1 +2\beta_1)\frac{\n_\mu
(x)}{x}\overset{(\beta)}{T_\nu^{\phantom {\nu}{\mu}}}\;. $$
Now
\bel{NEq}\Box_\m(\mcD^\beta  u)= \mcD^\beta  (\Box_\m u) +
[\Box_\m, \mcD^\beta]u =\mcD^\beta F + [\Box_\m, \mcD^\beta]u
\;, \ee
for any solution of the equation (\ref{E0}). Thus
\bel{E20} \nabla_\mu \overset{(\beta)}{T_\nu^{\phantom
{\nu}{\mu}}}= x^{-2\al-1+2\beta_1}\left\{\mcD^\beta F +
[\Box_\m, \mcD^\beta]u \right\}\nabla_\nu(\mcD^\beta  u
)+(-2\al -1 +2\beta_1)\frac{\n_\mu(
x)}{x}\overset{(\beta)}{T_\nu^{\phantom {\nu}{\mu}}}\;. \ee
Similarly to  the previous section, we set:
$$\overset{(\beta)}{T}\;^Y =\overset{(\beta)}{T_\tau^{\phantom
{\tau}{\tau}}}-\overset{(\beta)}{T_x^{\phantom {x}{\tau}}} \;, $$
\bel{E10}E^{\;\al}_k[u(\tau)] =
\sum_{|\beta|=0}^k\int_{{\bf{\Large{H}}}_{t}
}\;-\overset{(\beta)}{T}\;^Y\; dx\, d^{n-1}\nu_{t,x} \quad
\text{and}\quad E^{\;\al}_{k,\la} [u(t)] =
\sum_{|\beta|=0}^k\int_{{\bf{\Large{H}}}_{\la,\tau}
}-\overset{(\beta)}{T}\;^Y\; dx\, d^{n-1}\nu_{t,x}\;.\ee
\begin{Remark}
From (\ref{E47}) we deduce the following decomposition for
$\overset{(\beta)}{T}\;^Y$:
\bea\label{EE47} \overset{(\beta)}{T}\;^Y &=& -\frac12 \Bigg\{
-\m^{\tau\tau}\left(x^{-\al-\frac12+\beta_1}\mcD^\beta(\p_\tau
-\p_x)u\right)^2 \no
&&\qquad\qquad +
\la\left(x^{-\al-\frac12+\beta_1}\mcD^\beta(\p_x u)+
\frac{\left(\m^{xA}+\m^{A\tau}\right)}{\la}\p_A\left(x^{-\al-\frac12+\beta_1}\mcD^\beta
u\right)\right)^2\no
&&\qquad\qquad\qquad\qquad+ \left(\m^{AB}-\kappa^{AB}\right)
\p_A\left(x^{-\al-\frac12+\beta_1}\mcD^\beta
u\right)\p_B\left(x^{-\al-\frac12+\beta_1}\mcD^\beta
u\right)\Bigg\}\;.\qquad\qquad
\eea
Since the coefficients of the terms  arising in commutating
$\p_A$ and $\mcD^\beta$ are uniformly bounded, from the above
we find that the energy of order $k$ controls the
$\mcH^\al_k$-norms of the first order derivatives of the
unknown function $u$. That is:
\be\label{EE48} \|(\p_\tau
 -\p_x)u\|^2_{\mcH^\al_{k}({\bf{\Large{H}}}_{\la,\tau})}
 +\|\p_xu\|^2_{\mcH^\al_{k}({\bf{\Large{H}}}_{\la,\tau})}+
 \sum_{A}\|\p_Au\|^2_{\mcH^\al_{k}({\bf{\Large{H}}}_{\la,\tau})} \le
 C E^{\;\al}_{k, \la}[u(\tau)]\;. \ee
\end{Remark}
Let us set
\be
 \label{GC}
 {\Upsilon}^\nu := -\m^{\al\mu}\Gamma_{\al\mu}^\nu
 = \frac{1}{\sqrt{|\det\m|}}\p_\mu\left(\sqrt{|\det
 \m|}\m^{\mu\nu} \right)
  \;.
\ee

Let us define
\bean
 M(\tau) &:= &  \|F \|^2_{\mcB^\al_0({\bf{\Large{H_{\la,\tau}}}})}
 + \|(\m, (\p_\tau-\p_x)\m^\sharp)
 \|^2_{L^\infty({\bf{\Large{H_{\la,\tau}}}})}
\\
 &&
 + \|(\m^\sharp,\f^\sharp, {\Upsilon} )
\|^2_{\mcC^0_{\{x=0\},1}({\bf{\Large{H_{\la,\tau}}}})}
 \;.
\eeal{23VIII0.3}
We claim that:
\begin{Proposition}
 \label{pprop}
  Let  $\lambda>0$, $k\in \mathbb{N}$ and suppose that $\al \le
-\frac 12 $. Under hypotheses (\ref{G0})-(\ref{H2}) and
(\ref{hyphab}), there exists a function $C_2(\ep_0, C_0, \al,
k,n, M )$. monotonously increasing in $M$, which we write as
$C_2(M)$, such that for all
$$ \tau\in  \left[\tauz , \tauone \right] $$
and for all $u$ satisfying (\ref{E0})   we have
\bean
 \lefteqn{
  E^{\;\al}_{k,\la}[u(\tau)]
  }
  &&
\\
   &\le&E^{\;\al}_{k,\la}[
 u(\tauz )] +
  \int_{\tauz }^\tau C_2   (M(s))\Bigg\{
 E^\al_{k,\la}[u(s)]+
 \|F(s)\|^2_{\mcH^\al_k({\bf{\Large{H_{\la,\tau}}}})}
\no
&&
 +\|\big((\p_\tau-\p_x)u,\p_x u, \p_A u\big)
 \|^2_{\mcB^\al_1({\bf{\Large{H_{\la,\tau}}}})}
  \times
 \|\big(\m^\sharp,\f^\sharp, {\Upsilon} \big)
  \|^2_{\mcG^0_{k}({\bf{\Large{H_{\la,\tau}}}})} \Bigg\}ds
 \;.
\no \label{Pr}
 \eea
\end{Proposition}

%
\begin{Remark}
 The reader should note that  $C_2$  does not depend
 upon $\lambda$.
\end{Remark}
\proof
If   the right-hand side of \eq{Pr} is infinite there is
nothing to prove. Otherwise, the calculations that follow
should be done assuming smoothness of $u$, and the inequality
for general $u$'s can be obtained by a density argument.

The equivalent of (\ref{ST1}) for space-derivatives of the
solution of (\ref{E0}) reads:

\bea \label{E12}
\sum_{|\beta|=0}^k\int_{{\bf{\Large{H}}}_{\la,\tau}}\overset{(\beta)}{T_\nu^{\phantom
{\nu}{\mu}}}Y^\nu \eta_\mu dS
&+&\sum_{|\beta|=0}^k\int_{{\bf{\Large{H}}}_{\la,\tauz
}}\overset{(\beta)}{T_\nu^{\phantom {\nu}{\mu}}}Y^\nu \eta_\mu
dS
+\sum_{|\beta|=0}^k\int_{S_{\la,\tau}}\overset{(\beta)}{T_\nu^{\phantom
{\nu}{\mu}}}Y^\nu \eta_\mu dS
\nonumber\\
&+&
\sum_{|\beta|=0}^k\int_{S_{\theta,\tau}}\overset{(\beta)}{T_\nu^{\phantom
{\nu}{\mu}}}Y^\nu \eta_\mu dS
=\sum_{|\beta|=0}^k\int_{\myOmega_{\la,\tau}}
\nabla_\mu\left(\overset{(\beta)}{T_\nu^{\phantom
{\nu}{\mu}}}Y^\nu \right)dV\qquad\qquad\eea
which gives the following equation:
\bea\label{E13}-E^{\;\al}_{k,\la}[u(\tau)]+E^{\;\al}_{k,\la}[u(\tauz
)]&+&
\underbrace{\sum_{|\beta|=0}^k\int_{S_{\la,\tau}}\overset{(\beta)}{T_\nu^{\phantom
{\nu}{\mu}}}Y^\nu \eta_\mu dS
+\sum_{|\beta|=0}^k\int_{S_{\theta,\tau}}\overset{(\beta)}{T_\nu^{\phantom
{\nu}{\mu}}}Y^\nu \eta_\mu dS }_{\;:= \;\hat{L}_3 +
\hat{L}_{4}\; \le\; 0 }\nonumber\\ &&=
\sum_{|\beta|=0}^k\int_{\myOmega_{\la,\tau}}
\nabla_\mu\big\{\overset{(\beta)}{T_\nu^{\phantom
{\nu}{\mu}}}Y^\nu \big\}dx\, d\nu.\eea
Again as in the previous section we take $Y^\nu\p_\nu =
\p_\tau-\p_x$, then the divergence in the right-hand side of
(\ref{E13}) reads:

\bea \label{E22}
\nabla_\mu\big\{\overset{(\beta)}{T_\nu^{\phantom
{\nu}{\mu}}}Y^\nu \big\} &=& \nabla_\mu
\overset{(\beta)}{T_\nu^{\phantom {\nu}{\mu}}}Y^\nu
  + \overset{(\beta)}{T_\nu^{\phantom {\nu}{\mu}}}\n_\mu Y^\nu\nonumber\\
&=& x^{-2\al-1-2\beta_1}\left\{\mcD^\beta F + [\Box_\m,
\mcD^\beta]u
\right\}\left(\p_\tau-\p_x\right)(\mcD^\beta  u )\nonumber\\
&+&\overset{(\beta)}{T_\nu^{\phantom
{\nu}{\mu}}}\left(\Gamma_{\mu\tau}^\nu - \Gamma_{\mu
x}^\nu\right)+(-2\al -1 +2\beta_1)\frac{\n_\mu(
x)}{x}\left(\overset{(\beta)}{T_\tau^{\phantom
{\tau}{\mu}}}-\overset{(\beta)}{T_x^{\phantom
{x}{\mu}}}\right)\nonumber\\
&=:& \wh{R}_1+\wh{R}_2+\wh{R}_3. \eea
If we repeat the calculations in the previous section that led
to (\ref{E48}) and (\ref{E9}), we obtain that  there exists a
constant  $ C=C(n,k, C_0, \al,\epsilon_0)>0$ such that:
\bel{E14}   |\wh{R}_2| \le C |\m^\sharp|\left( |\m |^2+
|(\p_\tau -\p_x)\m^\sharp
|^2\right)|\overset{(\beta)}{T}\;^Y|\ee
and, keeping in mind that the term with the worst power of $x$
can be discarded because of a favorable sign,
\be\label{EEE14}\hat R_3 \ge - C |\f^\sharp|\left(1+ |\m|^2+
              |\m^\sharp|^2\right)|\overset{(\beta)}{T}\;^Y|\;.\ee
As far as the term $\wh{R}_1$ is concerned, from  the
inequality $ab \le \frac 12( a^2 + b^2)$, we have:
\bea \label{EE14} x^{2\al+1-2\beta_1}|\wh{R}_1| & =&|\{ \mcD^\beta F + [\Box_{\m}, \mcD^\beta]u \}\left(\p_\tau-\p_x\right)(\mcD^\beta u)| \nonumber\\
                           &\le& \frac 12 (\mcD^\beta F )^2 + \frac 12 \left(\left[\Box_{\m}, \mcD^\beta\right]u\right)^2 +
                            \left[\left(\p_\tau-\p_x\right)\left(\mcD^\beta u\right)\right]^2 \nonumber\\
                          & \le&  (\mcD^\beta F )^2 + C(\ep_0) |\overset{(\beta)}{T}\;^Y|
                          + \left(\left[\Box_{\m}, \mcD^\beta\right]u\right)^2 \;.
\eea
From   inequalities (\ref{E14})--(\ref{EE14}) and the fact that
$\hat{L}_3\le 0$ and $\hat{L}_4\le 0 $ we obtain that:
\bea \label{E15}E^{\;\al}_{k,\la}[u(\tau)]-E^{\;\al}_{k,\la}[
u(\tauz )]
  &\le&
  C\int_{\tauz }^\tau\Big[\left(1+\|\f^\sharp\|_{L^\infty}+\|\m^\sharp\|_{L^\infty}\right)\big(1+
  \|\m\|^2_{L^\infty}+\|\m^\sharp\|^2_{L^\infty}\nonumber \\
&& +  \|\left(\p_\tau-\p_x\right)\m^\sharp\|^2_{L^\infty}\big)
E^\al_{k,\la}[u(s)]+
\|F(s)\|^2_{\mcH^\al_k({\bf{\Large{H_{\la,s}}}})}\Big]ds\nonumber\\
&&+\;\sum_{|\beta|=0}^k\int_{\tauz
}^\tau\int_{{\bf{\Large{H}}}_{\la,s}}
x^{-2\al-1+2\beta_1}([\Box_{\m},
\mcD^\beta]u)^2(s)dx\, d\nu_{t,x}ds\nonumber\\
\eea
with $C = C(n,\al,k, C_0, \ep_0)$. Now, let us estimate the
last term of the right-hand side of the above inequality.  From
the definition \eqref{GC} of ${\Upsilon}^\nu$ we have
\bel{E46} \Box_\m= \m^{\mu\nu}\p^2_{\mu\nu} +
\Upsilon^\nu\p_\nu ,\ee
and then
\bea \label{E24} [\Box_\m, \mcD^\beta]u &=& \m^{\al\mu}[
\p_\al\p_\mu, \mcD^\beta]u - {\Upsilon}^\nu[\mcD^\beta,
\p_\nu]u -\left\{\mcD^\beta\left({\Upsilon}^\nu \p_\nu
u\right)-{\Upsilon}^\nu \mcD^\beta\left(\p_\nu
  u\right)\right\}\nonumber\\
   &&  - \left\{\mcD^\beta\left(\m^{\al\mu}\p_\al \p_\mu u\right)- \m^{\al\mu}\mcD^\beta\left(\p_\al \p_\mu
   u\right)\right\}\nonumber\\
&=:& A_1+A_2+A_3+A_4\;. \eea
To estimate the first and second terms, we use the explicit
form of the differential operator $\mcD:\; \mcD^\beta =
\p_x^{\beta_1}X_2^{\beta_2} \ldots X_r^{\beta_r}=
\p_x^{\beta_1} X_v^{\beta_v}$. Since  $\partial_\tau$ and
$\partial_x$ commute with $\mcD^\beta$, we have (see
(\ref{E45}))
$$
A_1=   \m^{\mu\al}[\p_\mu\p_\al,\mcD^\beta]u =2\m^{\tau
A}[(\p_\tau-\p_x)\p_A,\mcD^\beta]u+2(\m^{xA}+\m^{\tau
A})[\p_x\p_A,\mcD^\beta]u+\m^{AB}[\p_A\p_B,\mcD^\beta]u,
$$
and since
$$
\m^{\tau A}[(\p_\tau-\p_x)\p_A,\mcD^\beta]u= \m^{\tau
A}\p_x^{\beta_1} \p_AX_v^{\beta_v}[(\p_\tau-\p_x)u] - \m^{\tau
A}\p_x^{\beta_1}X_v^{\beta_v}\p_A[(\p_\tau-\p_x)u]\;
$$
we obtain that (see (\ref{EE48}):
\beaa\int_{{\bf{\Large{H}}}_{\la,\tau}}
x^{-2\al-1+2\beta_1}\left(\m^{\tau
A}[(\p_\tau-\p_x)\p_A,\mcD^\beta]u\right)^2dx\, d\nu &\le&
c\|\m^\sharp\|^2_{L^\infty({\bf{\Large{H_{\la,\tau}}}})}\|(\p_\tau-\p_x)
u\|^2_{\mcH^\al_k}\\
&\le&
c\|\m^\sharp\|^2_{L^\infty({\bf{\Large{H_{\la,\tau}}}})}E^{\;\al}_{k,\la}[u(\tau)]\;.
\eeaa
Similarly, we have
$$\left(\m^{xA}+\m^{\tau A}\right)[\p_x\p_A,\mcD^\beta]u=\left(\m^{xA}+\m^{\tau
A}\right)\left(\p_A \mcD^\beta(\p_x u) - \mcD^\beta\p_A(\p_x
u)\right)
\;,
$$
which leads to:
\beaa \int_{{\bf{\Large{H}}}_{\la,\tau}}
x^{-2\al-1+2\beta_1}\left\{\left(\m^{xA}+\m^{\tau
A}\right)[\p_x\p_A,\mcD^\beta]u\right\}^2(s)dx\, d\nu &\le&
C\|\m^\sharp\|^2_{L^\infty({\bf{\Large{H_{\la,\tau}}}})}\| \p_x
u\|^2_{\mcH^\al_{k}} \no
&\le&
c\|\m^\sharp\|^2_{L^\infty({\bf{\Large{H_{\la,\tau}}}})}E^{\;\al}_{k,\la}[u(\tau)]\;.
\eeaa
Similar calculations give:
\beaa  \int_{{\bf{\Large{H}}}_{\la,\tau}}
x^{-2\al-1+2\beta_1}(\m^{AB}[\p_A\p_B,\mcD^\beta]u)^2(s)dx\,
d\nu &\le&
c\|\m^\sharp\|^2_{L^\infty({\bf{\Large{H_{\la,\tau}}}})}\sum_A\|\p_A
u\|^2_{\mcH^\al_{k}}\no
&\le&
c\|\m^\sharp\|^2_{L^\infty({\bf{\Large{H_{\la,\tau}}}})}E^{\;\al}_{k,\la}[u(\tau)]\;.
\eeaa
We  obtain thus the following estimate for the first term of
the identity (\ref{E24}):
\bel{E25} \int_{{\bf{\Large{H}}}_{\la,\tau}}
x^{-2\al-1+2\beta_1}A^2_1dx\, d\nu \le
C\|\m^\sharp\|^2_{L^\infty({\bf{\Large{H_{\la,\tau}}}})}
E^{\;\al}_{k,\la}[u(\tau)]\;. \ee
Again since $\p_\tau$ and $\p_x$ commute with $\mcD^\beta$, if
we develop the second term of \eq{E24}, we find that:
$$ A_2 = {\Upsilon}^\nu[\mcD^\beta, \p_\nu]u =
{\Upsilon}^A[\mcD^\beta, \p_A]u$$
and    we then have the estimates:
\bel{E26}  \int_{{\bf{\Large{H}}}_{\la,\tau}}
x^{-2\al-1+2\beta_1}A^2_2dx\, d\nu \le \|
{\Upsilon}^A\|^2_{L^\infty}\|\p_Au\|^2_{\mcH^\al_{k-1}} \le \|
{\Upsilon}^A\|^2_{L^\infty} E^{\;\al}_{k,\la}[u(\tau)]\;.  \ee
As far as the third term is concerned, we write
\beaa A_3= \mcD^\beta\left({\Upsilon}^\nu \p_\nu
u\right)-{\Upsilon}^\nu \mcD^\beta\left(\p_\nu
  u\right)
&=& \mcD^\beta\left({\Upsilon}^\tau (\p_\tau-\p_x)
u\right)-{\Upsilon}^\tau \mcD^\beta\left((\p_\tau-\p_x)
  u\right)\\
&& +
  \mcD^\beta\left( ({\Upsilon}^x+{\Upsilon}^\tau)  \p_x
u\right) -({\Upsilon}^x+{\Upsilon}^\tau) \mcD^\beta\left(\p_x
  u\right)\\
&& +
  \mcD^\beta\left({\Upsilon}^A \p_A
u\right)-{\Upsilon}^A \mcD^\beta\left(\p_A
  u\right)\\
  &=:& I+II+III\;.\eeaa
Now we will use the weighted Moser-type  inequality (A.35) of
Proposition A.3 of~\cite{ChLengardnwe}  to estimate the terms
of $A_3$.  Its first term gives the following
\bean
 \lefteqn{
 \int_{{\bf{\Large{H}}}_{\la,\tau}}
x^{-2\al-1+2\beta_1}\left\{I\right\}^2dx\, d\nu
 }
 &&
\\
 \nonumber
  &=& \| x^{\beta_1}\mcD^\beta\left({\Upsilon}^\tau
(\p_\tau-\p_x) u\right)-x^{\beta_1}{\Upsilon}^\tau
\mcD^\beta\left((\p_\tau-\p_x)
 u\right)\|^2_{\mcH^{\al }_0({{\bf{\Large{H}}}_{\la,\tau}})}
\\
 \nonumber
&\le& C_s\left(\|(\p_\tau-\p_x)
u\|^2_{\mcB^\al_0}\|{\Upsilon}^\tau \|^2_{\mcG^0_{k}}+
\|(\p_\tau-\p_x) u\|^2_{\mcH^\al_{k-1
 }}\|{\Upsilon}^\tau \|^2_{\mcC^0_{\{x=0\},1}}\right)
\\
%
&\le& C\left(\|(\p_\tau-\p_x)
u\|^2_{\mcB^\al_0}\|{\Upsilon}^\tau \|^2_{\mcG^0_{k}}+
\|{\Upsilon}^\tau
 \|^2_{\mcC^0_{\{x=0\},1}}E^{\;\al}_k[u(\tau)]\right)
 \;.
 \label{19VIII0.4}
\eea
For the second term:
\beaa
 \lefteqn{
  \int_{{\bf{\Large{H}}}_{\la,\tau}}
x^{-2\al-1+2\beta_1}\left\{II\right\}^2dx\, d\nu
 }
 &&
\\
  &=&
\|x^{\beta_1}\mcD^\beta\left({\Upsilon}^x+{\Upsilon}^\tau
\right)\p_x u- x^{\beta_1} ({\Upsilon}^x+{\Upsilon}^\tau)\mcD^\beta( \p_x u)\|^2_{\mcH^{\al }_0({{\bf{\Large{H}}}_{\la,\tau}})}\\
&\le&
C_s\left(\|\p_xu\|^2_{\mcB^\al_0}\|{\Upsilon}^x+{\Upsilon}^\tau
\|^2_{\mcG^0_{k}}+ \|\p_x u\|^2_{\mcH^\al_{k-1
}}\|{\Upsilon}^x+{\Upsilon}^\tau\|^2_{\mcC^0_{\{x=0\},1}}
\right)\no
&\le& C\left(\| \p_x
u\|^2_{\mcB^\al_0}\|{\Upsilon}^x+{\Upsilon}^\tau
\|^2_{\mcG^0_{k}}+ \|{\Upsilon}^x+{\Upsilon}^\tau
\|^2_{\mcC^0_{\{x=0\},1}}E^{\;\al}_k[u(\tau)]\right)
\;.\eeaa
The same holds for the third term of $A_3$:
\beaa\int_{{\bf{\Large{H}}}_{\la,\tau}}
x^{-2\al-1+2\beta_1}\left\{III\right\}^2dx\, d\nu &=& \|
x^{\beta_1}\mcD^\beta\left({\Upsilon}^A\p_A
u\right)-x^{\beta_1}{\Upsilon}^A
\mcD^\beta\left(\p_A  u\right)\|^2_{\mcH^{\al }_0({{\bf{\Large{H}}}_{\la,\tau}})}\\\
&\le& C_s\left(\|\p_Au\|^2_{\mcB^\al_0}\|{\Upsilon}^A
\|^2_{\mcG^0_{k}}+ \|\p_A u\|^2_{\mcH^\al_{k-1
}}\|{\Upsilon}^A \|^2_{\mcC^0_{\{x=0\},1}}\right)\\
&\le& C\left(\| \p_A u\|^2_{\mcB^\al_0}\|{\Upsilon}^A
\|^2_{\mcG^0_{k}}+ \|{\Upsilon}^A
\|^2_{\mcC^0_{\{x=0\},1}}E^{\;\al}_k[u(\tau)]\right)
\; .\eeaa
We then obtain the following estimate for the third term of
equation (\ref{E24})
\bean
 \lefteqn{
  \int_{{\bf{\Large{H}}}_{\la,\tau}}
 x^{-2\al-1+2\beta_1}(A_3)^2dx\, d\nu
 }
 &&
\\
  &\le& \|(\p_\tau-\p_x) u\|_{\mcB^\al_0}
\|{\Upsilon}^\tau\|^2_{\mcG^0_k}+ \|\p_x u\|^2_{\mcB^\al_0}
\|{\Upsilon}^\tau+{\Upsilon}^x\|^2_{\mcG^0_k}
+\|\p_Au\|^2_{\mcB^\al_0} \|{\Upsilon}^A\|^2_{\mcG^0_k}\no
 &&+\left\{\|{\Upsilon}^\tau\|^2_{\mcC^0_{\{x=0\},1}}
 +\|{\Upsilon}^x+{\Upsilon}^\tau\|^2_{\mcC^0_{\{x=0\},1}} +
 \|{\Upsilon}^A\|^2_{\mcC^0_{\{x=0\},1}}
 \right\}E^\al_{k,\la}[u(\tau)]
 \;.
\no
  \label{E30}
\eea
In order to   estimate  the fourth term $A_4$ of (\ref{E24}),
we need to look separately at each of its components as we have
to make sure that every  $\p_x^2$ comes with a factor of $x$.
We write
\be \label{E36}A_4= A^{00}+ 2A^{\tau x}+2A^{\tau
A}+A^{xx}+2A^{xA}+A^{AB},\ee
where the labeling $A^{ab}$  corresponds to the terms obtained
when in $A_4$ we replace $\m^{\al\beta}\p^2_{\al\beta}$ with
its expression as in (\ref{E45}). Now we use again the weighted
Moser type inequality of Proposition A.3 of \cite{ChLengardnwe}
to estimate these terms. We have:
\bean
 \lefteqn{
 \int_{{\bf{\Large{H}}}_{\la,\tau}}
x^{-2\al-1+2\beta_1}\left\{A^{AB}\right\}^2dx\, d\nu
 }
 &&
\\
  &=& \| x^{\beta_1}\mcD^\beta\left(\m^{AB}\p_A \p_B u\right)-
x^{\beta_1}\m^{AB}\mcD^\beta\left(\p_A \p_B
u\right)\|_{\mcH^{\al }_0({{\bf{\Large{H}}}_{\la,\tau}})}\no
&\le&
C_s\sum_A\left(\|\p_Au\|^2_{\mcB^\al_1}\|\m^\sharp\|^2_{\mcG^0_{k}}+
\|\p_A u\|^2_{\mcH^\al_{k}}\|\m^\sharp
\|^2_{\mcC^0_{\{x=0\},1}}\right)\no
&\le&C\left(\sum_A\|\p_Au\|^2_{\mcB^\al_1}\|\m^\sharp\|^2_{\mcG^0_{k}}+
\|\m^\sharp
\|^2_{\mcC^0_{\{x=0\},1}}E^\al_{k,\la}[u(\tau)]\right)\;,
\phantom{xxxxx}
 \label{20VIII0.1}
\label{EE1}
\eea
and
\bean
 \lefteqn{
  \int_{{\bf{\Large{H}}}_{\la,\tau}}
 x^{-2\al-1+2\beta_1}\left\{A^{\tau A}\right\}^2dx\, d\nu }
  &&
\\
&=& \| x^{\beta_1}\mcD^\beta\left(\m^{\tau A}\p_A (\p_\tau
-\p_x )u\right)- x^{\beta_1}\m^{\tau A}\mcD^\beta\left(\p_A
(\p_\tau-\p_x) u\right)\|_{\mcH^{\al}_0}\no
&\le& C_s\left(\|(\p_\tau-\p_x)
u\|^2_{\mcB^\al_1}\|\m^\sharp\|^2_{\mcG^0_{k}}+
\|(\p_\tau-\p_x) u\|^2_{\mcH^\al_{k}}\|\m^\sharp
\|^2_{\mcC^0_{\{x=0\},1}}\right)\no
&\le& \label{EEE2}C_s\left(\|(\p_\tau-\p_x)
u\|^2_{\mcB^\al_1}\|\m^\sharp\|^2_{\mcG^0_{k}}+ \|\m^\sharp
\|^2_{\mcC^0_{\{x=0\},1}}E^\al_{k,\la}[u(\tau)]\right)
 .
 \phantom{xxx}
\eea
Continuing in this way  we have:
\bean
 \lefteqn{ \int_{{\bf{\Large{H}}}_{\la,\tau}}
x^{-2\al-1+2\beta_1}\left\{A^{xA}\right\}^2dx\, d\nu
 }
 &&
\\
   &\le& \| x^{\beta_1}\mcD^\beta\left\{\left(\m^{xA}+
\m^{\tau A}\right)\p_A\p_xu\right\}- x^{\beta_1}\left(\m^{xA}+
\m^{\tau A}\right)\mcD^\beta \p_A\p_x u\|_{\mcH^{\al}_0}\no
&\le& C\left(\|\p_A\p_x u\|^2_{\mcB^\al_0}\|\left(\m^{xA}+
\m^{\tau A}\right)\|^2_{\mcG^0_{k}} + \|\p_A\p_x
u\|^2_{\mcH^\al_{k-1}}\|\left(\m^{xA}+ \m^{\tau A}\right)
\|^2_{\mcC^0_{\{x=0\},1}}\right) \no
&\le& C\sum_A\left(\|\p_x u\|^2_{\mcB^\al_1}\|\left(\m^{xA}+
\m^{\tau A}\right)\|^2_{\mcG^0_{k}} + \|\p_x
u\|^2_{\mcH^\al_{k}}\|\left(\m^{xA}+ \m^{\tau A}\right)
\|^2_{\mcC^0_{\{x=0\},1}}\right) \no
&\le& \label{EEE3} C\sum_A\left(\|\p_x
u\|^2_{\mcB^\al_1}\|\left(\m^{xA}+ \m^{\tau
A}\right)\|^2_{\mcG^0_{k}} + \|\left(\m^{xA}+
\m^{\tau A}\right)
\|^2_{\mcC^0_{\{x=0\},1}}E^\al_{k,\la}[u(\tau)]\right) .
 \phantom{xxxxxx}
\eea
We recall that $\m^{\tau\tau}+\m^{x\tau}= x\f^1(\tau,x,v^A)$,
we then obtain the following expression for $A^{\tau x}$.
\beaa A^{\tau x}&=& \mcD^\beta\left[  \f^1x\p_x
(\p_\tau -\p_x)u\right]- x \f^1\mcD^\beta\left[\p_x(\p_\tau-\p_x) u\right]\\
&=&  \mcD^\beta\left[  \f^1x\p_x (\p_\tau-\p_x) u\right]-
\f^1\mcD^\beta\left[x\p_x(\p_\tau-\p_x)u\right]\\
&&\qquad\qquad\qquad+
\underbrace{\f^1\mcD^\beta\left[x\p_x(\p_\tau-\p_x)u\right]-x
\f^1\mcD^\beta\left[\p_x(\p_\tau-\p_x) u\right]}_{=\;\beta_1
\f^1\mcD^\beta(\p_\tau-\p_x)u  } \;.\eeaa
Since
$$\int_{{\bf{\Large{H}}}_{\la,\tau}}
x^{-2\al-1+2\beta_1}\left\{\f^1\mcD^\beta(\p_\tau-\p_x)u
\right\}^2dx\, d\nu \le
\|\f^1\|^2_{L^\infty}\|(\p_\tau-\p_x)u\|^2_{\mcH^\al_k} \;,$$
we have
\bea\lefteqn{\int_{{\bf{\Large{H}}}_{\la,\tau}}
x^{-2\al-1+2\beta_1}\left\{A^{\tau x}\right\}^2dx\, d\nu} \no
 &\le&
C_s\left(\| (\p_\tau-\p_x)
u)\|^2_{\mcB^\al_1}\|\f^1\|^2_{\mcG^0_{k}}+ \|
(\p_\tau-\p_x)u\|^2_{\mcH^\al_{k }}\|\f^1
\|^2_{\mcC^0_{\{x=0\},1}}\right)\no
&\le& \label{EEE4}C\left(\| (\p_\tau-\p_x)
u)\|^2_{\mcB^\al_1}\|\f^1\|^2_{\mcG^0_{k}}+ \|\f^1
\|^2_{\mcC^0_{\{x=0\},1}}E^\al_{k,\la}[u(\tau)]\right) .\eea
On  the other hand, since $\m^{\tau\tau}+2\m^{\tau x}+
\m^{xx}=1+ x\f$,  we have
\bea \int_{{\bf{\Large{H}}}_{\la,\tau}}
x^{-2\al-1+2\beta_1}\left\{A^{xx}\right\}^2dx\, d\nu &\le& \|
x^{\beta_1}\mcD^\beta\left(\f x\p_x [\p_xu]\right)-
x^{\beta_1}\f\mcD^\beta\left(x\p_x[\p_x u]
\right)\|^2_{\mcH^{\al }_0({{\bf{\Large{H}}}_{\la,\tau}})}\no
&&+\|\f
\underbrace{\left\{x^{\beta_1}\mcD^\beta\left(x\p_x[\p_x
u]\right)-x^{\beta_1}x\mcD^\beta \left(\p^2_x
u\right)\right\}}_{= \; \beta_1 x^{\beta_1}\mcD^\beta( \p_xu)
}\|^2_{\mcH^{\al }_0({{\bf{\Large{H}}}_{\la,\tau}})}\no
&\le& C_s\Big(\|x\p_x
[\p_xu]\|^2_{\mcB^\al_0}\|\f\|^2_{\mcG^0_{k}}+ \|x\p_x [\p_x
u]\|^2_{\mcH^\al_{k-1}}\|\f \|^2_{\mcC^0_{\{x=0\},1}}\no
&& \qquad\qquad\qquad\qquad\qquad +
\|\f\|^2_{L^{\infty}}\|\p_xu\|^2_{\mcH^\al_{k }}\Big)\nonumber\\
&\le&\label{E31}C_s\left(\|
\p_xu\|^2_{\mcB^\al_1}\|\f\|^2_{\mcG^0_{k}}+ \|\f
\|^2_{\mcC^0_{\{x=0\},1}}E^\al_{k,\la}[u(\tau)] \right) .\eea
We note that $\|x^j\p_x^j \Phi\|_{\mcH^\al_k} \le \|
\Phi\|_{\mcH^\al_{k+j}} $ which can be shown by induction.  In
order to estimate the term $A^{00}$, we proceed as follows:
\bea  A^{00}= \left[\mcD^\beta,\m^{\tau\tau}\left(\p_\tau-\p_x
\right)^2\right]u &=& \mcD^\beta\left([-1 +
x\f^0]\left(\p_\tau-\p_x \right)^2u\right)-  [-1 +
x\f^0]\mcD^\beta\left(\p_\tau-\p_x \right)^2u\nonumber\\
&=& \label{E35} \mcD^\beta\left([x\f^0]\left(\p_\tau-\p_x
\right)^2u\right)- [ x\f^0]\mcD^\beta\left(\p_\tau-\p_x
\right)^2u\;.\eea
Now  using equation (\ref{E0}), (\ref{E46}) and (\ref{E45}), we
obtain the following expression of $
\left(\p_\tau-\p_x\right)^2 u$:
\bea\label{E28} \left(\p_\tau-\p_x\right)^2 u &=&
-2\left(\hat\m^{\tau\tau}+\hat\m^{\tau
x}\right)\left(\p_\tau-\p_x\right)\p_x-\left(\hat\m^{\tau\tau}+2\hat\m^{\tau
x}+\hat\m^{xx}\right)\p^2_x - 2\hat\m^{\tau
A}\left(\p_\tau-\p_x\right)\p_A \nonumber\\&&
-2\left(\hat\m^{xA}+\hat\m^{\tau A}\right)\p_x\p_A
-\m^{AB}\p_A\p_B\ -\hat \Upsilon^\sigma \p_\sigma u + \hat
F\;.\eea
Here the hat means multiplication with $ 1/{\m^{\tau\tau}}$
(recall $|\m^{\tau\tau}|>\ep_0>0 \;).$  We will need the
following:

\begin{Lemma}\label{Lem0}
Let  \bel{hyplem}\tilde{\p}= ( x\p_x, \p_A), \quad k\in \N^*,
\quad \theta \in \R , \quad \hat{\psi} =
\frac{\psi}{\m^{\tau\tau}}\;, \quad \left|\frac
1{\m^{\tau\tau}}\right|\le \frac 1{\ep_0}\;.\ee
We have the following estimates:
\bel{CC0}\|\hat{\psi}\|_{\mcC^\theta_{\{x=0\},0}} \le \frac
1{\ep_0}\|\psi\|_{\mcC^\theta_{\{x=0\},0}} \;, \ee
\bel{CC1} \|\hat{\psi}\|_{\mcC^\theta_{\{x=0\},1}} \le \frac
1{\ep_0}\|\psi\|_{\mcC^\theta_{\{x=0\},1}}+\frac 1{\ep_0^2}
\|\tilde{\p}(x
\f^0)\|_{L^\infty}\|\psi\|_{\mcC^\theta_{\{x=0\},0}} \;, \ee
and
\bel{CC2} \|\hat{\psi}\|_{\mcH^\theta_k} \le \frac 1{\ep_0}
 \|\psi\|_{\mcH^\theta_k}
  + C(\ep_0,\|\f^0\|_{L^\infty})
 \|\psi\|_{\mcB^\theta_0}\left(1+\|\f^0\|_{\mcH^{-1}_k}\right)
 \;,
\ee
with identical estimates with $\mcC^\theta_{\{x=0\},0}$
replaced by $\mcB^\theta_0$ and $\mcH^\theta_k$ replaced by
$\mcG^\theta_k$.

\end{Lemma}

\proof The first inequality is obvious. Next:
\beaa \|\hat{\psi}\|_{\mcC^\theta_{\{x=0\},1}} &\leq&
\|x^{-\theta}\frac{1}{\m^{\tau\tau}}\psi\|_{L^\infty} +
\|x^{-\theta}\tilde{\p}\left\{\frac{1}{\m^{\tau\tau}}\psi\right\}\|_{L^\infty}\\
&\le& \frac 1{\ep_0}\|\psi\|_{\mcC^\theta_{\{x=0\},0}}+
\|x^{-\theta}\left\{\psi\tilde{\p}(\frac{1}{\m^{\tau\tau}}) +
\frac{1}{\m^{\tau\tau}}\tilde{\p}\psi\right\}\|_{L^\infty} \\
&\le&   \frac 1{\ep_0}\|\psi\|_{\mcC^\theta_{\{x=0\},0}}+ \frac
1{\ep_0^2}\|\psi\|_{\mcC^\theta_{\{x=0\},0}}\|\tilde{\p}(x
\f^0)\|_{L^\infty}+ \frac
1{\ep_0}\|\tilde{\p}\psi\|_{\mcC^\theta_{\{x=0\},0}} \\
&\le&\frac 1{\ep_0}\|\psi\|_{\mcC^\theta_{\{x=0\},1}}+\frac
1{\ep_0^2} \|\tilde{\p}(x
\f^0)\|_{L^\infty}\|\psi\|_{\mcC^\theta_{\{x=0\},0}}\;.
\eeaa
On the other hand, from inequality (A.27) of
\cite{ChLengardnwe} we have:
\bea\label{C}\|\hat{\psi}\|_{\mcH^\theta_k}&=&\|\frac{1}{\m^{\tau\tau}}\psi\|_{\mcH^\theta_k}
\le\|\psi\|_{\mcB^\theta_0}\|\frac{1}{\m^{\tau\tau}}\|_{\mcG^0_k}+
\|\psi\|_{\mcH^\theta_k}\|\frac{1}{\m^{\tau\tau}}\|_{\mcC^0_{\{x=0\},0}}\nonumber\\
&\le& \frac 1{\ep_0} \|\psi\|_{\mcH^\theta_k} +
\|\psi\|_{\mcB^\theta_0}\|\frac{1}{\m^{\tau\tau}}\|_{\mcG^0_k}\;.\eea
Now, by hypothesis we have,
\beaa \frac{1}{\m^{\tau\tau}(\tau,x,v^A)}&=& \frac{1}{-1+x
\f^0(\tau,x,v^A)}= -1 +
\frac{x\f^0(\tau,x,v^A)}{-1+x\f^0(\tau,x,v^A)}\\
&=& -1 + G(\tau,x,v^A, x\f^0) \;,\eeaa
where $G$ is any function which takes the correct values in the
range of interest, e.g.,
$$G(\tau,x,v^A, p) = \frac{p\chi(p)}{-1+p}\quad \text{with}\quad \chi \in C^{\infty}(\R)\quad \text{such that}\quad \chi(p) =
\left\{\begin{array}{l} 1 \quad \text{if}\quad p \le 1-\frac{3\ep_0}{4} \\
0 \quad \text{if}\quad p \ge
1-\frac{\ep_0}{4}\end{array}\right.\;.$$
Recall that hypothesis (\ref{G0}) reads $x\f^0 \le 1-\ep_0$. We
have (note that the space of functions $\mcG^\theta_k$ contains
constant functions)
\bel{C0}\|\frac{1}{\m^{\tau\tau}}\|_{\mcG^0_k} \le
\|1\|_{\mcG^0_k}+ \|G(.\;,x\f^0)\|_{\mcG^0_k}\le
C\left(1+\|G(.,\;x\f^0)\|_{\mcG^0_k}\right)\;.\ee
The function $G$ satisfies the following, for any $p\in \R$:
$$\|G(.\;, p)\|_{\mcC^0_{\{x=0\},k}} = \|G(.\;, p)\|_{\mcC^0_{\{x=0\},0}}\le C(\ep_0) $$
and for $i=0,1\;;$
$$  \left\|\frac{\p^i
G(.\;,p)}{\p p^i}\right\|_{\mcC^0_{\{x=0\},k-i}} \le
C(\ep_0)|p|^{1-i}\;.$$
These two inequalities show that $G$ has a uniform zero of
order $1$ at $p=0$. %
Therefore, we can apply inequality (A.31) of
\cite{ChLengardnwe} and obtain that
$$\|G(.\;, x\f^0)\|_{\mcG^0_k} \le C(\ep_0,\|\f^0\|_{L^\infty})\|h^0\|_{\mcH^{-1}_k}\;.$$
This implies (see (\ref{C0}))
\bel{C1} \|\frac{1}{\m^{\tau\tau}}\|_{\mcG^0_k} \le
C(\ep_0,\|\f^0\|_{L^\infty})\left(1+\|\f^0\|_{\mcH^{-1}_k}\right),
\ee
and (\ref{C}) leads to \eq{CC2}.
\qed

\medskip
%
If we insert (\ref{E28}) into equation (\ref{E35}), we obtain
seven commutators which we label $A_a^{00}\;, a = 1, \dots\;,
7.$ These terms can be estimated in the same way as we did
before, using (A.34), (A.35) of~\cite{ChLengardnwe} and
Lemma~\ref{Lem0}. They will be analyzed in the  order
$7-3-5-1-2-4-6$. %
Let us estimate  the term  $A^{00}_7$ containing the source
term $F$. We have
\bea\int_{{\bf{\Large{H}}}_{\la,\tau}}
x^{-2\al-1+2\beta_1}\left\{A_7^{00}\right\}^2dx\, d\nu & = &
\|x^{\beta_1}\mcD^\beta\left([x\f^0] \hat{F} \right)-
x^{\beta_1}[x\f^0]\mcD^\beta \hat{F}\|^2_{\mcH^\al_0}\no
&\le & C\left(\|
\hat{F}\|^2_{\mcB^\al_0}\|x\f^0\|^2_{\mcG^0_{k}}+
\|\hat{F}\|^2_{\mcH^\al_{k-1}}\|x\f^0\|^2_{\mcC^0_{\{x=0\},1}}\right)\no
&\le & \label{EEq1}C(\ep_0)\|
F\|^2_{\mcB^\al_0}\|x\f^0\|^2_{\mcG^0_{k}}+
C(\ep_0)\|x\f^0\|^2_{\mcC^0_{\{x=0\},1}}\no
&&\times\left\{\|F\|^2_{\mcH^\al_{k-1}} + \|
F\|^2_{\mcB^\al_0}C(\|\f^0\|_{L^\infty})
 \left(1+\|x\f^0\|_{\mcH^0_{k-1}}\right)\right\}
 .\phantom{xxxxx}
 \eea
The third term can be estimated as follows:
\bean
 \lefteqn{
 \int_{{\bf{\Large{H}}}_{\la,\tau}}
 x^{-2\al-1+2\beta_1}\left\{A_3^{00}\right\}^2dx\, d\nu
  }
   &&
\\
 &=&2\|\mcD^\beta\left(x\f^0 \hat{\m}^{\tau A}\p_A(\p_\tau-\p_x)
 u\right)- x\f^0\mcD^\beta\left(\hat{\m}^{\tau A}\p_A(\p_\tau
 -\p_x)u\right)\|^2_{\mcH^\al_0}\no
&\le& C\| \hat{\m}^{\tau A}\p_A(\p_\tau-\p_x) u
\|^2_{\mcB^\al_0}\|x\f^0\|^2_{\mcG^0_k} + C\|\hat{\m}^{\tau A}\p_A(\p_\tau -\p_x)u
\|^2_{\mcH^\al_{k-1}}\|x\f^0\|^2_{\mcC^0_{\{x=0\},1}}\no
&\le& C\| \hat{\m}^\sharp\|^2_{L^\infty}\|(\p_\tau-\p_x)
u\|^2_{\mcB^\al_1}\|x\f^0\|^2_{\mcG^0_k} \no
&&+
C\|x\f^0\|^2_{\mcC^0_{\{x=0\},1}}\left\{\|(\p_\tau-\partial_\tau)
u\|^2_{\mcB^\al_1}\|\hat{\m}^\sharp\|^2_{\mcG^0_{k-1}}+ \|(\p_\tau-\p_x)u \|^2_{\mcH^\al_{k}}\|
\hat{\m}^\sharp\|^2_{\mcC^0_{\{x=0\},0}}\right\}\no
& \le & C(\ep_0) \|\m^\sharp\|^2_{L^\infty}\|(\p_\tau-\p_x)
u\|^2_{\mcB^\al_1}\|x\f^0\|^2_{\mcG^0_{k}} +C(\ep_0)\|x\f^0\|^2_{\mcC^0_{\{x=0\},1}}\|
\m^\sharp\|^2_{L^\infty}\|(\p_\tau-\p_x)u
\|^2_{\mcH^\al_{k}}\no
&&+ \|x\f^0\|^2_{\mcC^0_{\{x=0\},1}}\|(\p_\tau-\tau)
u\|^2_{\mcB^\al_1}\left\{\|\m^\sharp\|^2_{\mcG^0_{k-1}}+
\|\m^\sharp\|^2_{L\infty}C(\|\f^0\|_{L^\infty})\left(1+\|x\f^0\|^2_{\mcG^0_{k-1}}\right)\right\}
 \;.
\no
%
  \label{Eq8}
\eeal{24VIII0.1}
%
%
A similar analysis applies to   $A^{00}_5$.

As far as the first term  $A^{00}_1$  is concerned, we have
$$ -\frac 12 A^{00}_1 =
\mcD^\beta\left(x\f^0\hat{\f}^1(x\p_x)(\p_\tau-\p_x) u\right) -
x\f^0\mcD^\beta\left(\hat{\f}^1(x\p_x)(\p_\tau-\p_x)u\right)\;.$$
Using again the weighted Moser-type  inequality of
\cite{ChLengardnwe}, we can evaluate the square of its norm as
follows:
\beaa\lefteqn{\int_{{\bf{\Large{H}}}_{\la,\tau}}
x^{-2\al-1+2\beta_1}(A^{00}_1)^2dx\, d\nu}
\\
 &=&
2\|x^{\beta_1}\mcD^\beta\left([x\f^0][\hat{\f}^1(x\p_x)(\p_\tau-\p_x)
u]\right)-x^{\beta_1}
[ x\f^0]\mcD^\beta\left( \hat{\f}^1 (x\p_x)(\p_\tau-\p_x) u\right)\|^2_{\mcH^{\al }_0}\\
& \le&  C\left(\|\hat{\f}^1 (x\p_x)(\p_\tau-\p_x)
u\|^2_{\mcB^\al_0}\| x\f^0\|^2_{\mcG^0_{k}}+ \| \hat{\f}^1 (x\p_x)(\p_\tau-\p_x)
u\|^2_{\mcH^\al_{k-1}}\|x\f^0
\|^2_{\mcC^0_{\{x=0\},1}}\right) \\
& \le&  C(\ep_0)\left(\|\f^1\|_{L^\infty}\|(\p_\tau-\p_x)
u\|^2_{\mcB^\al_1}\| x\f^0\|^2_{\mcG^0_{k}} +  \|
\hat{\f}^1 (x\p_x)(\p_\tau-\p_x) u\|^2_{\mcH^\al_{k-1}}\|x\f^0
\|^2_{\mcC^0_{\{x=0\},1}}\right)\;.\eeaa
Using now inequality (A.34)  of~\cite{ChLengardnwe}, we have
(the last inequality is obtained by using (\ref{CC2}):
\beaa
 \lefteqn{\| \hat{\f}^1 (x\p_x)(\p_\tau-\p_x) u\|^2_{\mcH^\al_{k-1}}}
 \\
 &\leq&
  C\left( \|x\p_x(\p_\tau-\p_x) u
 \|^2_{\mcB^\al_{0}}\|\hat{\f}^1\|^2_{\mcG^0_{k-1}}
 + \|x\p_x(\p_\tau-\p_x)
 u\|^2_{\mcH^\al_{k-1}}\|\hat{\f}^1\|^2_{\mcC^0_{0}} \right)
\\
&\le&   C\| (\p_\tau-\p_x) u
\|^2_{\mcB^\al_1}\|\hat{\f}^1\|^2_{\mcG^0_{k-1}}
  +
 C(\ep_0)\|\f^1\|^2_{L^\infty}\|
(\p_\tau-\p_x) u\|^2_{\mcH^\al_{k }}\\
&\le & C(\ep_0)\| \f^1\|^2_{L^\infty}\| (\p_\tau-\p_x)
u\|^2_{\mcH^\al_{k }} + C(\ep_0)\| (\p_\tau-\p_x) u
\|^2_{\mcB^\al_{1}}\\&&\times\left\{\|\f^1\|^2_{\mcG^0_{k-1}}
+C(\|\f^0\|_{L^\infty})\|\f^1\|^2_{L^\infty}(1+\|x\f^0\|^2_{\mcG^0_{k-1}})\right\},
\eeaa
which  gives
\bea \lefteqn{\int_{{\bf{\Large{H}}}_{\la,\tau}}
x^{-2\al-1+2\beta_1}(A^{00}_1)^2dx\, d\nu} \no &\leq&
 C(\ep_0)\|\f^1\|^2_{L^\infty}\|(\p_\tau-\p_x)
u\|^2_{\mcB^\al_1}\| x\f^0\|^2_{\mcG^0_{k}}+C(\ep_0)\|x\f^0\|^2_{\mcC^0_{\{x=0\},1}}\|
\f^1\|^2_{L^\infty}\| (\p_\tau-\p_x) u\|^2_{\mcH^\al_{k }}\no
&&+ C(\ep_0)\|x\f^0\|^2_{\mcC^0_{\{x=0\},1}}\| (\p_\tau-\p_x) u
\|^2_{\mcB^\al_{1}}\left\{\|\f^1\|^2_{\mcG^0_{k-1}}
+C(\|\f^0\|_{L^\infty})\|\f^1\|^2_{L^\infty}(1+\|x\f^0\|^2_{\mcG^0_{k-1}})\right\}\no
&\le& C(\ep_0)\|x\f^0\|_{\mcC^0_{\{x=0\},1}}\|
\f^1\|^2_{L^\infty}\| (\p_\tau-\p_x) u\|^2_{\mcH^\al_{k }}\no
&&+ C(\ep_0)\left(1+\|x\f^0\|_{\mcC^0_{\{x=0\},1}}\right)\|
(\p_\tau-\p_x) u \|^2_{\mcB^\al_{1}}\left\{\|\f^1\|^2_{\mcG^0_{k-1}}
+C(\|\f^0\|_{L^\infty})\|\f^1\|^2_{L^\infty}(1+\|x\f^0\|^2_{\mcG^0_{k}})\right\}
.
\no
\label{Eq1}
\eea
Since the terms $A^{00}_1$ and $A^{00}_4$  have same the
structure, to estimate the second one, we just have to replace
in the estimate on $A^{00}_1, $ $\|(\p_\tau -\p_x)
u\|^2_{\mcH^\al_{k}}$ by $\|\p_x u\|^2_{\mcH^\al_{k}}$ and
$\|x\f^1\|^2_{\mcG^0_{k-1}}$ by $\|\hat{\m}^{\tau
A}+\hat{\m}^{xA}\|^2_{\mcG^0_{k-1}}$. 
%

We continue with the most dangerous term $A^{00}_2$. We have
(recall that $\hat 1 = 1/ \m^{\tau\tau}$)
$$ - A^{00}_2 =
\mcD^\beta\left([x\f^0](\hat{1}+x\hat{\f})\p^2_x u\right) -
[x\f^0]\mcD^\beta\left([\hat{1}+x\hat{\f}]\p^2_xu\right)
\;,$$
\beaa
 \int_{{\bf{\Large{H}}}_{\la,\tau}}
 x^{-2\al-1+2\beta_1}\left\{A_2^{00}\right\}^2
 &=&
 \|x^{\beta_1}\mcD^\beta\left([\f^0][\hat{1}+x\hat{\f}
 ](x\p_x)\p_x u\right)
\\
 && -x^{\beta_1}
 [x\f^0]\mcD^\beta\left([\hat{1}+x\hat{\f} ]\p^2_xu\right)\|^2_{\mcH^\al_0}\\
&\le& \|x^{\beta_1}\mcD^\beta\left(x\f^0 (\hat{1}.\p^2_x
u)\right)- x^{\beta_1}[
x\f^0]\mcD^\beta\left( \hat{1}.\p^2_xu\right)\|^2_{\mcH^\al_0}\\
&&+\|x^{\beta_1}\mcD^\beta\left(x\f^0\hat{\f} x\p_x(\p_x
u)\right)- x^{\beta_1}[
x\f^0]\mcD^\beta\left(\hat{\f} x\p_x(\p_x u)\right)\|^2_{\mcH^\al_0}\\
&=:& (a)+(b)\;. \eeaa
Now, estimating these two expressions as we did with
$A^{00}_1$, we obtain the following
\beaa (b)&\le& C\left( \|\hat{\f}(x\p_x)\p_x
u\|^2_{\mcB^\al_0}\|x\f^0\|^2_{\mcG^0_k} +\|\hat{\f}(x\p_x)\p_x
u\|^2_{\mcH^\al_{k-1}}\|x\f^0\|^2_{\mcC^0_{\{x=0\},1}}\right)\no
&\le& C(\ep_0) \|\f\|^2_{L^\infty}\|
\p_xu\|^2_{\mcB^\al_1}\|x\f^0\|^2_{\mcG^0_k}
+C\|\hat{\f}(x\p_x)\p_x
u\|^2_{\mcH^\al_{k-1}}\|x\f^0\|^2_{\mcC^0_{\{x=0\},1}}
\;.\eeaa
Now equations (A.34) of~\cite{ChLengardnwe} and (\ref{CC2})
give
\beaa\|\hat{\f}(x\p_x)\p_x u\|^2_{\mcH^\al_{k-1}} &\le&
C\left(\|(x\p_x)\p_xu\|^2_{\mcB^\al_0}\|\hat{\f}\|^2_{\mcG^0_{k-1}}
+\|(x\p_x)\p_xu\|^2_{\mcH^\al_{k-1}}\|\hat{\f}\|^2_{\mcC^0_{\{x=0\},0}}
\right)\no
&\le&C(\ep_0)\|\f\|^2_{L^\infty}\|\p_xu\|^2_{\mcH^\al_{k}}\no
&&+C(\ep_0)
\|\p_xu\|^2_{\mcB^\al_{1}}\left\{\|\f\|^2_{\mcG^0_{k-1}}+
\|\f\|^2_{L^\infty}C(\|\f^0\|_{L^\infty})(1+\|x\f^0\|^2_{\mcG^0_{k-1}})\right\}
\;,\eeaa
which gives the desired estimate for $(b)$.

In order to estimate  the term $(a)$ we write   $\beta =
(\beta_1, \beta_v)$ and $\mcD^\beta =
\p_x^{\beta_1}X_v^{\beta_v}, $ with
$X_v^{\beta_v}=X_2^{\b_2}\ldots X_r^{\b_r}$:
\bea\mcD^\beta\left(\f^0 x(\hat{1}.\p^2_x u)\right)- [
x\f^0]\mcD^\beta\left( \hat{1}.\p^2_xu\right)&=&
\mcD^\beta\left(\f^0 x(\hat{1}.\p^2_x u)\right)-
\f^0\mcD^\beta\left( \hat{1}.x\p^2_xu\right)\no&&
+\f^0\mcD^\beta\left( \hat{1}.x\p^2_xu\right) -[
x\f^0]\mcD^\beta\left( \hat{1}.\p^2_xu\right) \no &=:&
\label{EEq2} (1)+(2)\;.\eea
We have
$$ x^{\beta_1}(2) = \beta_1\f^0 x^{\beta_1}\p_v^{\beta'}\p_x^{\beta_1-1}(\hat{1}.\p^2_xu)
$$
as well as
$$x^{-2\al-1+2\beta_1}(2)^2 = \beta_1^2(\f^0)^2 x^{-2(\al-1)-1+2(\beta_1-1)}
\left(\p_v^{\beta'}\p_x^{\beta_1-1}(\hat{1}.\p^2_xu)\right)^2\;.$$
This identity leads to
\beaa
 \lefteqn{
    \|x^{\beta_1}(2)\|^2_{\mcH^\al_0}
        \le
C\|\f^0\|^2_{L^\infty}\|\hat{1}.\p^2_x u\|^2_{\mcH^{\al-1}_{k-1}}
 } &&
\\
&\le& C\|\f^0\|^2_{L^\infty}\left(\|\p^2_x
u\|^2_{\mcB^{\al-1}_0}\|\hat{1}\|^2_{\mcG^0_{k-1}}+ \|\p^2_x
u\|^2_{\mcH^{\al-1}_{k-1}}\|\hat{1}\|^2_{\mcC^0_{0}}\right)\\
&\le& C\|\f^0\|^2_{L^\infty} \left(\|
\p_xu\|^2_{\mcB^{\al}_1}\|\hat{1}\|^2_{\mcG^0_{k-1}}+
\frac{1}{\ep_0}\|\p_xu\|^2_{\mcH^{\al}_{k}}\right) \;.\eeaa
Using again (\ref{CC2}) we have:
$$\|\hat{1}\|^2_{\mcG^0_{k-1}} \le \frac 1{\ep_0}\|1\|^2_{\mcG^0_{k-1}}
+ C(\|\f^0\|_{L^\infty})\left(1+\|x^2\f^0\|^2_{\mcG^0_{k-1}}\right),
$$
that is
\bel{Eq3}\|\hat{1}\|^2_{\mcG^0_{k-1}} \le
C(\|\f^0\|_{L^\infty})\left(1+\|x^2\f^0\|^2_{\mcG^0_{k-1}}\right)\;.
\ee
Thus,
\bea\label{Eq4} \|x^{\beta_1}(2)\|^2_{\mcH^\al_0} &\le&
C(\|\f^0\|_{L^\infty})\left\{\|
\p_xu\|^2_{\mcB^{\al}_1}\left(1+\|x\f^0\|^2_{\mcG^0_{k-1}}\right)
 + \frac{1}{\ep_0}E^\al_{k,\la}[u(\tau)]\right\}
 \;.
\no
%
\eea
As far as the first term of (\ref{EEq2}) is concerned, we have:
%
\bea\|x^{\beta_1}(1)\|^2_{\mcH^\al_0}
&=&\|x^{\beta_1}\mcD^\beta\left(\f^0 (\hat{1}.(x\p_x) \p_x
u)\right)- x^{\beta_1}\f^0\mcD^\beta\left( \hat{1}.(x\p_x)\p_x
u\right)\|^2_{\mcH^\al_0}\no
&\le&C\left\{\|\hat{1}.(x\p_x)\p_x
u\|^2_{\mcB^\al_0}\|\f^0\|^2_{\mcG^0_k} +\|\hat{1}.(x\p_x)\p_x
u\|^2_{\mcH^\al_{k-1}}\|\f^0\|^2_{\mcC^0_{\{x=0\},1}}\right\}\no
&\le& C(\ep_0)\Bigg\{\|\p_x
u\|^2_{\mcB^\al_1}\|\f^0\|^2_{\mcG^0_k}\no
&& +\|\f^0\|^2_{\mcC^0_{\{x=0\},1}}\left\{\|(x\p_x)\p_x
u\|^2_{\mcB^\al_0}\|\hat{1}\|^2_{\mcG^0_{k-1}} +\|(x\p_x)\p_x
u\|^2_{\mcH^\al_{k-1}}\|\hat{1}\|^2_{\mcC^0_{\{x=0\},0}}
\right\} \Bigg\}\no
&\le& C(\ep_0)\left\{\|\p_x
u\|^2_{\mcB^\al_1}\|\f^0\|^2_{\mcG^0_k}
+\|\f^0\|^2_{\mcC^0_{\{x=0\},1}}\left\{\|\p_x
u\|^2_{\mcB^\al_1}\|\hat{1}\|^2_{\mcG^0_{k-1}}
+\frac{1}{\ep_0^2}\|\p_xu\|^2_{\mcH^\al_{k}} \right\}
\right\}\no
&\le&
C(\ep_0)\left\{\|\p_xu\|^2_{\mcB^\al_1}\|\f^0\|^2_{\mcG^0_k}
+\|\f^0\|^2_{\mcC^0_{\{x=0\},1}}E^\al_{k,\la}[u(\tau)]
\right\}\no
&&+C(\ep_0)\|\f^0\|^2_{\mcC^0_{\{x=0\},1}}\|\p_xu\|^2_{\mcB^\al_1}\left\{C(\|\f^0\|_{L^\infty})
\left(1+\|x\f^0\|^2_{\mcG^0_{k-1}}\right)\right\}
 \;.
 \no \label{Eq5}
\eea
Equations (\ref{Eq4}) and (\ref{Eq5}) finish the proof of the
desired estimate for $(a)$, and hence for $A^2_{00}$.
%
%
%
%
%

Now let us consider the sixth term $A^{00}_6$ of $A^{00}$. We
have
$$\wh{\Upsilon}^\mu\p_\mu = \wh{\Upsilon}^\tau\left(\p_\tau-\p_x\right) +
\left(\wh{\Upsilon}^x+\wh{\Upsilon}^\tau\right)\p_x+\wh{\Upsilon}^A\p_A,$$
and we decompose $A^{00}_6$ as
\bel{Eq9}A^{00}_6 =\mya +\myab  +\myc  .\ee
We have
$$\mya := \mcD^\beta\left([x\f^0] \wh{\Upsilon}^\tau (\p_\tau -\p_x)u\right)- [
x\f^0]\mcD^\beta\left( \wh{\Upsilon}^\tau (\p_\tau -\p_x)u\right),
$$
and
\beaa
 \lefteqn{ \int_{{\bf{\Large{H}}}_{\la,\tau}}
x^{-2\al-1+2\beta_1}\mya ^2dx\, d\nu
 }
 &&
\\
  &=& \|x^{\beta_1}\mcD^\beta\left([x\f^0] \wh{\Upsilon}^\tau
(\p_\tau -\p_x)u\right)- x^{\beta_1}[
x\f^0]\mcD^\beta\left( \wh{\Upsilon}^\tau (\p_\tau -\p_x)u\right)\|^2_{\mcH^\al_0}\\
&\le& C\left(\| \wh{\Upsilon}^\tau (\p_\tau -\p_x)u
\|^2_{\mcB^\al_0}\|x\f^0\|^2_{\mcG^0_{k}}+ \|
\wh{\Upsilon}^\tau (\p_\tau -\p_x)u\|^2_{\mcH^\al_{k-1}}\|x\f^0
\|^2_{\mcC^0_{\{x=0\},1}}\right)\\
&\le& C(\ep_0)\| {\Upsilon}^\tau \|^2_{L^\infty}\|(\p_\tau
-\p_x)u \|^2_{\mcB^\al_0}\|x\f^0\|^2_{\mcG^0_{k}}  + \|x\f^0
\|^2_{\mcC^0_{\{x=0\},1}}\\
&& \qquad\quad \times \left\{(\p_\tau -\p_x)u
\|^2_{\mcB^\al_0}\|\wh{\Upsilon}^\tau\|^2_{\mcG^0_{k-1}}+ \|
(\p_\tau -\p_x)u\|^2_{\mcH^\al_{k-1}}\|\wh{\Upsilon}^\tau
\|^2_{\mcC^0_{\{x=0\},0}}\right\}\;.\eeaa
Now, from (\ref{CC0}) and (\ref{CC2}) we have
$$\|\wh{\Upsilon}^\tau
\|_{\mcC^0_{\{x=0\},0}}\le C(\ep_0)
\|{\Upsilon}^\tau\|_{L^\infty},
$$
and
$$
 \|\wh{\Upsilon}^\tau\|^2_{\mcG^0_{k-1}} \le C(\ep_0)\left\{
 \|{\Upsilon}^\tau\|_{\mcG^0_{k-1}}^2 +
 \|{\Upsilon}^\tau\|^2_{L^\infty}C(\|\f^0\|_{L^\infty})\left(1+\|x\f^0\|^2_{\mcG^{0}_{k-1}}\right)\right\}
  \;.
$$
Thus
\bea \int_{{\bf{\Large{H}}}_{\la,\tau}}
x^{-2\al-1+2\beta_1}\mya ^2dx\, d\nu &\le& C(\ep_0)\|
{\Upsilon}^\tau \|^2_{L^\infty}\|(\p_\tau -\p_x)u
\|^2_{\mcB^\al_0}\|x\f^0\|^2_{\mcG^0_{k}}\no
&&+ C(\ep_0)\|x\f^0\|^2_{\mcC^0_{\{x=0\},1}}\| {\Upsilon}^\tau
\|^2_{L^\infty}\|(\p_\tau -\p_x)u \|^2_{\mcH^\al_{k-1}}\no
&& + C(\|\f^0\|_{L^\infty})
\|x\f^0\|^2_{\mcC^0_{\{x=0\},1}}\|(\p_\tau -\p_x)u
\|^2_{\mcB^\al_{0}}\|{\Upsilon}^\tau \|^2_{\mcG^0_{k-1}}\no
&&+ C(\ep_0) \|x\f^0\|^2_{\mcC^0_{\{x=0\},1}}\|(\p_\tau -\p_x)u
\|^2_{\mcB^\al_{0}}\|{\Upsilon}^\tau
\|^2_{L^\infty}\left(1+\|x\f^0\|^2_{\mcG^{0}_{k-1}}\right)
 \;.
 \no
%
 \label{Eq10}
\eea
On the other hand,
$$\myab  :=  \mcD^\beta\left([x\f^0]
(\wh{\Upsilon}^\tau+\wh{\Upsilon}^x) \p_x u\right)- [
x\f^0]\mcD^\beta\left( (\wh{\Upsilon}^\tau+\wh{\Upsilon}^x) \p_x
u\right)$$
and we have
\bea
 \lefteqn{
  \int_{{\bf{\Large{H}}}_{\la,\tau}}
x^{-2\al-1+2\beta_1}\myab  ^2dx\, d\nu
 }
 &&
 \nonumber
\\
 &=& \|x^{\beta_1}\mcD^\beta\left([x\f^0]
(\wh{\Upsilon}^\tau+\wh{\Upsilon}^x) \p_x u\right)-
x^{\beta_1}x\f^0\mcD^\beta\left(
(\wh{\Upsilon}^\tau+\wh{\Upsilon}^x) \p_x
u\right)\|^2_{\mcH^\al_0}\no
&\le& C\|(\wh{\Upsilon}^\tau+\wh{\Upsilon}^x) \p_x
u\|^2_{\mcB^{\al}_0}
\|x\f^0\|^2_{\mcG^0_{k}}+C\|(\wh{\Upsilon}^\tau+\wh{\Upsilon}^x)
\p_x u\|^2_{\mcH^{\al}_{k-1}}
\|x\f^0\|^2_{\mcC^0_{\{x=0\},1}}\no
&\le&
C(\ep_0)\|{\Upsilon}^\tau+{\Upsilon}^x\|^2_{L^\infty}\|\p_x
u\|^2_{\mcB^{\al}_0}
\|x\f^0\|^2_{\mcG^0_{k}}+C\|x\f^0\|^2_{\mcC^0_{\{x=0\},1}}\no
&&\qquad \times\left\{\|\p_x
u\|^2_{\mcB^\al_0}\|\wh{\Upsilon}^\tau+\wh{\Upsilon}^x\|^2_{\mcG^0_{k-1}}
+\|\p_x
u\|^2_{\mcH^{\al}_{k-1}}\|\wh{\Upsilon}^\tau+\wh{\Upsilon}^x\|^2_{\mcC^0_{\{x=0\},0}}
\right\}\no
&\le& C(\ep_0)\|{\Upsilon}^\tau+{\Upsilon}^x\|^2_{L^\infty}\|
\p_xu\|^2_{\mcB^{\al}_0} \|x\f^0\|^2_{\mcG^0_{k}}\no
&&+C(\ep_0)\|x\f^0\|^2_{\mcC^0_{\{x=0\},1}}\|{\Upsilon}^\tau+{\Upsilon}^x\|^2_{L^\infty}E^\al_{k,\la}[u(\tau)]
\no
&&+C(\|\f^0\|_{L^\infty})\|x\f^0\|^2_{\mcC^0_{\{x=0\},1}}\|
\p_xu\|^2_{\mcB^\al_0}\no
&& \qquad \qquad
\qquad\times\left\{\|{\Upsilon}^\tau+{\Upsilon}^x\|^2_{\mcG^0_{k-1}}
 +\|{\Upsilon}^\tau+{\Upsilon}^x\|^2_{L^\infty}
 \left(1+\|x\f^0\|^2_{\mcG^{0}_{k-1}}\right)\right\}
 \;.
\no
%
 \label{Eq12}
\eea
The same holds for the term
$$\myc  :=  \mcD^\beta\left([x\f^0] \wh{\Upsilon}^A \p_A u\right)- [
x\f^0]\mcD^\beta\left( \wh{\Upsilon}^A \p_A u\right)$$
and we have
\bean
 \lefteqn{ \int_{{\bf{\Large{H}}}_{\la,\tau}}
x^{-2\al-1+2\beta_1}\myc^2dx\, d\nu
 }
 &&
\\
  &\le& C(\ep_0)\| {\Upsilon}^A \|^2_{L^\infty}\| \p_Au
\|^2_{\mcB^\al_0}\|x\f^0\|^2_{\mcG^0_{k}}\no
&&+ C(\ep_0)\|x\f^0\|^2_{\mcC^0_{\{x=0\},1}}\|
{\Upsilon}^A\|^2_{L^\infty}\| \p_Au \|^2_{\mcH^\al_{k-1}}\no
&& + C(\|\f^0\|_{L^\infty})
\|x\f^0\|^2_{\mcC^0_{\{x=0\},1}}\|\p_A u
\|^2_{\mcB^\al_{0}}\|{\Upsilon}^A \|^2_{\mcG^0_{k-1}}\no
&&+ C(\ep_0) \|x\f^0\|^2_{\mcC^0_{\{x=0\},1}}\| \p_Au
\|^2_{\mcB^\al_{0}}\|{\Upsilon}^A
\|^2_{L^\infty}\left(1+\|x\f^0\|^2_{\mcG^{0}_{k-1}}\right)
 \;.
\no
%
 \label{EEq10}
\eea
This provides the right estimate for  $A^{00}_6$, and hence for
of $A^{00}$. 

An identical estimate is obtained on the fourth term $A_4$ of
the commutator (\ref{E24}).
This finishes the estimation of the commutator $[\Box_{\m},
\mcD^\beta]u$ appearing in (\ref{E15}),
and the proof is complete.
\qed
\subsubsection{Conclusion}

The proof of the Proposition~\ref{pprop} used essentially
Stokes's theorem,  the weighted Moser type inequalities (A.34)
and (A.35) of Proposition A.3 of~\cite{ChLengardnwe},  and  the
weighted substitution inequality type (A.31) of the same
reference. One of the points there is that all the constants
appearing in these inequalities are independent of $x_2$
(recall that the sets $M_{x_2, x_1}$ there corresponds to the
sets ${\bf{\Large{H_{\la,\tau}}}}$ here) which is the distance
between the boundary of $M_{x_2, x_1}$ and the null
hypersurface $\mcN= \{x=0\}$. So, in our case, all the
constants involved in the proof of the previous proposition are
independent  of $\la$. This allows us to take the limit as
$\la$ goes to $0$ in \eq{Pr} and obtain an identical inequality
with $E^\al_{k,\la}[u(\tau)]$ there replaced with
$E^\al_{k}[u(\tau)]$. Therefore we have proved the following:
\begin{Proposition}
\label{ppropp}
Proposition~\ref{pprop} remains true with $\lambda =0$.
\end{Proposition}
Inequality (\ref{Pr}) with $\lambda=0$ is the key in deriving
an existence theorem for the Einstein-Maxwell equations with
data on a hyperboloid, singular near $\{x=0\}$. In this case,
we will show that all the $\mcH_k$ and $ \mcG_k$ norms
appearing in this inequality are controlled by the energy.
%

%
\newcommand{\mysigma}{\sigma}%
\newcommand{\mm}{\mathring \m}%
\newcommand{\mI}{\mathring I}%
\newcommand{\mA}{\mathring A}%
\newcommand{\mykappa}{\mykap}%
\newcommand{\mykap}{\dm}%
\newcommand{\dI}{{\delta I}{}}%
\newcommand{\dA}{{\delta A}{}}%
\newcommand{\dm}{{\delta\m}{}}%
\newcommand{\mUpsilon}{\mathring{\Upsilon}{}}%
\newcommand{\mUp}{\mUpsilon}%
\newcommand{\dmUp}{\delta{\Upsilon}{}}%
\newcommand{\dUpsilon}{\dmUp}%
It turns out that the proof, in Section~\ref{sPhgEM}, of global
polyhomogeneity of the geodesically complete metrics
constructed by Loizelet requires a slightly different
inequality. For this we need to split the metric into two parts
as
\bel{20VIII0.5}
 \m^{\alpha\beta} = \mm ^{\alpha\beta}+\mykap^{\alpha\beta}
 \;.
\ee
The rationale behind such a splitting is, that the Lorentzian
metric $\mm$ will be fixed (in fact, it will be the flat
Minkowski metric in our applications), while the correction
$\mykap$ will eventually depend on the fields. This leads to
the obvious corresponding decomposition of $\Upsilon$,
\bel{20VIII0.5x}
 \Upsilon^{\alpha } = \mUp ^{\alpha }+\dmUp^{\alpha}
 \;.
\ee
%

We  assume that there exist constants $\sigma$, $M$ and $N$
such that for $\tau \in [\tauz,\tauone ]$ we have
\bean
 M & \ge &  \|(\mathring \f^\sharp,\mathring
 \m^\sharp,\mUpsilon)\|_{\mcG^0_k({\bf{\Large{H_{\tau}}}})}
  + \|(\delta \f^\sharp,\delta
         \m^\sharp,\dUpsilon)\|_{\mcC^0_{\{x=0\},1}({\bf{\Large{H_{\tau}}}})}
\\
 &&
 + \|(\partial_x - \partial_\tau)
 \m^\sharp \|_{L^\infty({\bf{\Large{H_{\tau}}}})}
 \;,
  \label{1IX0.5}
\\
 N & \ge &
 \| ( \partial_\tau  u ,\partial_x u, \partial_A u)
          \|_{\mcB^\sigma_1({\bf{\Large{H_{\tau}}}})}
          +
          \|
 (\m^\sharp,\Upsilon) \|_{L^\infty({\bf{\Large{H_{\tau}}}})}
\nonumber
\\
 &&
  +
  \|(\mathring  \m^{\tau A} ,\mathring  \m^{x A})
         \|_{\mcG^{\alpha-\sigma}_{k-1}({\bf{\Large{H_{\tau}}}})}
  +
  \|( \delta \m^\sharp,\delta\Upsilon)
          \| _{\mcC^0_{\{x=0\},1}({\bf{\Large{H_{\tau}}}})}
 \;.
\eeal{23VIII0.5}
 We then have:

\begin{Proposition}
 \label{P19VIII0.1}
Let $k> n/2+1$, $\mysigma\in \R$, $\alpha\le - 1/2$.
There exist functions $ C_3(\ep_0, C_0, \al, k,n, M )$ and $
C_4(\ep_0, C_0, \al, \sigma, k,n, N )$, monotonously increasing
in $M$ and $N$, which we write as $C_3(M)$ and $C_4(N)$, such
that for all
$$ \tau\in  \left[\tauz , \tauone \right] $$
and for all $u$ satisfying (\ref{E0})   we have
\bean
  E^{\al}_{k}[u(\tau)]
    &\le&
    E^{\al}_{k}[ u(\tauz )]
 +
 \int_{\tauz }^\tau \Big \{ C _3(M) \left(
 E^\al_{k}[u(s)]+
 \|F(s)\|^2_{\mcH^\al_k({\bf{\Large{H_{\tau}}}})}
  \right)
\no
 &&
  +
 C_4 (N)\left(1+ \| (\dm^\sharp,\delta
 \f^\sharp,{\dUpsilon} )\|^2_{\mcG^{\alpha-\mysigma}_{k}({\bf{\Large{H_{\tau}}}})}
  \right)
  \Big\}ds
 \;.
%
\eeal{PPr2x}
\end{Proposition}

\proof
The result is obtained by calculations very similar to those of
Proposition~\ref{ppropp}. We follow that proof until \eq{E46},
which is rewritten as
\bel{E46x} \Box_\m= \mm^{\mu\nu}\p^2_{\mu\nu} +
\mykap^{\mu\nu}\p^2_{\mu\nu} + \mUp^\nu\p_\nu + \dmUp^\nu\p_\nu
 \;.
\ee
This leads to the following rewriting of \eq{E24}:
\bean
 [\Box_\m, \mcD^\beta]u &=& \underbrace{ \mm^{\al\mu}[
 \p_\al\p_\mu, \mcD^\beta]u}_{=:\mA_1}
  +  \underbrace{\dm^{\al\mu}[
 \p_\al\p_\mu, \mcD^\beta]u}_{=:\dA_1}
\\
 &&  \underbrace{- {\mUpsilon}^\nu[\mcD^\beta, \p_\nu]u}_{=:\mA_2}
   \underbrace{-\left\{\mcD^\beta\left({\mUpsilon}^\nu \p_\nu
 u\right)-{\mUpsilon}^\nu \mcD^\beta\left(\p_\nu
  u\right)\right\}}_{=:\mA_3}
  \nonumber
\\
 &&  \underbrace{- {\dmUp}^\nu[\mcD^\beta, \p_\nu]u
 }_{=:\dA_2}
  \underbrace{-\left\{\mcD^\beta\left({\dUpsilon}^\nu \p_\nu
 u\right)-{\dUpsilon}^\nu \mcD^\beta\left(\p_\nu
  u\right)\right\}}_{=:\dA_3}
  \nonumber
\\
   &&   \underbrace{- \left\{\mcD^\beta\left(\mm^{\al\mu}\p_\al \p_\mu u\right)
   - \mm^{\al\mu}\mcD^\beta\left(\p_\al \p_\mu
   u\right)
   \right\}}_{=:\mA_4}
 \nonumber
\\
   &&   \underbrace{- \left\{\mcD^\beta\left(\dm^{\al\mu}\p_\al \p_\mu u\right)
   - \dm^{\al\mu}\mcD^\beta\left(\p_\al \p_\mu
   u\right)\right\}}_{=:\dA_4}
   \;.
 \label{E24x}
\eea
The terms   $A_i:=\mA_i+\dA_i$, i=1,2  are estimated as   in
\eq{E25}-\eq{E26}. For $\mA_3$, instead of \eq{19VIII0.4}  the
estimates proceed as before, except that at the end one invokes
the weighted Sobolev
embedding~\cite[Proposition~A.1]{ChLengardnwe}; e.g.,
\bean
 \lefteqn{
 \int_{{\bf{\Large{H}}}_{\la,\tau}}
 x^{-2\al-1+2\beta_1}\left\{\mI\right\}^2dx\, d\nu
 }
\\
\nonumber
  &=& \|
x^{\beta_1}\mcD^\beta\left({\mUpsilon}^\tau (\p_\tau-\p_x)
u\right)-x^{\beta_1}{\mUpsilon}^\tau
\mcD^\beta\left((\p_\tau-\p_x)
 u\right)\|^2_{\mcH^{\al }_0({{\bf{\Large{H}}}_{\la,\tau}})}
\\
 \nonumber
&\le& C_s\left(\|(\p_\tau-\p_x)
u\|^2_{\mcB^\al_0}\|{\mUpsilon}^\tau \|^2_{\mcG^0_{k}}+
\|(\p_\tau-\p_x) u\|^2_{\mcH^\al_{k-1
 }}\|{\mUpsilon}^\tau \|^2_{\mcC^0_{\{x=0\},1}}\right)
\\
%
&\le& C\left( \|{\mUpsilon}^\tau \|^2_{\mcG^0_{k}}+
\|{\mUpsilon}^\tau
 \|^2_{\mcC^0_{\{x=0\},1}}\right)E^{\;\al}_k[u(\tau)]
 \;.
 \label{19VIII0.5}
\eea
For $\dA_3$,  the following version of the inequalities
of~\cite[Proposition~A.3]{ChLengardnwe} should be used, for any
$\alpha,\beta,\alpha'$  (the proof is identical to that given
there):
\bel{25VIII0.1}
 \|fg\|_{\mcH^{\alpha+\beta}_k}
 \le
   C\big(
        \|f \|_{\mcB^\alpha_{\{x=0\},0}}
        \|        g\|_{\mcG^{ \beta}_k} +
        \|g \|_{\mcC^{\alpha'}_{\{x=0\},0}}
        \|        f\|_{\mcH^{\alpha+\beta-\alpha'}_k}
         \big)
 \;,
\ee
and, for $\beta=(\beta_1,\beta')$,
\bean \lefteqn{
 \|x^{\beta_1}  \mcD^{\beta }(fg)
    - (x^{\beta_1} \mcD^{\beta }f)g)\|_{\mcH^{\alpha+\beta}_0}
 }
 &&
\\
  &\le &
   C\big(
        \|f \|_{\mcB^\alpha_{\{x=0\},0}}
        \|        g\|_{\mcG^{ \beta}_k} +
        \|(x\partial_xg, \partial_ A g) \|_{\mcC^{\alpha'}_{\{x=0\},0}}
        \|        f\|_{\mcH^{\alpha+\beta-\alpha'}_{k-1}}
         \big)
 \;.
  \nonumber
\\
 &&
\eeal{25VIII0.2}
Instead of  \eq{19VIII0.4} we then have
\bean
 \lefteqn{\int_{{\bf{\Large{H}}}_{\la,\tau}}
 x^{-2\al-1+2\beta_1}\left\{\dI\right\}^2dx\, d\nu
} &&
\\
 \nonumber
&&
 =\|
 x^{\beta_1}\mcD^\beta\left({\dUpsilon}^\tau (\p_\tau-\p_x)
 u\right)-x^{\beta_1}{\dUpsilon}^\tau
 \mcD^\beta\left((\p_\tau-\p_x)
 u\right)\|^2_{\mcH^{\al }_0({{\bf{\Large{H}}}_{\la,\tau}})}
\\
 \nonumber
&&\le C_s\left(\|(\p_\tau-\p_x)
u\|^2_{\mcB^\mysigma_0}\|{\dUpsilon}^\tau
\|^2_{\mcG^{\alpha-\mysigma}_{k}}+ \|(\p_\tau-\p_x)
u\|^2_{\mcH^\alpha_{k-1
 }}\|{\dUpsilon}^\tau \|^2_{\mcC^{0}_{\{x=0\},1}}\right)
\\%
%
&&\le C\left(\|(\p_\tau-\p_x)
u\|^2_{\mcB^\mysigma_0}\|{\dUpsilon}^\tau
\|^2_{\mcG^{\alpha-\mysigma}_{k}}+ \|{\dUpsilon}^\tau
 \|^2_{\mcC^{0}_{\{x=0\},1}}E^{\;\al}_k[u(\tau)]\right)
 \;.
 \label{19VIII0.5x}
\eea
An identical treatment applies to the remaining three displayed
equations following \eq{19VIII0.4}.

The term $A_4$ is split into $A^{\mu\nu}$'s  as in \eq{E36},
and then for  $\mu\nu \ne {00}$ we split
$A^{\mu\nu}=\mA^{\mu\nu}+\delta A^{\mu\nu}$ in the obvious way.
All the terms $\mA^{\mu\nu}$ with $\mu\nu \ne {00}$
are then treated as in the proof of Proposition~\ref{ppropp},
and at the end we invoke the inequality, for $k\ge n/2+1$,
$$
 \|f\|^2_{\mcB^\al_1}  \le C \|
f\|^2_{\mcH^\al_{k }}
 \;.
$$
The terms involving $\delta {A}^{\mu\nu}$ with $\mu\nu \ne
{00}$ are treated as in \eq{19VIII0.5x}; for example,
\eq{20VIII0.1} becomes
\bea
 \lefteqn{
 \int_{{\bf{\Large{H}}}_{\la,\tau}}
x^{-2\al-1+2\beta_1}\left\{\delta A^{AB}\right\}^2dx\, d\nu }
 &&
 \nonumber
\\
 &&= \| x^{\beta_1}\mcD^\beta\left(\delta\m^{AB}\p_A \p_B
 u\right)- x^{\beta_1}\delta\m^{AB}\mcD^\beta\left(\p_A \p_B
 u\right)\|_{\mcH^{\al }_0({{\bf{\Large{H}}}_{\la,\tau}})}
\no
 &&\le C_s
 \sum_A\left(\|\p_Au\|^2_{\mcB^{\al-\mysigma}_1}\|\delta\m^\sharp\|^2_{\mcG^\mysigma_{k}}+
\|\p_A u\|^2_{\mcH^\al_{k}}\|\delta\m^\sharp
 \|^2_{\mcC^0_{\{x=0\},1}}\right)
\no
 &&\le
 C\left(\sum_A\|\p_Au\|^2_{\mcB^{\al-\mysigma}_1}\|\delta\m^\sharp\|^2_{\mcG^\mysigma_{k}}+
\|\delta\m^\sharp
\|^2_{\mcC^0_{\{x=0\},1}}E^\al_{k,\la}[u(\tau)]\right)
 \;.
\phantom{xxxxx}
 \label{23VIII0.1}
\eea

In \eq{24VIII0.1} it is convenient  to use the splitting $\f =
\mathring \f + \delta \f$. The terms involving $\mathring \f$
are estimated, using the Sobolev embedding, by
$E^{\;\al}_k[u(\tau)]$,
 while for
those involving $\delta \f$ we write
\bean
 \lefteqn{ \int_{{\bf{\Large{H}}}_{\la,\tau}}
x^{-2\al-1+2\beta_1}\left\{\delta A_3^{00}\right\}^2dx\, d\nu}
 &&
\\
 &=&
2\|\mcD^\beta\left(x\delta \f^0 \hat{\m}^{\tau
A}\p_A(\p_\tau-\p_x) u\right)- x\delta
\f^0\mcD^\beta\left(\hat{\m}^{\tau A}\p_A(\p_\tau
 -\p_x)u\right)\|^2_{\mcH^\al_0}
\no
 &\le& C\| \hat{\m}^{\tau A}\p_A(\p_\tau-\p_x) u\|^2_{\mcB^\sigma_0}
 \|x\delta \f^0\|^2_{\mcG^{\alpha-\sigma}_k}
+ C\|\hat{\m}^{\tau A}\p_A(\p_\tau -\p_x)u
 \|^2_{\mcH^\al_{k-1}}\|x\delta \f^0\|^2_{\mcC^0_{\{x=0\},1}}
  \;.\no
\eea
The first line above is estimated as
\bean
 && C\| \hat{\m}^\sharp\|^2_{L^\infty}\|(\p_\tau-\p_x)
 u\|^2_{\mcB^\sigma_1}\| \delta \m^\sharp\|^2_{\mcG^{\alpha-\sigma}_k}
  \;,
\eea
as desired. The second is estimated as
\bean
\lefteqn{  C\| \delta \m^\sharp\|^2_{\mcC^0_{\{x=0\},1}}
  \left\{\|(\p_\tau-\partial_\tau) u\|^2_{\mcB^\sigma_1}
    \|\hat{\m}^{\tau A}\|^2_{\mcG^{\alpha-\sigma}_{k-1}}\right.
     +\left.\|(\p_\tau-\p_x)u \|^2_{\mcH^\al_{k}}\|
 \hat{\m}^\sharp\|^2_{\mcC^0_{\{x=0\},0}}\right\}
 }
&& \no
 & \le & C\| \delta \m^\sharp\|^2_{\mcC^0_{\{x=0\},1}}
  \left\{\|(\p_\tau-\partial_\tau) u\|^2_{\mcB^\sigma_1}
    \|\hat{\m}^{\tau A}\|^2_{\mcG^{\alpha-\sigma}_{k-1 }}\right.
     +\left.
       \|{\m}^\sharp\|^2_{\mcC^0_{\{x=0\},0}}
  E^\al_{k,\la}[u(\tau)]\right\}
 \;.
\no
 &&
\eea
%
%
To estimate the term $A^{00}_5$ (compare~\eq{Eq1}) we need to
split both $\f^\sharp$ and $\m^\sharp$ into two. The terms
there involving $\mathring\f^ \sharp$ and $\mathring \m^\sharp$
can be estimated by $E^{\;\al}_k[u(\tau)]$. The terms involving
$\delta \m^\sharp$ are estimated as in the analysis of  $\delta
A^{00}_3$. The mixed term involving $\mathring \m^\sharp$ and
$\delta \f$ is handled in the obvious way
%
\bea
 \lefteqn{
\| x^{\beta_1}\mcD^\beta\left([x\delta \f^0][\hat{\mathring\m}^{AB}\p_A\p_B u]\right)-  x^{\beta_1}[
 x\delta\f^0]\mcD^\beta\left([\hat{\mathring\m}^{AB}\p_A\p_B
 u]\right)\|^2_{\mcH^\al_0}
 } &&
\no
 &\le&
 C\big(\| \hat{\mathring\m}^{AB}\p_A\p_B
 u \|_{\mcB_0^\mysigma}\|\|x\delta
\f^0\|_{\mcG^{\al-\mysigma}_k}
 +\|x\delta
\f^0\|_{\mcC^0_{\{x=0\},1}}\|\hat{\mathring\m}^{AB}\p_A\p_B
 u\|_{\mcH^{\alpha}_{k-1}}\big)
 \no
 &\le&
 C\big(
 \| \hat {\mathring\m}^\sharp\|_{L^\infty}
  \| \partial _A u \|_{\mcB^{\sigma}_1 }
 \|x \delta \f^0
 \|_{\mcG^{\alpha-\sigma}_{k}}
 +
 \| \hat {\mathring\m}^\sharp\|_{\mcG^{0}_{k-1}}
 \|  \partial _A u \|_{\mcH^{\alpha}_k}
  \|x\delta
\f^0 \|_{\mcC^{0}_{\{x=0\},1}}
 \big)
\no
 &\le&
 C\big(
 \|   {\mathring\m}^\sharp\|_{L^\infty}
  \| \partial _A u \|_{\mcB^{\sigma }_1 }
   \|\delta \m^\sharp \|_{\mcG^{\alpha-\sigma}_{k}}
 +
 \|  {\mathring\m}^\sharp\|_{\mcG^{0}_{k-1}}
  \|\delta      \m^\sharp \|_{\mcC^{0}_{1}}
  E^\al_{k,\la}[u(\tau)]
 \big)
 \label{EEq9x}
  \;.
\eea
%
A similar analysis of the remaining terms proves the
proposition.
\qed

\section{Application to the Einstein-Maxwell Equations in wave coordinates and Lorenz gauge }
 \label{S1IV0.1}

\subsection{Change of coordinates}
 \label{ss1IV0.1}

\subsubsection{On the gauge condition}
 \label{sss1IV0.1}
\newcommand{\phih}{h}%
\newcommand{\PhiH}{H}%
%
Throughout this section,  the (unphysical) conformally rescaled
metric is denoted by $\m$, and the (physical) metric is denoted
by $g$; thus $\m_{\mu\nu} = \Omega^2 g_{\mu\nu}$.

Remember that in the original system of coordinates $(x^\mu)$
we have
$$\Box_{g}x^\mu = 0 \qquad \text{with} \qquad g  = \eta + \phih \;,$$
which leads to
\be \label{GC1} \p_\mu \big(g^{\mu\nu}\sqrt{|\det g|} \big) =
0\;.\ee
We want to rewrite the above equation in the new system of
coordinate $(y^\al)$ (see (\ref{T2})). We have
$$\sqrt{|\det g|} = 1+\frac 12  \eta^{\al\bet}\phih _{\al\bet}+ Q(\phih )\;,$$
%
%
where $Q$ has a uniform zero of order two in $\phih $. We set
\bel{19VIII0.2}
 g^{\mu\nu}= \eta^{\mu\nu} +
 \PhiH ^{\mu\nu}
 \;.
\ee
In what follows, we use a generic symbol $Q$ for functions
which have a uniform zero of order two. We have
\beaa \p_\mu \big(g^{\mu\nu}\sqrt{|\det g|} \big) &=& \p_\mu
\big[g^{\mu\nu}\{1+\frac 12  \eta^{\al\bet}\phih _{\al\bet}+Q(\phih )\}\big]\\
&=&\p_\mu \big[\{\eta^{\mu\nu}+\PhiH ^{\mu\nu}\}\{1+\frac 12\eta^{\al\bet}\phih _{\al\bet}+ Q(\phih )\}\big]\\
&=&\p_\mu \PhiH ^{\mu\nu}\{1+\frac 12\eta^{\al\bet}\phih
_{\al\bet}+ Q(\phih )\}+ \{\eta^{\mu\nu}+\PhiH ^{\mu\nu}\}
\{\frac 12\eta^{\al\bet}\p_\mu \phih _{\al\bet}+ \p_\mu Q(\phih
)\} \;.\eeaa
Using this identity, equation  (\ref{GC1}) takes the form:
\be \label{GC2} \p_\mu \PhiH ^{\mu\nu}+\frac
12\eta^{\mu\nu}\eta^{\al\bet}\p_\mu \phih _{\al\bet} = -\p_\mu
\PhiH ^{\mu\nu}\{\frac 12\eta^{\al\bet}\phih _{\al\bet}+
Q(\phih )\} -\PhiH ^{\mu\nu} \{\frac 12\eta^{\al\bet}\p_\mu
\phih _{\al\bet}+ \p_\mu
 Q(\phih )\}- \eta^{\mu\nu}\p_\mu Q(\phih )
 \;.
\ee
Let us rewrite this equation in the system of coordinates
$(\tau,x, v^A)$ where
\bel{T2} y^\mu= \frac{x^\mu}{\eta_{\al\bet}x^\al x^\beta},
\quad \tau = y^0 \le 0 ,\quad   x=-y^0-\rho\ge 0
\quad\text{and}\quad y^i=\rho\omega^i(v^A) \;.\ee
Recall that
\bel{TT2}\Om = -y_\al y^\al = \tau^2-\rho^2 = x(-\tau+\rho)\ge
0\;,\ee
and $\hat{f} = \Om^{-\frac{n-1}{2}}f$ (\emph{not} to be
confused with division by $\m^{\tau\tau}$, as used in the
previous section), so that
\bel{parthat} \frac{\p f}{\p x^\mu}   =
\Om^{\frac{n-1}{2}}\Big\{ -(n-1)y_\mu  - \Om \frac{\p}{\p
y^\mu}-2y_\mu y^\al\frac{\p}{\p y^\al}\Big\}\hat f \;, \ee
thus the left-hand-side of  (\ref{GC2}) can be rewritten as
%
$$-(n-1)\Om^{\frac{n-1}{2}}y_\mu\Big(\wh{\PhiH }^{\mu\nu}+ \frac 12
\eta^{\mu\nu}\eta^{\al\bet}\hat \phih _{\al\bet}\Big)-
\Om^{\frac{n-1}{2}}\Big\{  \Om \frac{\p}{\p y^\mu}+2y_\mu
y^\al\frac{\p}{\p y^\al} \Big\}\Big(\wh{\PhiH }^{\mu\nu}+ \frac 12
\eta^{\mu\nu}\eta^{\al\bet}\hat \phih _{\al\bet}\Big).$$
We want to analyze the structure of the right-hand side of
(\ref{GC2}). This expression is made of three terms which will
be labeled $R_1$, $R_2$, and $ R_3$. We have (see
(\ref{parthat}) and recall that $ y^\al\frac{\p \Om}{\p
y^\al}\; = 2 \Om$): 
%
\bea R_1 &=& \Om^{\frac{n-1}{2}}\left\{\frac
12\Om^{\frac{n-1}{2}}\tr_\eta(\hat{\phih })
+Q(\Om^{\frac{n-1}{2}}\hat{\phih })\right\}\left\{(n-1)y_\mu+
\Om\frac{\p}{\p y^\mu}+2y_\mu y^\al\frac{\p}{\p
y^\al}\right\}\wh{\PhiH }^{\mu\nu}\no
&=:&\label{rest1} Q(\Om^{\frac{n-1}{2}}\hat{\phih },
\Om^{\frac{n-1}{2}}y_\mu \wh{\PhiH
}^{\mu\nu})+Q(\Om^{\frac{n-1}{2}}\hat{\phih },
\Om^{\frac{n+1}{2}}\p_\mu\wh{\PhiH }^{\mu\nu})\no
&&+Q(\Om^{\frac{n-1}{2}}\hat{\phih }, \Om^{\frac{n-1}{2}}y_\mu
y^\al\frac{\p}{\p y^\al}\wh{\PhiH }^{\mu\nu}) \;. \eea Now,
since $\frac{\p Q}{\p \phih }$ has a uniform zero of order one,
we have
\beaa \frac{\p}{\p x^\mu} Q(\phih )&=&  \frac{\p Q}{\p \phih
}\frac{\p \phih }{\p x^\mu} =-\frac{\p Q}{\p \phih
}({\Om^{\frac{n-1}{2}}}\hat \phih ) \Om^{\frac{n-1}{2}}
\left\{(n-1)y_\mu+ \Om\frac{\p}{\p y^\mu}+2y_\mu
y^\al\frac{\p}{\p y^\al}\right\}\hat\phih
\\
&=:&Q(\Om^{\frac{n-1}{2}}\hat{\phih },
\Om^{\frac{n-1}{2}}y_\mu\hat{\phih
})+Q(\Om^{\frac{n-1}{2}}\hat{\phih },
\Om^{\frac{n+1}{2}}\frac{\p \hat{\phih }}{\p
y^\mu})+Q(\Om^{\frac{n-1}{2}}\hat{\phih },
\Om^{\frac{n-1}{2}}y_\mu y^\al\frac{\p}{\p y^\al}\hat{\phih })
\;. \eeaa
Thus $R_2$ reads:
\bea R_2 &=& \frac 12
 \eta^{\al\beta}\Om^{\frac{n-1}{2}}
 \wh{\PhiH}^{\mu\nu}\left\{\Om^{\frac{n-1}{2}}\right\}
\left\{(n-1)y_\mu+
 \Om\frac{\p}{\p y^\mu}+2y_\mu y^\al\frac{\p}{\p
 y^\al}\right\}\hat{\phih }_{\al\beta}
\no
 &&
  +\label{rest2} \Om^{\frac{n-1}{2}}\wh{\PhiH}^{\mu\nu}
  \left\{Q(\Om^{\frac{n-1}{2}}\hat{\phih },
 \Om^{\frac{n-1}{2}}y_\mu\hat{\phih})
 +Q(\Om^{\frac{n-1}{2}}\hat{\phih },
 \Om^{\frac{n+1}{2}}\p_\mu\hat{\phih })
  \right.
\no
 &&
  \left.
 +Q(\Om^{\frac{n-1}{2}}\hat{\phih }, \Om^{\frac{n-1}{2}}y_\mu
y^\al\frac{\p}{\p y^\al}\hat{\phih })\right\} \;.\no
&=:& Q(\Om^{\frac{n-1}{2}}\hat{\phih },
\Om^{\frac{n-1}{2}}y_\mu\wh{\PhiH
}^{\mu\nu})+Q(\Om^{\frac{n-1}{2}}\wh{\PhiH }^{\mu\nu},
\Om^{\frac{n+1}{2}}\p_\mu\hat{\phih
})+Q(\Om^{\frac{n-1}{2}}y_\mu\wh{\PhiH }^{\mu\nu},
\Om^{\frac{n-1}{2}} y^\al\frac{\p}{\p y^\al}\hat{\phih })
\;.\no \eea
Next
\bea R_3&=& -\eta^{\mu\nu}\p_\mu Q(\phih )\no
&=&\label{rest3}
\eta^{\mu\nu}\left\{Q(\Om^{\frac{n-1}{2}}\hat{\phih },
\Om^{\frac{n-1}{2}}y_\mu\hat{\phih
})+Q(\Om^{\frac{n-1}{2}}\hat{\phih },
\Om^{\frac{n+1}{2}}\p_\mu\hat{\phih
})+Q(\Om^{\frac{n-1}{2}}\hat{\phih }, \Om^{\frac{n-1}{2}}y_\mu
y^\al\frac{\p}{\p y^\al}\hat{\phih })\right\} \;.\qquad \eea
From this, we obtain the following form of the gauge condition
(\ref{GC2}):
\bea y_\mu\wh{\PhiH }^{\mu\nu} + \frac 12
y^\nu\eta^{\al\bet}\hat \phih _{\al\bet}&=&
\frac{1}{1-n}\Big\{\Om \frac{\p}{\p y^\mu}+2y_\mu
y^\al\frac{\p}{\p y^\al}\Big\}\Big(\wh{\PhiH }^{\mu\nu}+ \frac
12 \eta^{\mu\nu}\eta^{\al\bet}\hat
\phih _{\al\bet}\Big)\nonumber\\
&&\label{GC6}+\;\Om^{-\frac{n-1}{2}}(R_1+R_2+R_3)\;.\eea
Now we recall that
$$
 \PhiH ^{\mu\nu}
 := g^{\mu\nu}-\eta^{\mu\nu} = -\phih ^{\mu\nu} + \wt Q^{\mu\nu}(\phih ) \;,
$$
where $\phih^{\mu\nu} = \eta^{\mu\alpha}\eta^{\nu\beta}
\phih_{\alpha\beta}$. Therefore
%
$$\eta^{\al\bet}\hat \phih _{\al\bet} =
 - \eta_{\al\bet}\wh{\PhiH }^{\al\bet} +
  \Om^{-\frac{ n-1}{2}}\wt Q( \Om^{\frac{n-1}{2}}\wh{\PhiH }).$$
Equations (\ref{rest1})-(\ref{GC6}) lead finally to the
following form of the gauge condition (\ref{GC2}):
\bea y_\mu\wh{\PhiH }^{\mu\nu} - \frac 12 y^\nu
\tr_{\eta}(\wh{\PhiH })&=& \frac{1}{1-n}\Big\{\Om \frac{\p}{\p
y^\mu}+2y_\mu y^\al\frac{\p}{\p y^\al}\Big\}\Big(\wh{\PhiH
}^{\mu\nu}-
 \frac 12 \eta^{\mu\nu}\tr_{\eta}\wh{\PhiH }\Big)\nonumber\\
&&+\Om^{-\frac{n-1}{2}}Q (\Om^{\frac{n-1}{2}}\wh{\PhiH },
\Om^{\frac{n-1}{2}}\wh{\PhiH })\no
&&\label{GC7}+\Om^{-\frac{n-1}{2}}Q
(\Om^{\frac{n-1}{2}}\wh{\PhiH }, \Om^{\frac{n+1}{2}}\p\wh{\PhiH
})\no
&&+\Om^{-\frac{n-1}{2}}Q(\Om^{\frac{n-1}{2}}\wh{\PhiH },
\Om^{\frac{n-1}{2}}y^\al \frac{\p}{\p y^\al}\wh{\PhiH })\;. 
\eea
We will need the following consequence of this equation:
multiplying by $y_\nu$ and commuting derivatives one is led to
\bea
 (n-5) y_\nu y_\mu\wh{\PhiH }^{\mu\nu}
 & = & 2 y^\al\frac{\p}{\p y^\al} \left(y_\mu y_\nu \wh{\PhiH }^{\mu\nu}+
 \frac 12 \Om \tr_{\eta}\wh{\PhiH }\right)
 \nonumber
\\
 &&
 +\Omega\Big( \frac{n-5}2 \tr_{\eta}(\wh{\PhiH }) +y_\nu\frac{\p}{\p y^\mu}
 \big(\wh{\PhiH }^{\mu\nu}-\frac 12\eta^{\mu\nu}\tr_\eta(\wh{\PhiH }) \big)\Big)
 \nonumber
\\
 &&+\Om^{-\frac{n-1}{2}}Q (\Om^{\frac{n-1}{2}}\wh{\PhiH },
 \Om^{\frac{n-1}{2}}\wh{\PhiH })
\no
 &&+\Om^{-\frac{n-1}{2}}Q (\Om^{\frac{n-1}{2}}\wh{\PhiH },
 \Om^{\frac{n+1}{2}}\p\wh{\PhiH })\no
&&+\Om^{-\frac{n-1}{2}}Q(\Om^{\frac{n-1}{2}}\wh{\PhiH },
\Om^{\frac{n-1}{2}}y^\al \frac{\p}{\p y^\al}\wh{\PhiH })\;. 
\label{GC7x} \eea

\subsubsection{On the wave equation}

In wave coordinates $(x^\mu),$ we consider the following wave
equation
\begin{equation}
\label{T1}
\eta^{\alpha\beta}\frac{\partial^{2}f}{\partial x^{\alpha}\partial x^{\beta}%
}+H^{\alpha\beta}(f,\partial f)\frac{\partial^{2}f}{\partial
x^{\alpha}\partial x^{\beta}}=F(f,\partial f)\;.
\end{equation}
In order to check all the hypotheses made on components of the
metric in our  theorem on the energy estimate, we have to
rewrite this equation  with respect the system of coordinates
$(\tau, x,v^A)$ used there.
According to our previous calculations, equation (\ref{T1}) can
be written as
\begin{equation}
\label{T3}
\eta^{\la\mu}\frac{\partial^{2}\hat{f}}{\partial y^{\la}\partial y^{\mu}%
}+\Om^{-\frac{n+3}{2}} H^{\la\mu}(f,\partial
f)\frac{\partial^{2}f}{\partial x^{\la}\partial
x^{\mu}}=\Om^{-\frac{n+3}{2}}F(f,\partial f)\;,
\end{equation}
where
$$\hat{f} = \Om^{-\frac{n-1}{2}}f\;.$$
So, let us express the second term of the above equation in
terms of coordinates $y^\nu$. We already know the identity:
\bel{16VIII0.1}
  \frac{\p^2 f}{\p x^\la \p x^\mu}\circ \phi^{-1}
= \frac{\p^2( f\circ \phi^{-1})}{\p y^\al \p y^\beta }A^\al_\mu
A^\bet_\la + \frac{\p (f\circ \phi^{-1})}{\p y^\al}\frac{\p^2
y^\al}{\p x^\mu \p
  x^\la}\circ \phi^{-1} =:K_{\la\mu} + V_{\la\mu}\;,
\ee
with
$$\frac{\p^2 y^\al}{\p x^\mu\p x^\la}\circ \phi^{-1}= 2\Om\de^\al_\mu\et_{\la\si} y^\si +2\Om\de^\al_\la\et_{\mu\tau}  y^\tau
  +2\Om \eta_{\mu\la}y^\al +8\et_{\la\si}\et_{\mu\theta} y^\si y^\al y^\theta$$
and
$$A^\al_\mu A^\bet_\la =
\Om^2 \de^\al_\mu \de^\bet_\la + 4 y_\la y_\mu y^\al y^\bet +2\Om
(\de^\al_\mu y_\la y^\bet  + \de^\bet_\la y_\mu y^\al )\;.$$
These identities lead to
\bel{T4} H^{\la\mu}V_{\la\mu}=
H^{\la\mu}\left\{2\Om\de^\al_\mu\et_{\la\si} y^\si
+2\Om\de^\al_\la\et_{\mu\theta}  y^\theta
  +2\Om \eta_{\mu\la}y^\al +8\et_{\la\si}\et_{\mu\theta} y^\si y^\al y^\theta\right\}\frac{\p f}{\p y^\al}\;.\ee
Now we also know that
\be \label{T5}
  \frac{\partial f}{\partial
y^{\alpha}}=\Om^\frac{n-3}{2}\left\{\Om\frac
{\partial\hat{f}}{\partial
y^{\alpha}}-{(n-1)}y_\al\hat{f}\right\}\;. \ee
This implies that
\footnote{Note that in this equation, the term $y_\mu y_\la
H^{\mu\la} $ is the one which has the the smallest
multiplicative power of $\Om$.}
\bel{T6} H^{\la\mu}V_{\la\mu}=
2\Om^{\frac{n-1}{2}}H^{\la\mu}\Bigg\{(n-1)\left\{\Om\eta_{\la\mu}+2y_\mu
y_\la\right\}\hat{f}+ \left(2\Om\delta^\al_\mu y_\la
+\Om\eta_{\la\mu}y^\al+4y_\mu y_\la y^\al\right)\frac{\p \hat{f}}{\p
y^\al}\Bigg\}\;.\ee
On the other hand we have
\beaa \frac{\partial^{2}(f\circ\phi^{-1})}{\partial
y^{\alpha}\partial y^{\beta} } &=&
\Om^{\frac{n-5}{2}}\Bigg\{\Om^2\frac{\p^{2}\hat{f}}{\p y^{\alpha}\p
y^{\beta}}- (n-1) \Om\left(y_\beta\frac{\p\hat{f}}{\p y^{\alpha}
}+y_\al\frac{\p\hat{f}}{\p y^{\beta}}\right)
\\&&\qquad\qquad\qquad\qquad+ (n-1) \left[(n-3) y_\al y_\beta-\Om\eta_{\alpha\beta}
\right]\hat{f}\Bigg\}\;, \eeaa
which leads to the following expression of
$H^{\la\mu}K_{\la\mu}:$
\beaa
H^{\la\mu}K_{\la\mu}&=&\Om^{\frac{n-5}{2}}H^{\la\mu}\left\{\Om^2
\de^\al_\mu \de^\bet_\la + 4 y_\mu y_\la y^\al y^\bet +2\Om
y^\theta(\eta_{\la\theta}\de^\al_\mu y^\bet
+\et_{\mu\theta}\de^\beta_\la y^\al )
\right\}\nonumber\\
&\times& \left\{ \Om^2\frac{\partial^{2}\hat{f}}{\partial
y^{\alpha}\partial
y^{\beta}}-{(n-1)}\Om\left(y_\beta\frac{\partial\hat{f}}{\partial y^{\alpha}%
}+y_\al\frac{\partial\hat{f}}{\partial y^{\beta}}\right) +{(n-1) }
\left[(n-3) y_\al y_\beta-\Om\eta_{\alpha\beta}
\right]\hat{f}\right\}, \eeaa
%
and after simplifications, we find that
\bea \label{T7}
H^{\la\mu}K_{\la\mu}&=&\Om^{\frac{n-1}{2}}H^{\la\mu}\left\{\Om^2
\de^\al_\mu \de^\bet_\la + 4 y_\mu y_\la y^\al y^\bet +2\Om
(\de^\al_\mu y_\la y^\bet
+\de^\beta_\la y_\mu y^\al ) \right\}\frac{\p^{2}\hat{f}}{\p y^{\al}\p y^{\beta}}\nonumber\\
&+&(n-1)\Om^{\frac{n-1}{2}}H^{\la\mu}\left\{2\left(2y_\la y_\mu
y^\al+\Om\de^\al_\la y_\mu\right)\frac{\p\hat{f}}{\p y^{\al} } +
\left[(n-3) y_\mu y_\la-\Om\eta_{\la\mu}
\right]\hat{f}\right\}\;. \nonumber\\
\eea
With the expressions (\ref{T6}) and (\ref{T7}) and  writing
$H^{\la\mu} = \Om^{\frac{n-1}{2}}\wh{H}^{\la\mu}\;,$ equation
(\ref{T3}) reads after simplifications
\bea \label{T8} &&\left\{\eta^{\al\beta} +
\Om^{\frac{n-5}{2}}\wh{H}^{\la\mu}\left[ \Om^2\de^\al_\mu
\de^\bet_\la + 4 y_\mu y_\la y^\al y^\bet +2\Om (\de^\al_\mu y_\la
y^\bet +\de^\beta_\la y_\mu y^\al )\right]
\right\}\frac{\p^{2}\hat{f}}{\p y^{\al}\p
y^{\beta}}\nonumber\\
&+& 2\Om^{\frac{n-5}{2}}\wh{H}^{\la\mu}\Bigg\{\left\{2(n+1)y_\mu
y_\la y^\al +(n+1)\Om \de^\al_\mu y_\la+ \eta_{\la\mu}\Om y^\al
\right\}\frac{\p \hat{f}}{\p
y^\al}\nonumber\\&&\qquad\qquad\qquad\qquad\qquad\qquad \qquad+
(n-1)\left\{(n+1)y_\mu y_\la + \Om
\eta_{\la\mu}\right\}\hat{f}\Bigg\}\nonumber\\
& =&\Om^{-\frac{n+3}{2}}F\left(\Om^{\frac{n-1}{2}}\hat{f},
\Om^{(n-1)/2}\big\{- \Om \frac{\p}{\p y^\nu} - 2y_\nu
y^\al\frac{\p}{\p y^\al} -
(n-1)y_\nu\big\}\hat{f}\right) \nonumber\\
& =:&
\Om^{-\frac{n+3}{2}}\wt{F}\left(\Om^{\frac{n-1}{2}}\hat{f},\Om^{\frac{n-1}{2}}\frac{\p\hat{f}}{\p
y^\nu}\right)\;.\eea

We want to apply the energy estimates of Section~\ref{Energie} to
the equation considered here. So for consistency of notation in that
section, we write the above equation in the form (recall that $\Om=
x(\rho-\tau)$):
\bel{T13} \Box_\m u = \mathcal{F}(u,\p u)  \;,\ee
with
\bel{T10} u = \hat{f}\;,\ee
\bean
 \m^{\al\beta} &= &  \eta^{\al\beta} +
 \{x(\rho-\tau)\}^{\frac{n-5}{2}}\wh{H}^{\la\mu}
 \times
\\
 &&
 \underbrace{
 \left\{
 \{x(\rho-\tau)\}^2\de^\al_\mu \de^\bet_\la + 4 y_\mu y_\la
 y^\al y^\bet +2\{x(\rho-\tau)\} (\de^\al_\mu y_\la y^\bet
 +\de^\beta_\la y_\mu y^\al )\right\}
  }_{:= \psi^{\alpha\beta}{}_{\mu\lambda}}
 \;,
  \nonumber
\\
 &&
\eeal{T11}
(in order to reduce the typographical length of formulae we
will sometimes write $\psi^{\alpha \beta}_{\mu\nu}$ for
$\psi^{\alpha \beta}{}_{\mu\nu}$) and
\bean
\mathcal{F}\left(u , \frac{\p u}{\p y^\nu}\right)&=&
\Om^{-\frac{n+3}{2}}\wt{F}\left(\Om^{\frac{n-1}{2}}
u,\Om^{\frac{n-1}{2}}\frac{\p u}{\p y^\nu}\right)\nonumber\\
&&+\Bigg\{{\Upsilon}^\al-
2\Om^{\frac{n-5}{2}}\wh{H}^{\la\mu}\left\{2(n+1)y_\mu y_\la y^\al +
(n+1) \Om \de^\al_\mu y_\la+\eta_{\la\mu}\Om y^\al
\right\}\Bigg\}\frac{\p u}{\p y^\al}\nonumber\\
&& - 2(n-1)\Om^{\frac{n-5}{2}}\wh{H}^{\la\mu} \left\{(n+1)y_\mu
y_\la + \Om \eta_{\la\mu}\right\}u
 \;.
 \label{T12}
\eea
So, we have to check that the metric $\m$ defined by (\ref{T11}) and
the harmonicity functions
\bel{T14}{\Upsilon}^\mu = \frac{1}{\sqrt{|\det
\m|}}\p_\nu\left\{\sqrt{|\det \m|}\m^{\mu\nu}\right\}\ee
satisfy the hypotheses of our theorem.

The tensor $\psi^{\alpha \beta}{}_{\mu\nu}$ defined in \eq{T11}
has the  property
\bea
 \eta_{\alpha\beta} \psi^{\alpha \beta}{}_{\mu\nu} &= &
 \Omega^2 \eta_{\mu\nu}
 \;,
\eea
which implies that the contraction
$$
 \eta_{\alpha\beta}(\m^{\alpha\beta}- \eta^{\alpha\beta})= \Omega^{\frac{n-1}2} \tr_\eta \wh H
$$
gains two powers of $\Omega$, as compared to a direct
power-counting based on \eq{T11}. Furthermore, the structure
$y^\alpha y^\beta y_{\mu} y_{\nu}$ of the term without powers
of $\Omega$ in $ \psi^{\alpha \beta}{}_{\mu\nu} $ implies that
any contraction of the form $ \psi^{\alpha
\beta}{}_{\mu\nu}\eta_{\alpha\rho}
  \psi^{\rho \sigma}{}_{\gamma\delta} $ acquires an overall multiplicative factor
of  $\Omega$. So if we set
$$
 \delta \m^\alpha{}_\beta:= \m^{\alpha \mu}\eta_{\mu \beta} - \delta^\alpha_\beta
 \;,
$$
it follows that for $k\ge 2$ we have
$$
\left( (\delta\m)^k\right) {} ^\alpha{}_\beta
 := \delta \m^\alpha{}_{\alpha_1} \delta \m^{\alpha_1}{}_{\alpha_2}
 \cdots
 \delta \m^{\alpha_{k-1}}{}_\beta = \Omega^{k-1} Q_k(\Omega^{\frac{n-5}2} \wh H)\;,
$$
where we use the symbol $Q_k$ to denote a smooth function (in
this case,  a polynomial) with a uniform zero of order $k$, and
which may change from line to line. A similar analysis shows
that, again for $k\ge 2$, the trace
\bel{21VIII0.10}
 p_k(\delta \m):= \tr (\delta \m)^k =  \delta
\m^\alpha{}_{\alpha_1} \delta \m^{\alpha_1}{}_{\alpha_2}
 \cdots
 \delta \m^{\alpha_{k-1}}{}_\alpha =\Omega^{k} Q_k(\Omega^{\frac{n-5}2} \wh H)
%
\ee
(no summation over $k$) gains one more power of $\Omega$.

Set
\bel{21VIII0.11}
 A^\alpha{}_\beta:= \delta^\alpha_ \beta + \delta \m ^\alpha{}_\beta
 \;.
\ee
\Eq{21VIII0.10} implies
\bel{21VIII0.12}
 p_i (A) = \tr (I + \delta \m)^i = \sum_{j=0}^n C^j_ i p_j (\delta \m)
 = n+1 + i \tr \delta \m + \Omega^2 Q_2(\Omega^{\frac{n-5}2} \wh H)
 \;.
\ee
Let $W(\lambda)$ denote the characteristic polynomial of $A$,
$$
W(\lambda) = \det (A - \lambda I)= \det A + w_1 \lambda + \ldots
 +w_n \lambda^n+  (-\lambda)^{n+1}
 \;.
$$
Then the coefficients $w_i$ are homogeneous polynomials of
order $n+1-i$ in the entries of $A = I+\delta \m$, with $w_n =
(-1)^n\tr A = (-1)^n(n+1+\tr \delta \m)$.
It is a well known consequence of the Cayley-Hamilton theorem
(see, e.g., \cite[Theorem~1]{CayleyHamilton}) that both $\det
A$ and the $w_i$'s can be written as polynomials in the
$p_i$'s, and since each $p_i(A)$ has a factor $\Omega^2$ in
front of the $Q_2$ terms, we find that the $w_i$'s take the
form
\bel{21VIII0.13}
 w_i(A) = w_i(I) + \ell_ i (\tr \delta \m) + \Omega^{2} Q_2  (\Omega^{\frac{n-5}2} \wh H)
 \;,
\ee
where $\ell_i( \tr \delta \m) $ is linear in $\tr \delta \m $.

Now
\bel{21VIII0.15}
 \m^{\alpha \beta} = \m^{\alpha \sigma}\eta_{\sigma\rho} \eta^{\rho\beta}
 =
 \left( \delta^\alpha_\rho + \delta \m^\alpha{}_\rho \right)\eta^{\rho\beta}
 = A^\alpha{}_\rho \eta^{\rho\beta}
  \;,
\ee
hence
$$
 \det \m^\sharp = - \det (A)
 \;,
$$
which shows that
\bel{21VIII0.14}
 \det \m^\sharp = -1
 + \Omega^2\left(-\Omega^{\frac{n-5}2} \tr_\eta \wh H +  Q_2  (\Omega^{\frac{n-5}2} \wh H)\right)
 =
 -1 + \Omega^2 Q_1  (\Omega^{\frac{n-5}2} \wh H)
 \;.
\ee
%
From the Cayley-Hamilton theorem we have
$$
 A^{-1} = - \frac 1 {\det A}\left( w_1 I+ \cdots + w_{n }A^{n-1}+(-1)^{n+1}A^n\right)
 \;,
$$
and we conclude that $\m _{\alpha\beta} = (\eta^{-1}
A^{-1})_{\alpha\beta}$ takes the form
\bean
 \m _{\alpha\beta}
  &= & \frac 1 {1 + \Omega^2 Q_1  (\Omega^{\frac{n-5}2})} \left( \eta_{\alpha\beta} -\Omega^{\frac{n-5}2} \wh H^{\mu\nu}
 \psi_{\alpha\beta\mu\nu}
 + \Omega^2 Q_2  (\Omega^{\frac{n-5}2} \wh H)\right)
\\
  &= &  \eta_{\alpha\beta} -\Omega^{\frac{n-5}2} \wh H^{\mu\nu} y_\mu y_\nu
  y^\alpha y^\beta
 +   \Omega Q_1  (\Omega^{\frac{n-5}2} \wh H)
 \nonumber
\\
 &&
 + \Omega^2 Q_2  (\Omega^{\frac{n-5}2} \wh H)
 \;,
\eeal{21VIII0.18}
where the indices on $ \psi_{\alpha\beta\mu\nu}$ have been
lowered with the metric $\eta_ {\alpha\beta }$.

\subsubsection{On the components of the metric}
Recall that, to obtain energy inequalities, our hypotheses on
certain components of the metric were
\bel{Eq-h} \m^{00} = -1 +
x\f^0;\quad \m^{0\rho} = -x\f^1; \quad \m^{0 A}+\m^{\rho A}=
-x\f^A \quad \text{and} \quad \m^{\rho\rho}=1+ x\f\;,\ee
where the functions $\f,\;\f^0,\;\f^A$ are bounded on bounded sets.
Since (compare (\ref{T2}))
$$ \m^{0\rho} = \m^{0i}\omega_i,\quad \m^{0A} = \m^{0i}\frac{\p v^A}{\p y^i}, \quad \m^{\rho A} =
\m^{ij}\omega_j\frac{\p v^A}{\p y^i} \quad \text{and}\quad
\m^{\rho\rho} = \m^{ij}\omega_i\omega_j
\;,$$
from (\ref{T11}) we have (note that $y^i\omega_i = \rho,\; \rho
\omega_i\delta^i_\mu = y_\mu+ \tau\de^\tau_\mu $):
\bel{T18} \f^0 =
x^{\frac{n-7}{2}}(\rho-\tau)^{\frac{n-5}{2}}\wh{H}^{\la\mu}\left\{
\{x(\rho-\tau)\}^2\de^\tau_\mu \de^\tau_\la + 4\tau^2 y_\mu y_\la
+4\tau\{x(\rho-\tau)\}\de^\tau_\mu y_\la \right\}\;,\ee
\bel{T19} \f^1 =-
x^{\frac{n-7}{2}}(\rho-\tau)^{\frac{n-5}{2}}\wh{H}^{\la\mu}\left\{
\{x(\rho-\tau)\}^2\de^\tau_\mu \de^i_\la\omega_i + 4\tau\rho y_\mu
y_\la +2\rho\{x(\rho-\tau)\}y_\la(\de^0_\mu \rho
+\tau\de^i_\mu\omega_i)\right\}\;,\ee
\bel{T20} \f =
x^{\frac{n-7}{2}}(\rho-\tau)^{\frac{n-5}{2}}\wh{H}^{\la\mu}\Big\{
\{x(\rho-\tau)\}^2\de^i_\mu \de^j_\la \omega_i\omega_j+ 4\rho^2
y_\mu y_\la +4\{x(\rho-\tau)\}y_\la \rho\de^i_\mu y_\la\omega_i
\Big\}\;,\ee
\bel{TT20} \f^A
=-x^{\frac{n-3}{2}}(\rho-\tau)^{\frac{n-3}{2}}\left\{(\rho-\tau)\left(\wh{H}^{0i}+\omega_j\wh{H}^{ij}\right)-2y_\la\wh{H}^{\la
i}\right\}\frac{\p v^A}{\p y^i}\;. \ee
We see that the components of the metric  (\ref{T11}) have the
right structure (\ref{Eq-h}) if the space dimension $n$ is
greater then or equal to 7. We will see in
Section~\ref{sss1IV0.1} (see \eqref{GC7x}) that this can be
lowered to $n\ge 6$ using the harmonic coordinates condition.

We note the identities,
$$\eta^{ij}\omega_j\frac{\p v^A}{\p
y^i} = \sum_{j=1}^n \omega_j\frac{\p v^A}{\p y^j} = \frac{\p v^A}{\p
r} = 0\;,$$
which justify that  $\m^{0A}+\m^{\rho A}$ has the right
structure. In particular, for  this component the condition $n
\ge 4$ suffices to fulfill the structure condition.

We will also need
\begin{equs} \label{metric-x}\m^{\tau\tau}=-1+ O(x^{\frac{n-5}{2}})
,\quad
\m^{\tau x}=1+ O(x^{\frac{n-3}{2}}),\quad
\m^{xx}= O(x^{\frac{n-1}{2}}),\\
  \m^{xA}= O(x^{\frac{n-3}{2}}),\quad
\m^{AB}=\eta^{AB}+ O(x^{\frac{n-5}{2}})\;.\end{equs}
\subsubsection{On the harmonicity functions}
Now let us look at the harmonicity functions, defined as
$${\Upsilon}^ \mu:= \frac{1}{\sqrt{|\det \m|}}\p_\nu\left\{\sqrt{|\det
\m|}\m^{\mu\nu}\right\} \;.$$
Since our energy estimates have been established using the
coordinate system $(x,\tau,v^A)$ as defined in \eq{20VIII0.6},
we need to calculate $\Upsilon^\mu$ in that coordinate system.
But so far we only have the expression of the metric in the
$y^\mu$--coordinate system. To avoid confusion let us write
$\xUp$ for $\Upsilon$ associated to the coordinates
$(\tau,x,v^A)$ and $\yUp$ for that associated to the
coordinates $y^\mu$. To understand the behaviour of $\Upsilon$
under coordinate changes, it is useful to write the Christoffel
symbols $\Gamma^\alpha _{\beta\gamma}$ of the metric $\m$ in
the form
$$
 \Gamma^\alpha _{\beta\gamma} = \mathring \Gamma^\alpha _{\beta\gamma}
 + C^\alpha _{\beta\gamma}
 \;,
$$
where the $\mathring \Gamma^\alpha _{\beta\gamma}$'s are the
Christoffel symbols of the Minkowski metric $\eta$, and $
C^\alpha _{\beta\gamma}$ is a tensor. Then, in the coordinate
system $y^\mu$ we have
\bel{21VIII0.1}
 \yUp^\alpha = - \underbrace{\m^{\beta\gamma}C^\alpha _{\beta\gamma}}_{=:C^\alpha}
  \;,
\ee
since the $ \mathring \Gamma^\alpha _{\beta\gamma}$'s vanish in
the $y^\mu$--coordinates. Note that $C^\alpha$ as defined in
\eq{21VIII0.1} is a vector field, being the contraction of two
tensors. In the coordinates $(\tau,x,v^A)$ we have
\bel{xUp21-08-10}
 \xUp^\alpha = - \m^{\beta\gamma}\left(\mathring \Gamma^\alpha _{\beta\gamma}
 +C^\alpha _{\beta\gamma}\right)
   =
 - \m^{\beta\gamma} \mathring \Gamma^\alpha _{\beta\gamma}
 -
 C^\alpha
 \;.
\ee
Thus, to calculate $\xUp$ we need to vector-transform
$C^\alpha$ to the $(\tau,x,v^A)$ coordinates, and calculate the
missing term $ \m^{\beta\gamma} \mathring \Gamma^\alpha
_{\beta\gamma}$ above.
We start by calculating the vector field $C^\mu$.  We set
\bel{T23}\;\m^{\al\beta} =: \eta^{\al\beta} +
 \Om^{\frac{n-5}{2}}K^{\al\beta}\;,
\ee
thus
$$
 K^{\alpha\beta} = \wh H^{\mu\nu}\psi^{\alpha\beta}_{\mu\nu}
$$
as in (\ref{T11}); we hope that the clash of notation with the
completely different  $K_{\alpha\beta}$ appearing in
\eq{16VIII0.1} will not confuse the reader.

 From
\eq{21VIII0.14} we have (recall that $Q$ means $Q_2$)
$$
 \left(\sqrt{|\det\m|}\right)^{\mp 1}
 = 1\pm\frac 12 \Om^{\frac{n-1}{2}} \tr_{\eta}(\wh{H}) +\Om^2
 Q(\Om^{\frac{n-5}{2}}\wh{H})
 \;.
$$
Thus in the coordinate system $y^\mu$,
\bea\label{T17}\m^{\mu\nu}\sqrt{|\det\m|}&=& \eta^{\mu\nu}
\left(1-\frac 12 \Om^{\frac{n-1}{2}} \tr_{\eta}\wh{H}\right)+
\Om^{\frac{n-5}{2}}K^{\mu\nu}\no
&& + \Om^2Q^{\mu\nu}(\Om^{\frac{n-5}{2}}\wh{H})
 \;,
\eea
\beaa \p_\nu \left(\m^{\mu\nu}\sqrt{|\det \m|}\right) &=&\frac
12\Om^{\frac{n-3}2}\left\{(n-1)y^\mu \tr_\eta\wh{H}
-\eta^{\mu\nu}\Om \p_\nu \tr_\eta\wh{H}\right\}\\
&&+  \Om^{\frac{n-7}2}\big\{(5-n)y_\nu K^{\mu\nu} +\Om \p_\nu
K^{\mu\nu}
\big\}+\p_\nu\left\{\Om^2Q^{\mu\nu}(\Om^{\frac{n-5}{2}}\wh{H})\right\}\eeaa
and since
\bel{K1} \partial_\nu \Omega K^{\mu\nu} \sim y_\nu K^{\mu\nu}=-
\Om y_\beta \wh{H}^{\al\beta}\left\{\Om\de^\mu_\al +2y_\al
y^\mu\right\} \;,\ee
and
\bel{K2}\p_\nu K^{\mu\nu} = \p_\nu
\wh{H}^{\al\beta}\psi^{\mu\nu}_{\al\beta}+
2(n+3)y_\beta\wh{H}^{\al\beta}\left\{\Om\de^\mu_\al  +2y_\al
y^\mu\right\}+ 2\Om y^\mu \tr_\eta{\wh{H}}\;,\ee
we obtain
\beaa \p_\nu \left(\m^{\mu\nu}\sqrt{|\det \m|}\right) &=&\frac
12\Om^{\frac{n-3}2}\left\{(n+3)y^\mu \tr_\eta\wh{H}
-\eta^{\mu\nu}\Om \p_\nu \tr_\eta\wh{H}\right\}\\
&&+  \Om^{\frac{n-5}2}\bigg\{\p_\nu\wh{H}^{\al\beta}\psi^{\mu\nu}_{\al\beta}
+(3n+1)y_\beta\wh{H}^{\al\beta}(\Om\delta^\mu_\al+2y_\al y^\beta)
\bigg\}\\
&&+\Om^2 Q_2^{\mu}(\Om^{\frac{n-5}{2}}\wh{H})
+\Om^2 Q_2^{\mu\nu}(\Om^{\frac{n-5}{2}}\wh{H},\Om^{\frac{n-5}{2}}\p_\nu\wh{H} )\;.\eeaa
Multiplying this last identity with $\left(\sqrt{|\det\m|}\right)^{-1}$ 
%
we then obtain the following expression for the vector field
$C^\mu$:
\bea C^\mu = {\yUpsilon}^\mu &=&\frac 12\Om^{\frac{n-3}2}\left\{(n+3)y^\mu
\tr_\eta\wh{H} -\eta^{\mu\nu}\Om \p_\nu \tr_\eta\wh{H}\right\}\nonumber\\
&&+  \Om^{\frac{n-5}2}\big\{ \p_\nu
\wh{H}^{\al\beta}\psi^{\mu\nu}_{\al\beta}+(3n+1)y_\beta\wh{H}^{\al\beta}\left\{\Om\de^\mu_\al
+2y_\al y^\mu
\right\} \big\}\nonumber\\
&& + \Om^2 Q^{\mu}(\Om^{\frac{n-5}{2}}\wh{H})
+\Om^2 Q^{\mu\nu}(\Om^{\frac{n-5}{2}}\wh{H},\Om^{\frac{n-5}{2}}\p_\nu\wh{H} )
 \;.
 \label{T25}
\eea
Now writing the vector field $C$ as
$$
C=C^\mu\p_\mu=:C^\tau\p_\tau+C^x\p_x+C^A\p_A\;,
$$
one is led to:
\beaa C^\tau = C^0\;,\quad
 C^\tau+C^x = -\omega_i(v)C^i\;,\quad
 C^A = \frac{\p v^A}{\p y^i}C^i
  \;.
\eeaa
In order to have all the harmonicity functions in the
$(\tau,x,v^A)$-coordinates,  it remains to calculate the term
$\m^{\beta\gamma} \mathring \Gamma^\alpha _{\beta\gamma}$ of
the formula (\ref{xUp21-08-10}). In these coordinates the
Christoffell's symbol of the Minkowski metric $\mathring
\Gamma^\alpha _{\beta\gamma}$ read:
\beaa
\mathring\Gamma^\tau_{\al\beta}&=&  0,
 \\
\mathring\Gamma^x_{\tau\mu}&=&\mathring \Gamma^x _{x\mu}=0,
\;\;\mathring\Gamma^x _{AB}= \rho \chi_{AB}\\
\mathring\Gamma^A_{\tau\tau }&=&\Gamma^A_{\tau x}= \mathring\Gamma^A _{x x}=0,
\;\;\mathring\Gamma^A _{\tau B}=\mathring\Gamma^A _{x B}=-\frac 1\rho\delta^A_B, \;\;
\mathring\Gamma^A _{BC}=\gamma^A_{BC}
 \;,
\eeaa
where we have denoted the round metric on the sphere by $\chi$,
and its corresponding Christoffel symbols $\gamma^A_{BC}.$
These identities lead to the following (see identity
(\ref{T23})):
\begin{deqarr}
\m^{\beta\gamma} \mathring \Gamma^\tau _{\beta\gamma}&=&0\phantom{xxxxx}\arrlabel{RxUp}\\
\m^{\beta\gamma} \mathring \Gamma^x _{\beta\gamma}&=&\rho\m^{AB}\chi_{AB}
=\frac{n-1}\rho+\rho\Om^{n-5\over 2}\wh{H}^{\mu\nu}\psi^{AB}_{\mu\nu}\chi_{AB}\qquad\\
\m^{\beta\gamma} \mathring \Gamma^A _{\beta\gamma}&=&-\frac 2\rho (\m^{\tau A}+  \m^{xA})+\m^{BC}\gamma^A_{BC}\\
&=&\frac{1}{\rho^2} \mathring C^A+ \Om^{n-5\over
2}\wh{H}^{\mu\nu}\left(2\psi^{iA}_{\mu\nu}\frac{\omega_i}{\rho}
 + \psi^{BC}_{\mu\nu}
 \gamma^A_{BC}\right)
 \;;
\end{deqarr}
where $\mathring C^A= \chi^{BC}\gamma^A_{BC}$ is minus the
harmonicity function on the unit sphere. Finally, we obtain
that the harmonicity functions of the metric $\m$ in the
$(\tau,x,v^A)$-coordinates read:
\begin{deqarr}
 \xUp^\tau &=& - C^0\phantom{xxxxx}\arrlabel{xUpsilon}
\\
 \xUp^\tau+\xUp^x
 &=&
 \omega_i(v)C^i
 -\frac{n-1}\rho
 -
 \rho\Om^{n-5\over 2}\wh{H}^{\mu\nu}\psi^{AB}_{\mu\nu}
\\
 \xUp^A
 &=& -
 \frac{\p v^A}{\p y^i}C^i
 -\frac{1}{\rho^2} \mathring C^A
 \nonumber
 \\
 &&
 -
 \Om^{n-5\over 2}\wh{H}^{\mu\nu}
 \left(2\psi^{iA}_{\mu\nu}\frac{\omega_i}{\rho}
 +
 \psi^{BC}_{\mu\nu}
 \gamma^A_{BC}\right)
  \;.
  \phantom{xxxxx}
 \end{deqarr}
We revert now to the notation $\Upsilon$ for what was denoted
by $\xUp$ above.

\subsubsection{The source term $\mathcal{F}$}

Recall that the source term  in $y^\mu$coordinates reads:
\bea\label{Sterm} \mathcal{F}\left(u , \frac{\p u}{\p
y^\nu}\right)&=&
\Om^{-\frac{n+3}{2}}\wt{F}\left(\Om^{\frac{n-1}{2}}
u,\Om^{\frac{n-1}{2}}\frac{\p u}{\p y^\nu}\right)\nonumber\\
&&+\Bigg\{{\yUpsilon}^\al-
2\Om^{\frac{n-5}{2}}\wh{H}^{\la\mu}\left\{2(n+1)y_\mu y_\la y^\al +
(n+1) \Om \de^\al_\mu y_\la+\eta_{\la\mu}\Om y^\al
\right\}\Bigg\}\frac{\p u}{\p y^\al}\nonumber\\
&& - 2(n-1)\Om^{\frac{n-5}{2}}\wh{H}^{\la\mu} \left\{(n+1)y_\mu
y_\la + \Om \eta_{\la\mu}\right\}u\;.\eea
From (\ref{T25})  we have
\beaa \lefteqn{{\yUpsilon}^\al-
2\Om^{\frac{n-5}{2}}\wh{H}^{\la\mu}\left\{2(n+1)y_\mu y_\la y^\al +
(n+1) \Om \de^\al_\mu y_\la+\eta_{\la\mu}\Om y^\al
\right\}}\nonumber\\
&=&\frac 12\Om^{\frac{n-3}2}\left\{(n-1)y^\al
\tr_\eta\wh{H} -\eta^{\al\nu}\Om \p_\nu \tr_\eta\wh{H}\right\}\nonumber\\
  &&+\Om^{\frac{n-5}{2}}\Big\{\psi_{\mu\la}^{\al\nu}\p_\nu \wh{H}^{\la\mu}
+(n-1) y_\la\wh{H}^{\mu\la}\left\{\Om
\de^\al_\mu+2y_\mu  y^\al \right\}\Big\}\\
&& +\Om^2 Q^{\al}(\Om^{\frac{n-5}{2}}\wh{H},\Om^{\frac{n-5}{2}}\wh{H})
+\Om^2 Q^{\al\beta}(\Om^{\frac{n-5}{2}}\wh{H},\Om^{\frac{n-5}{2}}\p_\beta\wh{H} ) \;. \eeaa
This shows that the source term takes the following form:
\bea \mathcal{F}\left(u , \frac{\p u}{\p y^\nu}\right)&=&
\Om^{-\frac{n+3}{2}}\wt{F}\left(\Om^{\frac{n-1}{2}}
u,\Om^{\frac{n-1}{2}}\frac{\p u}{\p y^\nu}\right)\nonumber\\
&& - 2(n-1)\Om^{\frac{n-5}{2}}\wh{H}^{\la\mu} \left\{(n+1)y_\mu
y_\la + \Om \eta_{\la\mu}\right\}u\nonumber\\
&&+\frac 12\Om^{\frac{n-3}2}\left\{(n-1)y^\al
\tr_\eta\wh{H} -\eta^{\al\nu}\Om \p_\nu \tr_\eta\wh{H}\right\}\frac{\p u}{\p y^\al}\nonumber\\
  &&+\Om^{\frac{n-5}{2}}\Big\{\psi_{\mu\la}^{\al\nu}\p_\nu \wh{H}^{\la\mu}
+ (n-1)y_\la\wh{H}^{\mu\la}\left\{\Om
\de^\al_\mu+2y_\mu  y^\al \right\}\Big\}\frac{\p u}{\p y^\al}\nonumber\\
&& + \Big\{\Om^2 Q^{\al}(\Om^{\frac{n-5}{2}}\wh{H},\Om^{\frac{n-5}{2}}\wh{H})
+\Om^2 Q^{\al\beta}(\Om^{\frac{n-5}{2}}\wh{H},\Om^{\frac{n-5}{2}}\p_\beta\wh{H}\Big\}\frac{\p
 u}{\p y^\al}\;.\nonumber
\\
 \label{Sterm1}
\eea
%

\subsection{The Einstein-Maxwell case}
\subsubsection{Existence of a solution} \label{ssSource}

The  Einstein-Maxwell equations, in  harmonic and Lorenz gauge, take
the form (\ref{T1}) (see
\cite{LindbladRodnianski2,CCL,LoizeletCRAS}) with the following
replacements there:
\bel{E-M} f = (\underbrace{g_{\mu\nu}- \eta_{\mu\nu}}_{\;:= \;
h_{\mu\nu}}
  , A_\mu) \quad \text{and} \quad   H^{\al\beta}= g^{\al\beta}-\eta^{\al\beta}  \;.
\ee
Recall that, if $v$ is an arbitrary function, then
\bel{18VIII0.1}
 \hat  v = \Om^{-\frac{n-1}{2}}v
 \;.
\ee
Therefore, we have
$$\hat f =(\hat h_{\mu\nu},\;\wh{A}_\mu):=
(\Om^{-\frac{n-1}{2}} h_{\mu\nu} ,\; \Om^{-\frac{n-1}{2}}A_\mu)\quad
\text{and} \quad \wh{H}^{\al\beta}= \Om^{-\frac{n-1}{2}}H^{\al\beta}
\;.
$$
For consistency of notation with Section~\ref{Energie}   we set
$$\hat f \equiv
u\;.
$$
In this notation
$$
\|\hat h_{\mu\nu}\|_{\mcH^\theta_k}\le \|u\|_{\mcH^\theta_k}\;,
$$
and, since
$$
\wh{H}^{\al\beta}= -\eta^{\al\mu}\eta^{\beta\nu}\hat
h_{\mu\nu}
 +\Omega^{-(n-1)/2} Q^{\al\beta} \left(\Omega^{(n-1)/2}\hat h_{\mu\nu}\right)\;,
$$
where $Q^{\al\beta}$ has a uniform zero of order two, from
\cite[Proposition~A.2]{ChLengardnwe} we obtain that
\bean
  \|\wh{H}^{\al\beta}\|_{\mcH^\theta_k}
   & \le &
 \|\eta^{\al\mu}\eta^{\beta\nu}\hat
h_{\mu\nu}\|_{\mcH^\theta_k}+ \|\Omega^{-\frac{n-1}{2}} Q^{\al\beta}
\left(\Omega^{ n-1 \over2}\hat
 h_{\mu\nu}\right)\|_{\mcH^\theta_k}
\\
 & \le &  C\left(\|\hat
 h_{\mu\nu}\|_{L^\infty}\right)\|u\|_{\mcH^{\theta-(n-1)/2}_k}
  \nonumber
\\
 & \le &  C\left(\|\hat
 h_{\mu\nu}\|_{L^\infty}\right)\|u\|_{\mcH^{\theta }_k}
 \;.
\eeal{H-u}
We define the energy $E^\al_{k,\la}[u(\tau)]$ as in Equation
(\ref{E10}) of Section (\ref{Energie}), the metric being
defined by (\ref{T11}). Recall (see Equation (\ref{EE48}) of
Section (\ref{Energie})) that this quantity controls the
$\mcH^\al_k$-norms of $\p \hat f$. Now,
$$ \|\p\wh{H}^{\al\beta}\|^2_{\mcH^\theta_k}\le
\|\p(\eta^{\al\mu}\eta^{\beta\nu}\hat
h_{\mu\nu})\|^2_{\mcH^\theta_k}+ \|\p \left(\Omega^{-\frac{n-1}{2}} Q^{\al\beta}
(\Omega^{(n-1)/2}\hat
 h_{\mu\nu})\right)\|^2_{\mcH^\theta_k}\;.$$
Since
\beaa
 \p\left(\Omega^{-\frac{n-1}{2}} Q^{\al\beta}(\Omega^{(n-1)/2}\hat h_{\mu\nu})\right)
 &=& \Om^{-\frac{n+1}{2}} Q^{\al\beta}(\Om^{\frac{n-1}{2}}\hat h)
     +\Om^{-\frac{n-1}{2}} Q^{\al\beta}(\Om^{ \frac{n-1}{2}}\hat h,\Om^{ \frac{n-1}{2}}\p\hat h)\\
 &=&
\Om^{-\frac{n+1}{2}} Q^{\al\beta}(\Om^{\frac{n-1}{2}}\hat h)
     +\Om^{-\frac{n-1}{2}+\al} Q^{\al\beta}(\Om^{ \frac{n-1}{2}}(\hat h,
      x^{-\al}\p\hat h))
,\eeaa
we have the estimate:
\beaa  \|\p \left(\Omega^{-\frac{n-1}{2}} Q^{\al\beta}
(\Omega^{(n-1)/2}\hat h_{\mu\nu})\right)\|^2_{\mcH^\theta_k}
&\leq&
 \|\Om^{-\frac{n+1}{2}} Q^{\al\beta}(\Om^{\frac{n-1}{2}}\hat h)\|_{\mcH^\theta_k}\\
 &&+ \|\Om^{-\frac{n-1}{2}+\al} Q^{\al\beta}(\Om^{ \frac{n-1}{2}}(\hat h,
   x^{-\al}\p\hat h)) \|_{\mcH^{\theta}_k} \\
 &\le& C (\|\hat h\|_{L^\infty})\|\hat h\|_{\mcH^{\theta-{(n-1)/2}}_k} \\
 &&+ C(\|\hat h,\;x^{-\al}\p\hat h)\|_{L^\infty} \| (\|\hat h,\;x^{-\al}\p\hat
    h)\|_{\mcH^{\theta-\al-(n-1)/2}_k}  \\
 &\le& C(\|\hat h,\;x^{-\al}\p\hat h)\|_{L^\infty}
        \left(\|\hat h\|_{\mcH^{\theta}_k}+\|\hat h \|_{\mcH^{\theta-\al}_k}+\| \p\hat
    h \|_{\mcH^{\theta }_k}  \right) \\
 &\le& C(\|\hat h,\;x^{-\al}\p\hat h)\|_{L^\infty}
        \left(\|\hat h \|_{\mcH^{\theta-\al}_k}+\| \p\hat
    h \|_{\mcH^{\theta }_k}  \right)\;.
\eeaa
Thus,
$$
 \|\p\wh{H}^{\al\beta}\|^2_{\mcH^\theta_k}\le C\left(\|\hat h,\;x^{-\al}\p\hat
 h)\|_{L^\infty}\right)\left(\| u\|_{\mcH^{\theta-\al}_k}+\|\p u\|_{\mcH^{\theta}_k}\right)
\;.
$$

To continue, we suppose that at $x=x_1>0$ the maximal globally
hyperbolic development of the data exists for $\tau \in [\tauz,
\tau_1]$, with
$$
 M_1:= \| \hat f|_{\{x=x_1\}}\| _{L^\infty }< \infty
 \;.
$$
We define (compare \eq{23VIII0.3})
\bean
 \hat   M(\tau) &:= &  \|{\cal F} \|^2_{\mcB^\al_0({\bf{\Large{H_{\tau}}}})}
 + \|(\m, (\p_\tau-\p_x)\m^\sharp)
 \|^2_{L^\infty({\bf{\Large{H_{\tau}}}})}
 + \|(\m^\sharp,\f^\sharp, {\Upsilon} )
\|^2_{\mcC^0_{\{x=0\},1}({\bf{\Large{H_{\tau}}}})}
\\
 &&
+\|\big((\p_\tau-\p_x)\hat f ,\p_x \hat f, \p_A \hat f\big)
 \|^2_{\mcB^\al_1({\bf{\Large{H_{\tau}}}})} + \| \hat f(\tau)|_{\{x=x_1\}}\| _{L^\infty
 }
 \;,
\eeal{23VIII0.3x}
with  the functions $ \m^\sharp,\; \f^\sharp,$
${\Upsilon}^\mu\equiv \xUp$ and $\cal F$   defined by equations
(\ref{T11}), (\ref{T18})-(\ref{TT20}), (\ref{xUpsilon}) and
(\ref{Sterm1}).

For any positive function $N(\tau)$ we set
\bel{28VIII0.6}
 \underline{ N(\tau)} := \sup_{s\in [\tauz,\tau]} N(s)
 \;.
\ee

We then have the following:
\begin{Proposition}
\label{prop-EM} Let $k\in \N $, $\alpha\in (-1,-1/2]$. Consider
the Einstein-Maxwell equations (\ref{T1}) in space-time
dimension $ 1+n \ge7$ if $\al= -\frac 12$, and $1+n \ge 8$
otherwise. Let $f$ be defined  in  (\ref{E-M}), suppose that
$t_0>0$ and assume that the initial data, given on the
hyperboloid
\bel{defmcS0} \mcS_0 = \left\{(x^\mu): x^0- t_0 =
\sqrt{t^2_0+|\vec x |^2}\;\right\}  \ee
in Minkowski space-time, are such that:
\bel{Ini-data}\hat{f}\left|_{\phi(\mcS_0)} \in\left(
\mcH^\al_{k+1}\cap L^\infty\right)(\phi(\mcS_0))\right., \quad \text
and\quad \left((\p_\tau-\p_x)\hat{f}, \p_x\hat{f}, \p_A \hat{f}
\right)\left|_{\phi(\mcS_0)} \in
\mcH^\al_k(\phi(\mcS_0))\right.\;.\ee
There exists  functions $\hat C_3( n,  k, \epsilon_0,
C_0,\alpha,\hat M)$ and $\hat C_4( n,  k, \epsilon_0,
C_0,\alpha,\hat M$), monotonously increasing in $\hat M$, which
we write as $\hat C_3(\hat M)$ and $\hat C_4(\hat M)$, such
that the energy of the system as defined in (\ref{E10}),
Section~\ref{Energie} satisfies the inequality
\bea\label{In-EM}
 \|\hat f(\tau)\|^2_{L^\infty}
 +E^\al_{k }[\hat f(\tau)]
 &\le&
  2\Big\{  M_1^2 +E^\al_{k }[\hat f(\tauz )] \no
 &&
 + \int_{\tauz }^\tau \hat C_3(\hat M(s))
 E^\al_{k }[\hat f(s)] ds\Big\}
\;,\qquad \quad \eea
where  $\tauz  = -\frac{1}{2t_0}$. Furthermore, for $n+1\ge 7$
and $\alpha=-1/2$
one has
\bea\label{In-EMx}
 \underline{ \|\hat f(\tau)\|^2_{L^\infty}}
 + \underline{E^\al_{k }[\hat f(\tau)]}
 &\le&
  2\Big\{  M_1^2 +  {E^\al_{k }[\hat f(\tauz )] }
  +  \|x^{(n-7)/2}\wh{H}^{\mu\nu}y_\mu
  y_\nu(\tauz)\|_{\mcG^{0}_k}^2
   \no
 &&
 + \int_{\tauz }^\tau \hat C_4( \underline{\hat M(s)})
 \underline{ E^\al_{k }[\hat f(s)]} ds\Big\}
\;.\qquad \quad \eea
\end{Proposition}
\begin{Remark} \label{Remark2}
For $n\ge 7$, a prefactor $\Om^{\frac{n-7}{2}}$ in the fourth line%
\footnote{The fall-off of the component of this term with the lowest
power of $\Om$ can be improved using the gauge condition. }
of the nonlinear term in (\ref{Sterm1})  still leads to the
estimates here. This remark is important for the estimation of the
time derivatives in Section~\ref{Stime} below.
\end{Remark}

\proof
%
%
For all $0<x <x_1$ the trivial identity
$$ \hat f(\tau, x) = \hat f(\tau , x_1)- \int_{x
}^{x_1} \p_x\hat f(\tau,s)ds   $$
leads to the estimate (recall that $\alpha>-1$)
\beaa
  \|\hat f(\tau)\|_{L^\infty}  &\le&  M_1
 +  \int_{x  }^{x_1} \|\p_x\hat f(\tau )\|_{\mcC^\alpha_{\{x=0\},0}} s^{\alpha} ds
\\
   &\le&  M_1
 +    \|\p_x\hat f(\tau )\|_{\mcG^\alpha_k}
  \;.
\eeaa
From this one easily concludes
\bel{ng}
 \|\hat f (\tau)\|_{\mcG^0_k}
 \leq C\big(
 { M_1} +
 \|(\p_x \hat f, \p_A \hat f)(\tau)\|_{\mcG^\a_{k-1}}\big)
 \;.
\ee
Now we apply Proposition~\ref{ppropp} of Section~\ref{Energie}.
To obtain (\ref{In-EM}) we will show first that, in the
Einstein-Maxwell case, the $\mcH^{\al}_k $-norm of the source
term, the $\mcG^0_k $-norms of $\m^\sharp$, $\f^\sharp$ and
${\Upsilon}^\mu$ are controlled by the energy. Let us start
with the $\mcG^0_k $-norm of $\m^\sharp$.  From the expression
of $\m$ given by (\ref{T11}) and the estimate (\ref{ng}), if
$n\geq 5$ then
\bean\label{norm1}
 \|\m^\sharp(\tau)\|^2_{\mcG^0_k }&\le&  C  \left(M_1+ \|\p\wh{H}\|^2_{\mcG^\a_k}\right)
\\
 &\le&  C  \left(M_1+E_{k,\la}^{\al}[u(\tau)]\right)
 \;.
\eea
The same holds for $\f^\sharp$ but with the constraint that the
space dimension $n$ is larger than or equal to $7$. We will
return later to the question how to improve on the dimension on
this term when $\alpha=-1/2$.

To estimate the  harmonicity functions $\xUp$ given by
(\ref{xUpsilon}),  we start by estimating the functions
$C^\mu$. We decompose ${C}^\mu = {C}^\mu_1 +{C}^\mu_2 +
{C}^\mu_3 $, each corresponding to a line in (\ref{T25}). The
first and second terms are estimated as we did for $\m^\sharp$
and $\f^\sharp$:
\bea
 \label{norm3}\|{C}^\mu_1 \|^2_{\mcG^0_k } &\le&
 C (\|x^{\frac{n-3}{2}}\wh{H}\|^2_{\mcG^{0}_k }
  +\|x^{\frac{n-1}{2}} \p \wh{H}\|^2_{\mcG^0} ) \no
&\le& C \big(M_1+E^\al_{k,\la}[u(\tau)] \big)\quad \text{for}
\quad n\ge 3 \;, \eea
and
%
\bea\label{norm4}\|{C}^\mu_2\|^2_{\mcG^0_k } &\le&C \big(
\|x^{\frac{n-5}{2}}\wh{H}\|^2_{\mcG^{0}_k }
 +\|x^{\frac{n-5}{2}}\p \wh{H}\|^2_{\mcG^0_k }\big)
\no
&\le&C \big(M_1+ E^\al_{k,\la}[u(\tau)]\big)\quad \text{for}
\quad
  n\ge 5-2\alpha \;.\eea
To estimate ${C}^\mu_3$ we recall that its components have  a
uniform zero of order two in $  \wh{H}$ and $(\wh{H},
x^{-\al}\p \wh{H})$ respectively, with the second term linear
in $\partial \wh H$, thus we can apply inequality (A.31)
of~\cite{ChLengardnwe} on the $\mcG$-norm with $\ell = 2,\;
\beta = \frac{n-5}{2}.$ We obtain:
\beaa\|{C}^\mu_3 \|^2_{\mcG^0_k } &\le&
\| Q^{\mu}(\Om^{\frac{n-5}{2}}\wh{H}) \|^2_{\mcG^{-2}_k }+
\|  Q^{\mu\nu}(\Om^{\frac{n-5}{2}}(\wh{H},x^{-\a}\p_\nu\wh{H}) ) \|^2_{\mcG^{-2-\a}_k }\\
&\le&C (\| (\wh{H}  \|_{L^\infty})
  \|\wh{H}\|^2_{\mcG^{3-n}_k }+ C (\| (\wh{H}, x^{-\a}\p\wh{H})  \|_{L^\infty})
  \|(\wh{H}, x^{-\a}\p\wh{H})\|^2_{\mcG^{3-\a-n }_k }\\
&\le&
  C\left(\| \wh{H}, x^{-\alpha} \p\wh{H}) \|_{L^\infty}\right)
 \left( \|\wh{H}\|^2_{\mcG^{0}_k }+\|\p \wh{H}\|^2_{\mcG^{\al}_k
 }\right)\quad \text{for}\quad n \ge 3- \al
  \;.
\eeaa
Thus we have:
\be
 \label{norm5}
 \|{C}^\mu_3\|^2_{\mcG^0_k }
  \le
  C\left(\|\wh{H}, x^{-\alpha} \p\wh{H}) \|_{L^\infty}\right)
 \left(\|u\|^2_{L^\infty}+E^\al_{k,\la}[u(\tau)]\right)\quad
 \text{for} \quad n\ge 4
 \;.
\ee
Note that the function $ C\left(\|\wh{H}, x^{-\alpha} \p\wh{H})
\|_{L^\infty}\right)$ will give a contribution to the function
$C_2(M(s))$ of (\ref{In-EM}). The remaining terms of $\xUp$ as
given by (\ref{RxUp}) are estimated in a similar way. They are
controlled by
\bel{norm55}
C\left(1+ E_{k,\la}^{\al}[u(\tau)]\right)
 \qquad\text{for}\quad n\geq 5  \;.
\ee
%
%

We continue by writing the source term $\mathcal{F}$ (see
(\ref{Sterm1})) as a sum of terms, each of the following form
\bel{Form}
 x^{p_i}\mathcal{F}_i\left(\;.\;,x^{q_i}(\hat f,\;x^{-\al}\p
 \hat f)\right)
 \;.
\ee
Note that all terms are polynomial in $\partial f$, at most
quadratic in $\partial f$. For instance, the first term $\tilde
F$ arises from products of the Christoffels in the Ricci
tensor, and from the products of the derivatives $\partial A$
of the vector potential $A$  in the energy-momentum tensor. We
then write, for example, in the $x^\mu $ coordinates,
$$
 \Gamma^2 \sim (g^\sharp \partial g)^2 \sim F(g^\sharp) \partial g \partial g
  = x^{2\alpha} F(g^\sharp) (x^{-\alpha} \partial g) (x^{-\alpha} \partial g)
 \;;
$$
we then express this in term of $h_{\mu\nu}$,    transform the
whole expression to the $y^\mu$--coordinates, and finally
reexpress $h_{\mu\nu}$ in term of $\hat H_{\mu\nu}$. This
formula shows that the $\Gamma^2$ in the Einstein equations
have a uniform zero of order two in $(\hat f ,x^{-\alpha}\hat
f)$. A similar analysis applies to the contribution of the
Maxwell fields to the Einstein-Maxwell equations.

We use the following estimate to show that the
$\mcH^{\al}_k$-norm of $\mathcal{F}$ is controlled by the
energy of the system: Suppose that $\mathcal{F}_i$ has a
uniform zero of order $\ell_i$ in $(u,\;x^{-\al}\p u),$ then
applying to this function the second part of lemma A.2
of~\cite{ChLengardnwe}, for
\bel{cond-plq}
 p_i+\ell_i q_i>\alpha
 \;.
\ee
We choose  $\epsilon >0$ so that $
 p_i+\ell_i q_i>\alpha +\epsilon$, and write
\beaa
 \lefteqn{
  \|x^{p_i}\mathcal{F}_i\left(\;.\;,x^{q_i}(u,\;x^{-\al}\p
 u)\right)\|^2_{\mcH^{\al}_k}
  }
   &&
\\
   &=&
 \|\mathcal{F}_i\left(\;.\;,x^{q_i}(u,x^{-\al}\p
 u)\right)\|^2_{\mcH^{{\al}-p_i}_k}
\\
%
%
 & \le&
 C\big( \|(u,\;x^{-\al}\p u)\|_{L^{\infty}}\big)
 \|(u,\;x^{-\al}\p u)\|^2_{\mcH^{{\al} -p_i-\ell_i
 q_i}_k}
\\
 &  \le&  C\big( \|(u,\;x^{-\al}\p u)\|_{L^{\infty}}\big)
  \left(\|u\|^2_{\mcH^{{-\epsilon}}_k}+\|x^{-\al}\p
 u\|^2_{\mcH^{{-\epsilon}}_k}\right)
\\
 &  \le&
  C\big( \|(u,\;x^{-\al}\p
 u)\|_{L^{\infty}}\big)
 \left(\|u\|^2_{L^\infty}+E^\al_{k,\la}[u(\tau)]\right)
\;.\eeaa
The analysis of the nonlinear terms (\ref{Sterm1}) along those
lines gives the following table:
 \begin{table}
\begin{center}
\begin{tabular}{||c||c||c||c||c||c||}
  \hline
    $\quad\mathcal{F}_i\quad$ &
    $p_i\quad\quad$  & $\quad  q_i\quad$ &$\quad \ell_i\quad$ & constraint& $ \;n \ge  $ \\
   \hline
   \hline
  $\mathcal{F}_1 $ &$-\frac{n+3}2$  &  $\frac{n-1}2+\al $& $2$ & $n>5-2\al$ & $7 [6] $  \\
      \hline
  $\mathcal{F}_2$ &$-\frac{n-5}{2}$  & $\frac{n-5}{2}$  & $2$  &$n>5$& $6 $ \\
  \hline
$\mathcal{F}_3$    & $ \frac{n-3}{2}+\al$  &$0$  & $2$ &$n>3 $& $4 $  \\
    \hline
$\mathcal{F}_4$   &$ \frac{n-5}{2}+2\al$  & $0$  & $2$&$n>5-2\al$  & $ 7 [6] $ \\
     \hline
   $\mathcal{F}_5$  &$2 - \frac {n-5}2 + \alpha$  & $\frac{n-5}{2}$  & $3$
&$n>3$  & $4$
\\
   \cline{2-6}
   &$2-\frac{n-5}{2}+2\al$  & $\frac{n-5}{2}$
    & $3$ &$n>3- \alpha $  & $4$
\\
    \hline\hline
   \end{tabular}
\end{center}
\caption{Restrictions on the dimension from the source terms.
\label{TFi}}
 \end{table}
%
%
%
Here the $\mathcal {F}_i$'s, $i=1,\ldots,4$, correspond to the
$i$-th line of \eq{Sterm1}, while the two rows for
$\mathcal{F}_5$ correspond to the two respective terms in the
last line of (\ref{Sterm1}).
In the last column the number in square bracket is obtained by
estimating below the non-linearity in a more efficient way.

It turns out that the threshold on the space dimension $n$ can
be lowered to $n=6$ for the components $\mathcal{F}_1$ and
$\mathcal{F}_4$ of the source term  $\mathcal{F}$. The
quadratic terms in those expressions with the lowest powers of
$\Om$ are  of the form $\Om^{\frac{n-5}{2}}
G(\Om^{\frac{n-1}{2}} \hat f)\p \hat f\p \hat f$ for
$\mathcal{F}_1$ and $\Om^{\frac{n-5}{2}}\wh{H}\p u $ and $
\Om^{\frac{n-5}{2}}\p\wh{H}\p u $ for $\mathcal{F}_4.$
One can estimate the $\mcH^\al_k$-norm of  $
\Om^{\frac{n-5}{2}}\wh{H}\p u $ using instead \eq{25VIII0.1}:
\bean
 \|\Om^{\frac{n-5}{2}}\wh{H}\p u\|^2_{\mcH^\al_k}
  &\le&\|\wh{H}\p u\|^2_{\mcH^{\al-\frac{n-5}{2}}_k}
   \nonumber
\\
 &\le&
C\left(\|\wh{H}\|^2_{\mcC^0_0}\|\p
u\|^2_{\mcH^{\a-\frac{n-5}{2}}_k}+\|x^{\frac{n-5}{2}}\wh{H}\|^2_{\mcG^0_k}\|\p
u\|^2_{\mcB^{\a}_0}\right)
   \nonumber
\\
 &\le&
 C(\|u\|^2_{L^\infty}+\|\p u\|^2_{\mcB^\al_0})\left(\| u\|^2_{\mcG^0_k}+\|\p
 u\|^2_{\mcH^\al_k}\right)
 \quad \text{if}\quad   n \ge 5
   \nonumber
\\
 &\underbrace{\le}_{\mbox{\scriptsize see} \ \eq{ng}}&
 C(\|u\|^2_{L^\infty}+\|\p u\|^2_{\mcB^\al_0})\left(  1+\|\p
 u\|^2_{\mcH^\al_k}\right)
   \nonumber
\\
 & {\le}  &
 C(\|u\|^2_{L^\infty}+\|\p u\|^2_{\mcH^\al_k})\left(  1+\|\p
 u\|^2_{\mcH^\al_k}\right)
 \;,
\eeal{31VIII0.3}
for $k>n/2$.
%
Next,
\beaa \|\Om^{\frac{n-5}{2}}\p\wh{H}\p u\|^2_{\mcH^\al_k}&\le&\|\p\wh{H}\p u\|^2_{\mcH^{\al-\frac{n-5}{2}}_k}\\
&\le&C\left(\|\p\wh{H}\|^2_{\mcC^\al_0}\|\p
u\|^2_{\mcH^{-\frac{n-5}{2}}_k}+\|\p\wh{H}\|^2_{\mcH^\al_k}\|\p
u\|^2_{\mcC^{-\frac{n-5}{2}}_0}\right)\\
&\le&C\|\p u\|^2_{\mcC^\al_0}\|\p u\|^2_{\mcH^\al_k}
\quad\text{if}\quad -\frac{n-5}{2}-\al\le 0\quad \text{i.e.}\quad n
 \ge 5-2\al
\\
 &\le&
  C\|\p\wh{H}\|^2_{\mcC^\al_0}E^\al_{\la,k}[u(\tau)]
 \;,
\eeaa
and so the last inequality will be true provided that
\beaa
\left\{\begin{array}{l} \;n\ge 6 \quad \text{if} \quad \al= -\frac 12\\
                         \;n\ge 7 \quad   \text{if}  \;-1 < \al< -\frac
                         12 \qquad
       \end{array} \right.\;.
\eeaa
A similar calculation  applies to ${\cal F}_1$.
%
%

These estimates and the table show that
\bel{esti-Sterm} \|\mathcal{F}\left(u, \p
u\right)\|^2_{\mcH^{\al}_k} \le C(\|u\|_{L^\infty},\|\p u
   \|_{\mcC^{\al}_0})\left(1+  E^\al_{k,\la}[u(\tau)]\right)
\ee
for
$$
\left\{\begin{array}{l} \;n\ge6 \quad \text{if} \quad \al= -\frac 12\\
                         \;n\ge 7 \quad   \text{if}  \;-1 < \al< -\frac
                         12 \qquad
       \end{array} \right. \;.$$
%
Inserting inequalities (\ref{norm1})-(\ref{norm4}) and
(\ref{norm5})-(\ref{esti-Sterm}) in (\ref{Pr}) of Section
\ref{Energie} gives (\ref{In-EM}).

Now, at several places of the calculations above the term
$$
 \psi:= y_{\alpha} y_\beta \hat H^{\alpha \beta}
$$
is the one that occurs with the lowest power of $\Omega$. It
follows from the wave-coordinates conditions that this term
solves equation \eq{GC7x}, which can be written in the form
\bel{28VIII0.5}
 - y^\alpha \partial_\alpha \psi  + \frac {n-5} 2 \psi= \zeta
  \;,
\ee
where
\bea \zeta
 &: = &
\Omega\Big( \frac{n-1}2 \tr_{\eta}(\wh{\PhiH })
+y_\nu\frac{\p}{\p y^\mu}
 \big(\wh{\PhiH }^{\mu\nu}+\frac 12\eta^{\mu\nu}\tr_\eta(\wh{\PhiH }) \big)\Big)
 \nonumber
\\
 &&+\Om^{-\frac{n-1}{2}}Q (\Om^{\frac{n-1}{2}}\wh{\PhiH },
 \Om^{\frac{n-1}{2}}\wh{\PhiH })
\no
 &&+\Om^{-\frac{n-1}{2}}Q (\Om^{\frac{n-1}{2}}\wh{\PhiH },
 \Om^{\frac{n+1}{2}}\p\wh{\PhiH })\no
&&+\Om^{-\frac{n-1}{2}}Q(\Om^{\frac{n-1}{2}}\wh{\PhiH },
\Om^{\frac{n-1}{2}}y^\al \frac{\p}{\p y^\al}\wh{\PhiH }) 
\no
  &=:&
   \zeta_1 + \zeta_2 + \zeta_3 + \zeta_4
 \;,
  \label{GC7xx}
\eea
where $\zeta_i$ corresponds to the $i$-the line. The point is
that all terms in $\zeta$ contain effectively multiplicative
powers of $\Omega$.

Solutions of \eq{28VIII0.5} take the form, for $\tauz\le \tau
\le \tau_1< 0$,
\bel{28VIII0.1}
 \psi(\tau,x) = (-\tau)^{- (n-5)/2} \left(\int_{\tau_0}^\tau (-s)^{(n-7)/2}
 \zeta  \left(s,\frac {sx}\tau \right) ds
   +
     (-\tauz)^{(n-5)/2} \psi\left(\tauz,\frac{x\tauz}\tau\right)
  \right)
 \;.
\ee
%
This gives immediately, for any $\gamma$,
\bel{1IX0.1}
 \|\psi(\tau)\|_{\mcG^{\gamma}_k}
  \le
   \|\psi(\tauz)\|_{\mcG^{\gamma}_k}
    +
     C(\tauz,\tau_1) \int_{\tauz}^\tau \|\zeta (s)\|_{\mcG^{\gamma}_k} ds
     \;,
\ee
similarly for  $\mcH^{\gamma}$- or $\mcC^\gamma$-norms. In the
notation of \eq{28VIII0.6} one thus finds
\beaa
 \underline{ \|\psi(\tau)\|_{\mcG^{\gamma}_k}}
  &\le &
   \|\psi(\tauz)\|_{\mcG^{\gamma}_k}
    +
     C(\tauz,\tau_1) \int_{\tauz}^\tau  \underline{ \|\zeta (s)\|_{\mcG^{\gamma}_k}} ds
\\
  &\le &
   \|\psi(\tauz)\|_{\mcG^{\gamma}_k}
    +
     C(\tauz,\tau_1)(\tau_1 - \tauz)  \underline{
     \|\zeta (\tau)\|_{\mcG^{\gamma}_k}}
     \;.
\eeaa
Using this to estimate $\f^0$ we obtain
\beaa
  \underline{ \|\f^0(\tau)\|_{\mcG^{0}_k}}
   & \le &
    C\big ( \underline{\|x^{(n-7)/2}  \psi(\tau) \|_{\mcG^{0}_k}}
    +
     \underline{\|x^{(n-5)/2}  \wh{H}(\tau) \|_{\mcG^{0}_k}}
\\
   & \le &
    C\big (  \|x^{(n-7)/2}\psi(\tauz)\|_{\mcG^{0}_k} + \underline{\|x^{(n-7)/2}  \zeta(\tau) \|_{\mcG^{0}_k}}
    +
     \underline{\|x^{(n-5)/2}  \wh{H}(\tau) \|_{\mcG^{0}_k}}
     \big)
 \;.
\eeaa
We have, for example, 
\beaa
 \underline{\|x^{(n-7)/2}  \zeta_1 (\tau) \|_{\mcG^{0}_k}}
  & \le &
   C
    \left(
 \underline{\|x^{(n-5)/2}  \wh{H} (\tau) \|_{\mcG^{0}_k}}
  +
 \underline{\|x^{(n-5)/2} \partial \wh{  H}  (\tau) \|_{\mcG^{0}_k}}
 \right)
 \;,
\eeaa 
which, for $n-5\ge -2\alpha$, can  be controlled by $
\underline{\|  \wh{H} (\tau) \|_{L^\infty}}$ and
$\underline{E^\alpha _k [ u(\tau)]}$ in view of \eq{ng}. This
requires $n\ge 6$ if $\alpha=-1/2$, or $n\ge 7$ if $\alpha \in
(-1,-1/2)$. An estimation of the remaining $\zeta_i$'s along
the lines of those already done above presents no difficulties.

The functions $\f ^1$ and $\f $ have the same structure and so
the same estimate applies; the function $\f ^A$ has a higher
multiplicative power of $\Omega$ so that the original
straightforward estimate applies.

The final inequality (\ref{In-EMx}) follows immediately from
this and from an obvious version of the estimate (\ref{In-EM})
for the remaining terms in the equation.

We finish this proof by noting that the above treatment of
$y_\alpha y_\beta \hat H^{\alpha \beta}$ can be used to improve
the threshold on dimension for some of the entries of
Table~\ref{TFi}; this will, however, not improve the threshold
on $n$ of the theorem.
%
\qed

We are now ready to prove existence of solutions in weighted
Sobolev spaces. For $s>0$ consider the family of hyperboloids:
\bel{defmcS} \mcS_s = \left\{(x^\mu): x^0- s = \sqrt{s^2+|\vec
x |^2}\;\right\}\;. \ee
Let $\phi$ be defined in \eqref{ytox}. We have the following
\begin{Theorem}[Propagation of weighted Sobolev regularity]
\label{theo-EM} Suppose that $k > \left[\frac n 2\right]+ 1$,
with $n=6$ and $\alpha=-1/2$, or $n\ge 7$ with $\alpha\in
(-1,1/2]$, and let $t_0>0$. Suppose that
\bel{Ini-datax}
 \hat{f}\left|_{\phi(\mcS_0)} \in\left( \
 \mcH^\al_{k+1}\cap L^\infty\right)(\phi(\mcS_0))\right., \quad \
 \quad \left(\p_\tau\hat{f}, \p_x\hat{f}, \p_A
 \hat{f} \right)\left|_{\phi(\mcS_0)} \in
 \mcH^\al_k(\phi(\mcS_0))\right.\;,
\ee
where $f$ and $\hat f$ are defined by
(\ref{E-M})-\eq{18VIII0.1}. In the case $\alpha=-1/2$ and $n=6$
assume moreover that
\bel{31VIII0.1}
  x^{-1/2} y_\alpha y_\beta \wh{H}^{\alpha\beta}\big|_{\phi(\mcS_0)} \in \mcG^0_k
  \;.
\ee
Then there exists $t_*> t_0$ and a solution of (\ref{T1})
defined on $\underset{s\in [t_0,\;t_*]}{\cup}\mcS_s$ such that,
$\forall \tau \in
[-\frac{1}{2t_0},-\frac{1}{2t_*}]=:[\tauz,\tau_*] $ we have:%
%
\bel{th31} \hat f  \in L^{\infty}\Big([\tauz ,\tau_*],\;
\mcH^\al_k({\bf{\Large{H}}}_{\tau})\cap
L^\infty({\bf{\Large{H}}}_{\tau})\Big)\;, \ee
\bel{th41} \left( \p_\tau\hat f,\; \p_x\hat f , \;\p_A\hat
f\right)\in L^\infty\Big([\tauz
,\tau_*],\;\mcH^\al_k({\bf{\Large{H}}}_{\tau})\Big)\;. \ee
Moreover, any solution for which $\hat M(\tau)$, as defined in
\eq{23VIII0.3x}, is bounded on $[\tauz,\tau_1]$ satisfies
\eq{th31}-\eq{th41} with $\tau_*=\tau_1$.
\end{Theorem}
\begin{Remark}
\label{Remark1} Using the weighted Sobolev embedding theorem we
conclude
\bel{R1} \hat f(\tau) \in \left(\mcC^\al_{ k-\left[\frac
n2\right]-1}\cap L^\infty\right)({\bf{\Large{H}}}_{\tau})\;, \ee
\bel{R2}\left( \p_\tau\hat f(\tau),\p_x\hat f(\tau),\p_A\hat
f(\tau)\right)  \in \mcC^\al_{ k-\left[\frac
n2\right]-1}({\bf{\Large{H}}}_{\tau})\;. \ee 
when the prescribed data are   as in Theorem~\ref{theo-EM}.
\end{Remark}
\proof In order to apply the Gronwall-type Lemma 5.2
of~\cite{ChLengardnwe}, we need to prove that all the norms in
$\hat M$ (see \eq{23VIII0.3x} and (\ref{In-EM})) are controlled
by the energy or the $L^\infty$-norm of $u$.
Since $k
> \left[\frac n 2\right]+ 1$, from the weighted Sobolev's
inequality, we have:
\bel{th1} \| (\p_\tau-\p_x, \p_x,\p_A)\hat
f\|^2_{\mcB^{\al}_1}\le \| (\p_\tau-\p_x, \p_x,\p_A)\hat
 f\|^2_{\mcH^{\al}_k}\le E^\al_{k,\la}[u(\tau)]\;.
\ee
Let us look at the
$L^\infty$-norm of $(\p_\tau-\p_x) \m^\sharp$. Recall that the
expression of $\m^\sharp$ is given by (\ref{T11}). We estimate
here only its worse term which is of the form
$\Om^{\frac{n-5}{2}} \wh{H}.$ We have:
\beaa\| (\p_\tau-\p_x)(\Om^{\frac{n-5}{2}} \wh{H}) \|^2_{L^\infty}
&\le& C\left(\|\Om^{\frac{n-5}{2}} (\p_\tau-\p_x)\wh{H}
\|^2_{L^\infty} +\|
\Om^{\frac{n-7}{2}} \wh{H}  \|^2_{L^\infty}\right)\\
&\le&  C\left(\|u\|^2_{L^\infty}+ E^\al_{k,\la}[u(\tau)]\right)
\qquad \text{for} \quad n\ge 7 \;.\eeaa
Thus,
\bel{g-sharp}
\|(\p_\tau-\p_x) \m^\sharp\|\le
 C\left(\|u\|^2_{L^\infty}+ E^\al_{k,\la}[u(\tau)]\right)
\qquad \text{for} \quad n\ge 7    \;.
\ee
Next, since $k>n/2+1, $ as in (\ref{norm1}), we have
\bean\label{th2} \|\m^\sharp\|^2_{\mcC^0_{1}}    &\le&
 \|\m^\sharp\|^2_{\mcG^0_{k}}
\no
 &\le& C\left(M_1+
 E^\al_{k,\la}[u(\tau)]\right), \quad \text{for}\quad n \geq 5
 \;.
\eea
Similarly,
\bel{h-sharp}
\|\f^\sharp\|^2_{\mcC^0_{1}}\le\|\f^\sharp\|^2_{\mcG^0_{k}}\le
C\left(M_1+ E^\al_{k,\la}[u(\tau)]\right)\quad \text{for}\quad
n\ge 7 \;.
 \ee

If $\alpha=-1/2$ the threshold $n=7$ in \eqref{g-sharp} and
\eqref{h-sharp} can be lowered to $n=6$ by using the estimate
\eqref{1IX0.1} on the slowest decaying term $\psi$.

 To estimate the $\mcC^0_1$-norms
of the harmonicity functions, we use again as in the previous
estimate the Sobolev inequality and obtain a control of these
norms by the energy with the same constrains as in
(\ref{norm3})-(\ref{norm55}).
Let us estimate now the $L^\infty$-norm of $u$. Integrating
backward along the integral curve of the vector field
$Y^\nu\p_\nu = \p_\tau-\p_x$ we can write the identity (here we
omit the variable $v^A$)
\bel{infty} u(\tau, x) - u(\tauz , \tau-\tauz +x)= \int_{\tauz
}^\tau \left(\p_\tau-\p_x\right)u(s, \tau-s+x)ds \;. \ee
Thus we have
\beaa |u(\tau, x)| &\le& |u(\tauz , \tau-\tauz +x)|+
\int_{\tauz }^\tau
|\left(\tau-s+x\right)^{-\al}\left(\p_\tau-\p_x\right)u(s,
\tau-s+x)|\left(\tau-s+x\right)^\al ds \\
&\le& |u(\tauz , \tau-\tauz +x)|+\int_{\tauz }^\tau
\|\left(\p_\tau-\p_x\right)u(s)\|_{\mcC^\al_{0}}\left(\tau-s+x\right)^{\al}ds\\
&\le& \|u(\tauz )\|_{L^\infty}+\int_{\tauz }^\tau
\|\left(\p_\tau-\p_x\right)u(s)\|_{\mcC^\al_{0}}\left(\tau-s\right)^{\al}ds
\;.\eeaa
Since $k>\frac n2$ we can now write ($-1<\al \le -1/2$):
\bea \label{th4} \|u(\tau)\|_{L^\infty} &\le& \|u(\tauz
)\|_{L^\infty} + \int_{\tauz }^\tau
\|\left(\p_\tau-\p_x\right)u(s)\|_{\mcH^{\al}_k}\left(\tau-s\right)^{{\al}}ds\nonumber\\
&\le&  \|u(\tauz )\|_{L^\infty} + \int_{\tauz }^\tau
\sqrt{E^\al_{k,\la}[u(s)]}\left(\tau-s\right)^{{\al}}ds\eea
Inequalities (\ref{th1})-(\ref{th4}) show that from (\ref{In-EM}) we
have the following:
\bea\label{In-EM1} \|u(\tau)\|^2_{L^\infty} +
E^\al_{k,\la}[u(\tau)]&\le&C\left(\|u(\tauz )\|_{L^\infty} +
E^\al_{k,\la}[u(\tauz )]\right)\no
&&
 +\int_{\tauz }^\tau
 \Phi\left(E^\al_{k,\la}[u(s)],\;\|u(s)\|_{L^\infty}\right)
 \left(1+ (\tau-s)^{\al}\right)ds \;,
 \quad\qquad
\eea
where $\Phi$ is  bounded on bounded sets. We can now apply
Lemma 5.2 of~\cite{ChLengardnwe} and obtain that there exists a
time $\tauz < \tau_*<0$ depending on $\|u(\tauz )\|_{L^\infty}
+ E^\al_{k,\la}[u(\tauz )]$ and on the function $\Phi$ such
that $\forall \tau \in [\tauz ,\tau_*]\;,$
\bel{En} \quad \|u(\tau)\|_{L^\infty} +
E^\al_{k,\la}[u(\tau)]\le 1+ C\left(\|u(\tauz
 )\|_{L^\infty} + E^\al_{k,\la}[u(\tauz )]\right)\;,
\ee
which provides the desired bounds.

If one knows a priori that $\hat M(\tau)$ is bounded,
\eqref{In-EM1} becomes effectively a linear inequality, and the
claimed global bound immediately follows.

\begin{figure}[h]
\begin{center} {
\psfrag{Vtau}{\huge $\mcV_{+}$} \psfrag{Omegastar}{\huge
$\bf{{\myOmega_{\tau_*}}}$}
\resizebox{4in}{!}{\includegraphics{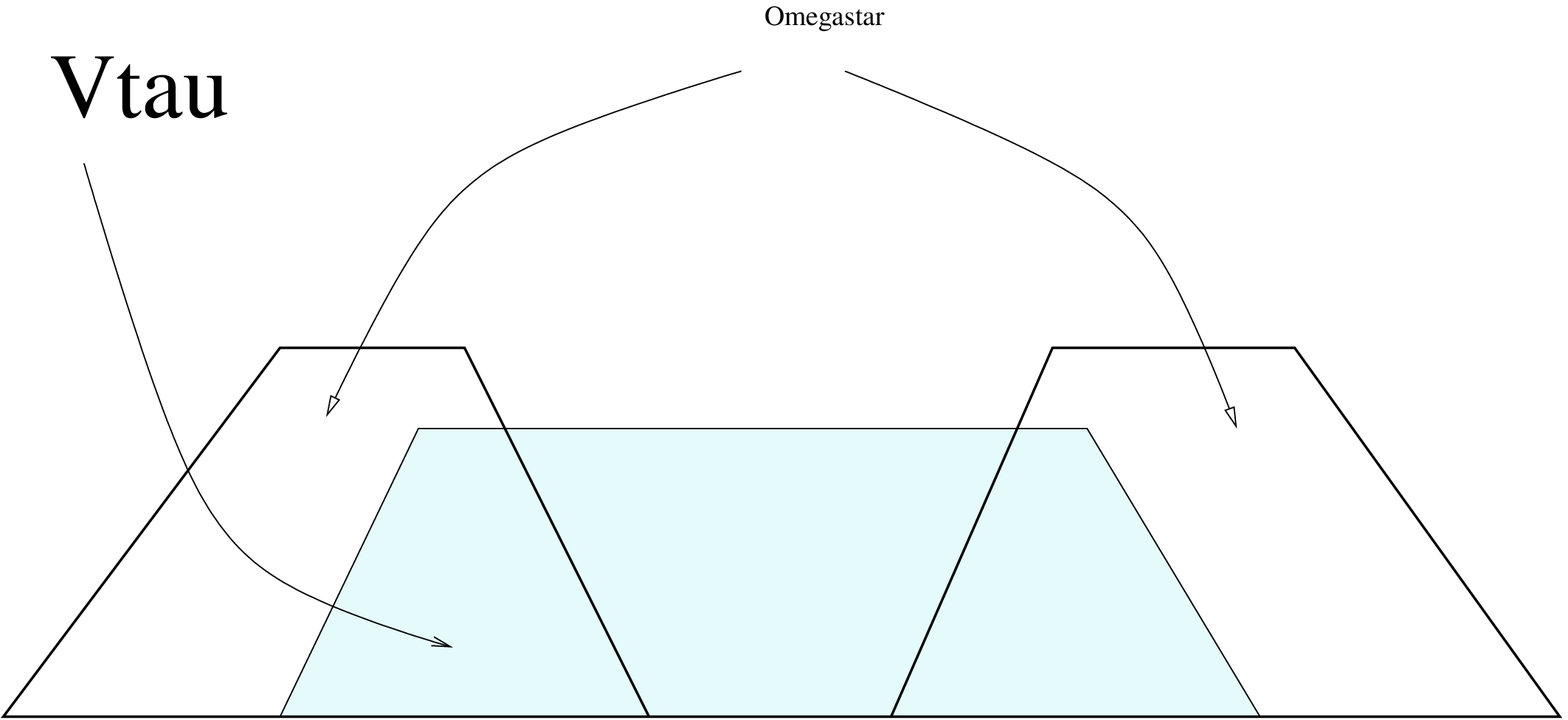}} } \caption{The sets
$\mcV_+$ and $\myOmega_{\tau_*}$. \label{V-set}}
\end{center}
\end{figure}

%
Actually, the solution constructed here is defined on
$\myOmega_{\tau_*}$ (see Figure~\ref{V-set}). In order to
obtain a solution in a whole neighborhood of the hyperboloid
$\mcS_0$, we proceed as follows: Let  $R>0$ be a real positive
number such that the level set $r=R$ lies in the region where
the energy estimates above apply.  We consider the Cauchy
problem for (\ref{T1}) with initial data obtained by
restriction on
$$\mcS_0(R)= \mcS_0\cap\{(x^\mu): 0 \le |\vec x |\le R\}\;.$$
We thus obtain a Cauchy problem on a compact region.   We can
now apply to this problem the conclusion of Proposition 3.2,
p.~378 of~\cite{Taylor}: there exists a time $\tau_+ \in]\tauz
, 0[$ and a smooth solution  on (see Figure~\ref{V-set})
$$
\mcV_+= \underset{t \in [t_0,
-\frac{1}{2\tau_+}]}{\cup}\phi(\mcS_t(R)) \cap \mcD^+ (\phi(\mcS_0(R)))
 \;,
$$
where $\mcD^+ $ denotes the domain of dependence, and where
$$\mcS_t(R)= \mcS_t\cap\{(x^\mu): 0 \le |\vec x |\le R\}\;.$$
%
%
From uniqueness in Proposition 3.2, p.~378 of~\cite{Taylor}, we
conclude that the solutions constructed on $\mcV_+$ and
$\myOmega_{\tau_*}$ coincide on$\mcV_+\cap\myOmega_{\tau_*}$
which is not empty for $R$ large enough. We thus obtain a
solution of (\ref{T1}) with (\ref{E-M}) in a whole neighborhood
of $\mcS_0$. %
\qed

\subsubsection{Space-regularity of the solution}

For smooth initial data the solution constructed in the
previous section is in $C^\infty(\mcV_+\cup
\myOmega_{\tau_*})$. In this section we want to show that, for
data given in the space $\underset{k\in \N}{\cap}\mcH^\al_k,$
we can  control the growth, near $x=0$, of all space
derivatives of the corresponding solution. We have the
following:

\begin{Theorem}
\label{regu} Under the hypotheses of Theorem~\ref{theo-EM},
suppose moreover that the initial data given on the hyperboloid
$\mcS_0$ satisfy
\bel{ini-data-regu} \hat f\left|_{\phi(\mcS_0)} \in
\left(\mcH^\al_\infty \cap
L^\infty\right)({\bf{\large{H}_{\tauz }}})\right. \ \text{and}\
\p \hat f\left|_{\phi(\mcS_0)} \in
\mcH^\al_\infty({\bf{\large{H}_{\tauz }}})\right.
 \;.
\ee
If $\alpha=-1/2$ and $n=6$ we also suppose  that \eq{31VIII0.1}
holds for all $k$. Let $\tau_*$ be as in  Theorem~\ref{theo-EM}
with $k=k_0$, where $k_0$ is the smallest integer larger than
$[n/2]+1$. Then
\bel{31VIII0.2}
 \forall \tau \in [\tauz , \tau_*] \qquad  \hat f(\tau) \in (\mcH^\al_{\infty}\cap
 L^\infty) ({\bf{\large{H}_\tau}})
 \;,
 \quad  \partial \hat f(\tau) \in \mcH^\al_{\infty}  ({\bf{\large{H}_\tau}})
 \;.
\ee
Furthermore, any solution with smooth initial data as above for
which $\hat M(\tau)$, as defined in \eq{23VIII0.3x}, is bounded
on $[\tauz,\tau_1]$ satisfies \eq{31VIII0.2} with
$\tau_*=\tau_1$.
\end{Theorem}
\proof
We provide the details for   $n>6$; the treatment of the case
$n=6$ is similar. From Theorem~\ref{theo-EM} there exists a
time $\tau_*$ and a constant $C^*$ depending on $k_0$ such that
$\forall \tau \in [\tauz ,\;\tau_*[\;,$
\bel{et1} \|u(\tau)\|^2_{L^\infty} + E^{\al}_{k_0,\la}[u(\tau)]
\le C^* \; .\ee
Now let $ k \in \N, \; k \ge k_0\;,$ since
$\hat f\left|_{\phi(\mcS_0)} \in \left(\mcH^\al_k \cap
L^\infty\right)({\bf{\large{H}_{\tauz }}})\right.$
inequality (\ref{In-EM}) holds. Now the function $C_3(\hat M(s))$ appearing in
this inequality is controlled by $E^{\al}_{k_0,\la}[u(\tau)]$ and
thus by $C^*$, therefore, from (\ref{et1}) we have:
$$
E^\al_{k,\la}[u(\tau)] \le C(C^*)\left(1+\int_{\tauz }^\tau
E^\al_{k,\la}[u(s)]ds\right)\;.
$$
Applying Gronwall's inequality we obtain:
$$E^\al_{k,\la}[u(\tau)]\le Ce^{C\tau_*} \;.$$
This inequality shows that, for all $k$,
\bel{th5}\p u \in \mcH^\al_k \;, \ee
as desired.%
\qed

\subsubsection{Estimates on time derivatives of the solution}
\label{Stime}

In order to  estimate the time derivatives of the solution, we
introduce a new set of variables $(y, \tilde x)$ (compare
Figure~\ref{F1}):
$$\left\{\begin{array}{l} \tau= \frac{y-\tilde x}2+\tauz   \\ x=\tilde x \end{array} \right.
\qquad \text{which implies that}
\qquad \left\{\begin{array}{l} \p_y= \frac 12\p_\tau\\ \p_{\tilde
x}= \p_x-\frac 12 \p_\tau
\end{array} \;.\right.
$$
Note that in these new coordinates, the hyperboloid $\mcS_0$ is
represented by the set $\{y=\tilde x\}$. Since we are
interested in the behavior of solution in a neighborhood of the
set $\{x=0\} $, as in~\cite{ChLeski} we restrict our attention
on the subset $\mcU$ of $\myOmega_{\tau_*}$ defined by:
\begin{figure}[ht]
\setlength{\unitlength}{1cm} \noindent
\begin{picture}(10,4)(-1,-1)
\thicklines \put(0,0){\line(1,1){2}} \put(0,0){\circle*{.15}}
\thinlines \put(2,2){\line(1,0){3}} \put(2,2){\circle*{.15}}
\put(-0.43,1.4){$x=0$} \put(0.57,1.1){$\searrow$}
\put(7,0){\line(-1,0){7}} \put(7,0){\circle*{.15}}
\put(5,2){\line(1,-1){2}} \put(5,2){\circle*{.15}}
\put(-1,-0.5){$(x=0, \tau=\tauz)$} \put(6,-0.5){$(x=x_0, \tau=\tauz)$}
\put(3,1){$\myOmega_{\tau_*}$} \put(0.5,2.3){$(x=0, \tau=\tau_*)$}
\put(4.3,2.3){$(x=\sigma(\tau_*), \tau=\tau_*)$}
\put(2.3,-0.5){$\mcS_0$} \put(2.75,-0.35){$\uparrow$}
\put(8,0.5){\vector(1,1){1}} \put(9,1.7){$\partial_\tau$}
\put(8,0.5){\vector(1,0){1}} \put(9,0.7){$\partial_x$}
\put(10,1){\vector(1,1){0.5}} \put(10.7,1.7){$\partial_y$}
\put(10,1){\vector(1,-1){0.5}} \put(10.7,0.7){$\partial_{\tilde{x}}$}
\end{picture}
\noindent
\begin{picture}(7,4)(-1,-1.4)
\thicklines \put(0,0){\line(1,1){2}} \put(0,0){\circle*{.15}}
\thinlines \put(4,0){\line(-1,0){4}} \put(4,0){\circle*{.15}}
\put(-0.43,1.4){$x=0$} \put(0.57,1.1){$\searrow$}
\put(2,2){\line(1,-1){2}} \put(2,2){\circle*{.15}}

\put(-1,-0.5){$(x=0, \tau=\tauz)$}

\put(2.7,-0.5){$(x=2T, \tau=\tauz)$}

\put(1.8,0.5){$\myOmega$}
\put(5,1){$
{\longrightarrow}$} \put(0.5,2.3){$(x=0, \tau=\tau_*)$}
\put(1.8,-0.5){$\mcS_0$} \put(2.35,-0.35){$\uparrow$}
\end{picture}
\begin{picture}(10,5)(-1,-1)
\thicklines \put(0,0){\line(0,1){2.8}} \put(0,0){\circle*{.15}}
\thinlines \put(0,2.8){\line(1,0){2.8}} \put(0,2.8){\circle*{.15}}
\put(2.8,2.8){\line(-1,-1){2.8}} \put(2.8,2.8){\circle*{.15}}
\put(-1,-0.5){$(\tilde{x}=0, y=0)$}

\put(-2.2,3.1){$(\tilde{x}=0, y=2T)$}

\put(2.1,3.1){$(\tilde{x}=2T,y=2T)$} \put(0.7,1.9){$\mcU$}
\put(-1.5,2.2){$\tilde{x}=0$} \put(-0.5,1.9){$\searrow$}
\put(1.42,1.08){$\nwarrow\mcS_0$}
\put(3.4,1.0){\vector(0,1){0.71}} \put(3.4,1.85){$\partial_y$}
\put(3.4,1.0){\vector(1,0){0.71}}
\put(4.3,1.1){$\partial_{\tilde{x}}$}
\end{picture}
\caption{\label{F1}The variables $(x,\tau)$ and
$(\tilde{x},y)$, with $T:=\tau_*-\tau_0$. The function $\sigma$
has been introduced in \eqref{7X0.1}. We hope that the reader
will not get confused by the fact that the boundary $x=0$, at
the left-hand sides of the figures here, is depicted at the
right-hand side of Figure~\ref{U-set}.}
\end{figure}
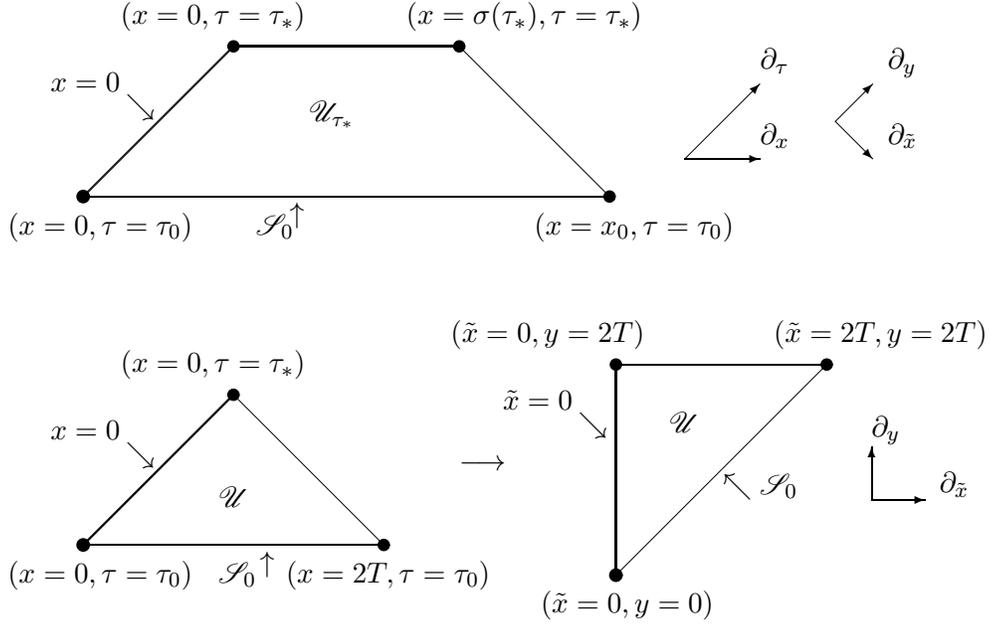

$$\mcU=\left\{(y, \tilde x, v^A): 0 < x< y,\; v\in\mcO\;, 0<y<2(\tau_*-\tauz )\right\}\;.$$
Recall that the definitions of the spaces
$$
\fCx{\alpha}{k}(\mcU) ,\quad
\fCy{\sigma}{k}(\mcU) ,\quad
\cDak(\mcU) ,\quad \text{and} \quad
\cDask(\Omega) \;, $$
can be found in Appendix A.1 page 116 of~\cite{ChLeski} with
$\p_x$ there
corresponding to $\p_{\tilde x}$ here. 
%
\begin{Remark}
In the coordinates $(y,\tx)$ the components of the inverse of the
metric read (compare~\ref{metric-x}):
\bel{metric-xy1} \m^{yy}= 4\left(\m^{\tau\tau}+\m^{x\tau} \right)+
\m^{xx} = \text{O}(x^{\frac{n-5}{2}}) \ee
\bel{metric-xy2} \m^{y\tx}= 2\m^{x\tau} + \m^{xx}\ee
\bel{metric-xy3} \m^{yA}= 2\m^{\tau A} + \m^{xA}\ee
\bel{metric-xy4} \m^{\tx\tx}= \m^{xx}=\text{O}(x^{\frac{n-1}{2}})\ee
\bel{metric-xy5} \m^{\tx A}=  \m^{xA}\;.\ee
\end{Remark}
Recall that the hypersurfaces $\mcS_s$ have been defined in
\eq{defmcS}. As a first step towards proving propagation of
polyhomogeneity, we obtain some information about the
$\partial_y$-derivatives of the fields:

\begin{Theorem}
\label{theo-EM1} Suppose that $k > \left[\frac n 2\right]+ 1$.
Under the hypotheses of Proposition~\ref{prop-EM}, there exists
$t_*> t_0$ and a solution of (\ref{T1})   defined on
$\underset{s\in [t_0,\;t_*]}{\cup}\mcS_s$ such that:
\bel{tthh1} \hat f  \in \left(\fCxy{\al}{ k-\left[\frac
n2\right]-1}\cap L^\infty\right)\left(\mcU\right)\;. \ee
\bel{tthh2} \left(\p_\tau\hat f,\p_x\hat f,\p_A\hat f \right) \in
\fCxy{\al}{ k-\left[\frac n2\right]-1}\left(\mcU\right)\;, \ee 
%
where $f$ and $\hat f$ are defined by
(\ref{E-M})-\eq{18VIII0.1}.
\end{Theorem}
\proof
The proof of existence is given by Theorem~\ref{theo-EM} and we have
$  \hat f \in L^\infty(\mcU),\;\p \hat f \in \fCtx{\al}{k-\left[\frac
n2\right]-1}(\mcU).$
We note that from (\ref{T2}) and (\ref{TT2}) we have:
\bel{part-Om} \Om= \tilde{x}(-y-2\tauz ), \qquad y\p_y \Om = - \tilde{x}y ,
\qquad \tilde{x}\p_{\tilde{x}} \Om= \Om  \qquad \text{and}\qquad \p_A \Om=
0\;.\ee
Identities (\ref{part-Om}) show that if we apply to (\ref{T13})
the operator $( \p_A , {\tilde{x}}\p_{\tilde{x}},y\p_y),$  then we obtain a wave
equation with $(u,   \p_A u, y\p_y u, {\tilde{x}}\p_{\tilde{x}} u\;)$ as the new
unknown functions in which the coefficients have the same
powers of $x$ as in the original equation, and the source term
the same structure. More precisely, set
\bel{U} U= \begin{pmatrix} u\\ \p_A u \\ {\tilde{x}}\p_{\tilde{x}} u\\
y\p_y u\end{pmatrix},\qquad
\text{we thus obtain} \qquad
\begin{pmatrix} U\\ \p U \end{pmatrix}= \begin{pmatrix} u\\ \p_A u \\ {\tilde{x}}\p_{\tilde{x}} u\\
y\p_y u\\ \p u\\ \p(\p_A u) \\\p( {\tilde{x}}\p_{\tilde{x}} u)\\
\p\left(y\p_y u\right)\end{pmatrix}\;, \ee
and let us derive a wave equation on $U$. Straightforward
calculations lead to the following identity (here we write the
source term as a function of variables $p_1$ and $p_2^\sigma$,
that is $\mathcal{F}= \mathcal{F}(\cdot, \; p_1, p_2^\sigma)$
):
\bea\label{par-tau} \Box_\m (y\p_y u) &=& -(y\p_y
\m^{\al\beta})\p^2_{\al\beta}u + 2\m^{\al y}\p_\al\p_y u -(y\p_y
{\Upsilon}^\al) \p_\al u+ {\Upsilon}^y\p_y u\no
&& +(y\p_y \mathcal{F})(\cdot, u, \p u)+  (y\p_y u)\frac{\p
 \mathcal{F}}{\p p_1}(\cdot, u,\p u)
\no &&
 + \left(\p_y(y\p_\sigma u)- \de^y_\sigma \p_y
u\right)\frac{\p \mathcal{F}}{\p p^\sigma_2}(\cdot, u,\p u)\;.
\eea
We write
$$(y\p_y\m^{y y})\p^2_y u = \p_y\m^{yy}\left( \p_y(y\p_y u) -\p_y u\right)
 \sim  \Omega^{\frac{n-5}2 }(\partial U + U )\partial U
 \;,
$$
$$(y\p_y\m^{{\tilde{x}} y})\p_y\p_{\tilde{x}} u = \p_y\m^{yy}  \p_y({\tilde{x}}\p_{\tilde{x}} u)
 \sim  \Omega^{\frac{n-5}2 }\partial U \partial U
 \;,
$$
$$(y\p_y\m^{{\tilde{x}}{\tilde{x}}})\p^2_{\tilde{x}} u = O({\tilde{x}}^{\frac{n-3}{2}})
\left( \p_{\tilde{x}}({\tilde{x}}\p_{\tilde{x}} u) -\p_{\tilde{x}}
u\right)
 \sim \Omega^{\frac{n-3}2} U
 \partial U
\quad\text{see (\ref{metric-x})},$$
$$\m^{y y}\p^2_y u
 = \text{O}({\tilde{x}}^{\frac{n-7}{2}})\frac{{\tilde{x}}}{y}\left(\p_y(y\p_y u)-\p_y u\right)
 \sim  \Omega^{\frac{n-7}2} U
 \partial U
 \;,$$
and
%
%
%
$$ 2\m^{\al y}\p_\al\p_y u =  \m^{y y}\p^2_y u -\left\{ \m^{\tx
\tx}\p^2_{\tx} u +2\m^{ \tx A}\p_{\tx}\p_A u +\m^{AB}\p_A\p_B u +
 {\Upsilon}^\sigma\p_\sigma u- \mathcal{F}(u, \p u)\right\}\;.
$$
All the terms arising above have a structure similar to
\eq{Sterm1}. A similar comparison of the remaining terms shows
that we have
\bel{we-ypy}
 \Box_\m (y\p_y u) = \mathcal{F}_1(U, \p U)\;,\ee
where the source term $\mathcal{F}_1$ is of the general form as
in (\ref{Sterm1}) with the difference that it has a term
$\Omega^{\frac{n-7}2} U
 \partial U$ with
a multiplicative $\Om^{\frac{n-7}{2}}$; this term can be
estimated as in \eq{31VIII0.3} as long as $n\ge 7$. Moreover,
it is easily checked that this remains compatible with the
estimate of Proposition~\ref{prop-EM} (see
Remark~\ref{Remark2}). Note that the procedure above introduces
into the coefficients of the source terms the function $(y,{\tilde{x}})
\longmapsto \frac {\tilde{x}}y$, which is bounded on $\mcU$; furthermore,
${\tilde{x}}\p_{\tilde{x}}\frac {\tilde{x}}y= -y\p_y\frac {\tilde{x}}y = \frac {\tilde{x}}y$, which implies that
we will not loose the regularity of the source terms, as needed
for the problem at hand, when iterating the process.

From  the identities,
\bea\label{par-x} \Box_\m ({\tx}\p_{\tx} u) &=& -({\tx}\p_{\tx}
\m^{\al\beta})\p^2_{\al\beta}u + 2\m^{\al {\tx}}\p_\al\p_{\tx} u
-({\tx}\p_{\tx} {\Upsilon}^\al) \p_\al u+ {\Upsilon}^{\tx}\p_{\tx}
u\no
&& +(\tx \p_{\tx} \mathcal{F})(\cdot, u, \p u)+
\p_{\tx}({\tx}\p_{\tx} u)\frac{\p \mathcal{F}}{\p
 p_1}(\cdot,u,\p u)
\no
 &&
  + \left(\p_{\tx}({\tx}\p_\sigma u)- \de_\sigma^{\tx}
\p_{\tx} u\right)\frac{\p \mathcal{F}}{\p
p^\sigma_2}(\cdot,u,\p u)
 \;,
\eea
\be\label{par-A} \Box_\m (\p_A u) = -(\p_A
\m^{\al\beta})\p^2_{\al\beta}u  -(\p_A{\Upsilon}^\al) \p_\al u
+ \p_A u \frac{\p \mathcal{F}}{\p p_1}(u,\p u) +   \p\p_A u\frac{\p
\mathcal{F}}{\p p_2}(u,\p u)\;,\ee
we deduce that the same analysis holds for $\Box_\m (\p_A u)$
and $\Box_\m ({\tx}\p_{\tx} u)$. Therefore we have derived for
the new unknown function $U$ a wave equation of the form
(\ref{T13}), i.e.:
\bel{WE}\Box_\m U = \mathfrak{F}(U,\;\p U)\;.\ee
%
%
In order to apply to this equation Theorem~\ref{theo-EM}, we
have to check that the initial data for $U$ are in the right
spaces. Note that the initial data are prescribed on the subset
$\{x=y\}$ of $\mcU$. We denote this hypersurface by $\Sigma_0$,
thus $\Sigma_0\;= \; \phi(\mcS_0)\cap \mcU\;,$ and we set
\bel{Sigma-set}
 \Sigma_s\;= \; \phi(\mcS_s)\cap \mcU \subset {\Large\bf  H}_{-1/2s}\;.\ee
%
We want to
prove the following.
\begin{Lemma}
Under the hypotheses of Proposition~\ref{prop-EM} we have:
\bel{ini-data1} \left.(u, \p_Au, {\tilde{x}}\p_{\tilde{x}} u, y\p_y
u)\right|_{{\Sigma_0}}\in \left(\mcH^\al_k\cap
L^\infty\right)({\Sigma_0})\;,\ee
 \bel{ini-data2} \left.\left(\p u,
\p\p_Au, \p ({\tilde{x}}\p_{\tilde{x}} u), \p (y\p_y
u)\right)\right|_{{\Sigma_0}}\in
\mcH^\al_{k-1}({\Sigma_0})\;.\ee
\end{Lemma}
\proof
By assumption, we have
\bel{CH}\;u|_{\Sigma_0} \in\left( \mcH^\al_k\cap
L^\infty\right)({\Sigma_0}),\quad \text{and}\quad  \left. \left(\p_A
u,
\p_{\tilde{x}} u\;,
  \p_y u\right)\right|_{\Sigma_0} \in
\mcH^\al_k({\Sigma_0})\;.\ee
Now, using Sobolev's embedding theorem, we have \be
\left.{\tilde{x}}^{-\al} \left(\p_A u, \p_{\tilde{x}} u\;,
  \p_y u\right)\right|_{\Sigma_0} \in
L^\infty({\Sigma_0})\;. \ee
This leads to the following estimates:
\beaa \left.|{\tilde{x}}\p_{\tilde{x}} u|_{{\Sigma_0}}\right|&=&
{\tilde{x}}^{1+\al}\left|{\tilde{x}}^{-\al}\p_{\tilde{x}} u|_{\Sigma_0}\right|<\infty,\\
\left|y\p_y u|_{{\Sigma_0}}\right|&=& \left|{\tilde{x}}\p_y
u|_{{\Sigma_0}}\right|={\tilde{x}}^{1+\al}\left|{\tilde{x}}^{-\al}\p_y
u|_{\Sigma_0}\right|<\infty\;.\eeaa
To see that $ \p_A u(\tauz ) $ is in $L^\infty(\mcS_0),$ we
proceed as follows: integrating $\p_A u(\tauz )$ in $x$ until
$x_0$ gives the inequality
$$\p_A u(\tauz , x_0,v^A)-\p_A u(\tauz , {\tilde{x}}, v^A)= \int_{\tilde{x}}^{x_0}\p_{\tilde{x}}\p_A u(\tauz ,s,v^A)ds\;,$$
which leads to the estimate
\beaa|\p_A u(\tauz , {\tilde{x}},v^A)|&\le& |\p_A u(\tauz , x_0, v^A)|+
\|\p_{\tilde{x}}
u(\tauz )\|_{\mcC^{\al}_{\{{\tilde{x}}=0\},1}}\int_{\tilde{x}}^{x_0}s^{\al}ds \\
&\le&|\p_A u(\tauz , x_0, v^A)|+ \|\p_{\tilde{x}} u(\tauz
)\|_{\mcH^{\al}_{k }}\int_{\tilde{x}}^{x_0}s^{\al}ds\eeaa
(recall $ k-1 > \frac n2$). Since $$ \|\p_A u(\tauz , x_0,
v^A)\|_{L^\infty(\mcO)} < \infty, \qquad \|\p_{\tilde{x}} u(\tauz
)\|_{\mcH^{\al}_{k }}\le E^\al_{k }[u(\tauz )]< \infty
 \;,
$$
and
$$ \int_{\tilde{x}}^{x_0}s^{\al}ds=
\frac{1}{\al+1}\left(x_0^{\al+1}-{\tilde{x}}^{\al+1}\right)<\infty\;,
$$
we conclude that $ \|\p_A u(\tauz )\|_{L^\infty} <\infty $.
Thus $\left(\p_A u\right)|_{{\Sigma_0}}\in L^\infty(\Sigma_0)$
and we then obtain (\ref{ini-data1}).
On the other hand we have
\beaa\left.\|\p_\nu ({\tilde{x}}\p_{\tilde{x}}
u) |_{\Sigma_0}\right\|_{\mcH^\al_{k-1}({\Sigma_0})}
 &\le&
\left.\|{\tilde{x}}\p_{\tilde{x}}(\p_\nu
u)|_{\Sigma_0}\right\|_{\mcH^\al_{k-1}({\Sigma_0})}
+ \left.\|\de^{\tilde{x}}_\nu\p_{\tilde{x}}  u |_{\Sigma_0}\right\|_{\mcH^\al_{k-1}({\Sigma_0})}\\
&\le& \left.\|\p_\nu u |_{\Sigma_0}\right\|_{\mcH^\al_{k}({\Sigma_0})} \\
&<& \infty \qquad \text{see (\ref{CH})}\;. \eeaa
Similarly,
we have $\p(y\p_y u)|_{\Sigma_0}, \; \p \p_A u|_{\Sigma_0}\;  \in
\; \mcH^\al_{k-1}({\Sigma_0})$. We thus obtain (\ref{ini-data1})
and the
proof of the lemma is complete. \qed\\
Now, we apply Theorem~\ref{theo-EM} to (\ref{WE}) and obtain that
%
%
\bel{in1}  (u, \p_Au, {\tilde{x}}\p_{\tilde{x}} u, y\p_y
u)   \in
L^{\infty}\Big([0,2(\tau_*-\tauz )],\; \left(\mcH^\al_{k}\cap
L^\infty\right)(\Sigma_\ep)\Big)\;, \ee
\bel{in2}   \left(\p u, \p\p_Au, \p ({\tilde{x}}\p_{\tilde{x}} u), \p (y\p_y
 u)\right)  \in
 L^\infty\Big([0,2(\tau_*-\tauz )],\;\mcH^\al_{k}(\Sigma_\ep)\Big)
 \;.
\ee
Using once more the Sobolev embedding theorem,
we obtain that $\forall \ep\in [0,2(\tau_*-\tauz )]$
$$\left.(u, \p_Au, {\tilde{x}}\p_{\tilde{x}} u, y\p_y
u)\right|_{{\Sigma_\ep}}\in \fCtx{\al}{k-\left[\frac
n2\right]-1}({\Sigma_\ep})\;.$$
\beaa \|(u, \p_Au, {\tilde{x}}\p_{\tilde{x}} u, y\p_y u)\|_{L^\infty(\mcU)}&=&
\sup_{\tau\in[\tauz ,\tau_*]}\|\left.(u, \p_Au, {\tilde{x}}\p_{\tilde{x}} u, y\p_y
u)\right|_{\mcS_\tau}\|_{L^\infty(\mcS_\tau)} \\
&\le& \sup_{\tau\in[\tauz ,\tau_*]}\|\left.(u, \p_Au, {\tilde{x}}\p_{\tilde{x}} u,
y\p_y
u)\right|_{\mcS_\tau}\|_{\mcH^{\al}_k(\mcS_\tau)}\\
&\underbrace{<}_{\; \text{see (\ref{in1})  }}& \infty\;.\eeaa
Using now (\ref{in2}) instead of (\ref{in1}) we have
$$ \|(u, \p_A\p u, {\tilde{x}}\p_{\tilde{x}} \p u, y\p_y\p u)\|_{L^\infty(\mcU)}<
\infty\;.$$
This allows us to conclude that $(u, \;\p u)$ is in
$\mcC^\a_{\{0\leq\tilde{x}\leq y\}, 1}(\mcU) $. Now, if we repeat this process $j$ times
with $j = k-\left[\frac n2\right] -1$ then we obtain that $u$
is in $\mcC^\a_{\{0\leq\tilde{x}\leq y\}, k-\frac n2 -1}(\mcU)$. This completes the
proof of Theorem~\ref{theo-EM1}. \qed
\begin{Corollary}
\label{coro1} Under the hypotheses of Theorem~\ref{regu} we
have the following:
$$ \hat f \in \left(\mcC^\a_{\{0\leq\tilde{x}\leq y\}, \infty}(\mcU)\cap L^\infty\right)(\mcU)  \qquad\text{and}
\qquad  \p\hat f \in  \mcC^\a_{\{0\leq\tilde{x}\leq y\}, \infty}(\mcU)\;.$$
\end{Corollary}
\proof
The result is a combination of Theorems~\ref{regu} and
\ref{theo-EM1}.
\qed

\section{Polyhomogeneous solutions of the Einstein-Maxwell equations}
\label{sPhgEM}

Let $\delta$ be a positive real number. We recall that the
spaces of polyhomogeneous functions $\Ax,$ $\Axd,$ $\Axy$ and
$\Axyd$ are defined in~\cite[Equations (A.1)-(A.2)]{ChLeski}.
We consider the Cauchy problem for the Einstein-Maxwell
equations (\ref{T1}) with (\ref{E-M}) in wave coordinates
$(x^\mu)$ and Lorenz gauge with prescribed data on the
hyperboloid $\mcS_0$ (see (\ref{defmcS0})) at the interior of
the future light-cone with vertex the origin of coordinates.
The coordinate $x$ in which the polyhomogeneous expansion is
taken is $x= \frac{1}{t+r}$ where $t= x^0$ and $r= |\vec x
|=\sum\limits_{i=1}^n(x^i)^2).$ Indeed we have (see
(\ref{T2})):
\beaa x = -\tau-\rho &=& -\frac{t}{-t^2+r^2}-\left(
\sum\frac{(x^i)^2}{(-t^2+r^2)^2}\right)^{1/2}\\
&=& -\frac{t}{-t^2+r^2}- \frac{r}{t^2-r^2}\\
&=& \frac{1}{t+r}\;.\eeaa
We want to prove that, polyhomogeneous initial data for  the above
Cauchy problem lead to polyhomogeneous solution. We have the
following:
\begin{Theorem}
 \label{TEMphg}
Consider the Einstein-Maxwell equations on $\R^{1+n},\;n \ge
8$. Let $\de \in \R$ be such that $1/(2\de)\in\N$ when $n$ is
even and $1/\de\in\N$ when $n$ is odd. Suppose that the initial
data for (\ref{T1}) in wave coordinates and Lorenz gauge are
polyhomogeneous on the hyperboloid $\mcS_0:$
\bel{poly-data} f\big|_{\mcS_0} \in x^{\frac{n-1}{2}}\Axd \cap
L^\infty, \qquad  \p_\tau f\big|_{\mcS_0} \in
x^{\frac{n-1}{2}}\Axd\;,\ee
with $f=(g_{\mu\nu}-\eta_{\mu\nu},A_\mu)$.
There exists a time
$t_+>t_0$ and a solution
defined on $\underset{t\in[t_0,t_+]}{\cup}\mcS_t$ such that
$\forall t \in [t_0, t_+] $ we have:
\bel{poly-solution} f(t)= f\big|_{\mcS_t} \in
x^{\frac{n-1}{2}}\Axd \qquad \text{and} \qquad \p_\tau
f(t)=\p_\tau f\big|_{\mcS_t} \in x^{\frac{n-1}{2}-1}\Axd \;.\ee
Moreover, the solution is polyhomogeneous at $\scri$, in the
above polyhomogeneity class, as long as it remains in
$\mcH^{\alpha}_k(\bf{\Large{H}}_\tau)$, for some $\alpha\in
(-1,-1/2]$.
\end{Theorem}
\proof
Choose any $  \alpha < 0 $; we then have the inclusion
$\Axd(\overline{\phi(\mcS_0)}) \;\subset \;
\mcH^\alpha_\infty(\overline{\phi(\mcS_0)})$. It follows from
(\ref{poly-data}) that we have:
\bel{poly-data1} \hat f\left|_{\phi(\mcS_0)} \in
\left(\mcH^\al_\infty \cap L^\infty\right)(\phi(\mcS_0))\right.
\quad \text{and}\quad \p \hat f\left|_{\phi(\mcS_0)} \in
\mcH^\al_\infty(\phi(\mcS_0))\right.\;. \ee
For definiteness set $\alpha=-1/2$. From Theorem~\ref{regu},
there exists a time $\tau_*$ and a smooth solution $\hat f$ of
(\ref{T1})-(\ref{E-M})-(\ref{poly-data1}) defined on
$\myOmega_{\tau_*}$ such that $\forall \tau \in [\tauz ,
\tau_*], \; \hat f(\tau) \in
\mcC^\al_{j}({\bf{\large{H}_\tau}}).$
Next, applying   Corollary~\ref{coro1} one  obtains  that
$$ \hat f \in \left(\fCxy{\al}{\infty}\cap L^\infty\right)(\mcU)  \qquad\text{and}
\qquad  \p\hat f \in  \fCxy{\al}{\infty}(\mcU)\;.$$
From Theorem~\ref{Twavephg} of Section~\ref{SaEMe}, with
$$\psi_1 = \hat f\;, \quad \psi_2=(\p_y\hat f, \p_A \hat f)\;, \quad
\varphi= \p_x f\;,$$ we obtain (\ref{poly-solution}), and the
proof is completed.
\qed

It is natural to find conditions which guarantee that solutions
remain in weighted Sobolev spaces on hyperboloids, and hence
remain polyhomogeneous if the initial data are. One such
criterion is provided by the following:

\begin{Theorem}
 \label{TEMphg2}
Suppose that $k > \left[\frac n 2\right]+ 1$, with $n=6$ and
$\alpha=-1/2$, or $n\ge 7$ with $\alpha\in (-1,1/2]$. Solutions
of the Einstein-Maxwell equations remain in $\mcH^{\alpha}_k$,
$\alpha \in (-1,-1/2]$ as long as $\hat f$ remains in
$\mcC^{\kappa}_{\{x=0\},1}$, with
\bel{1IX0.6}
 \kappa>-\frac{(n-7)} 2
  \;.
\ee
The same is true for
\bel{1IX0.7}
 \kappa>-\frac{(n-5)} 2
  \ \mbox{provided that}\ \| x^{\frac {n-7} 2}y_\mu y_\nu \hat H^{\mu \nu}(\tauz)\|_{L^\infty}<\infty
  \;.
\ee
In particular, in dimensions $n+1\ge 9$ the small data
solutions of \cite{Loizelet:these,LoizeletCRAS} evolving out
from data stationary outside of a compact set are
polyhomogeneous.
\end{Theorem}

\proof
We want to use   Proposition~\ref{P19VIII0.1} to show that
solutions as above remain in $\mcH^{\alpha}_k$, $\alpha \in
(-1,-1/2]$. For this,  consider first the right-hand side  of
\eq{1IX0.5}. For $\kappa\ge  -(n-5)/2$ one immediately finds
that $\|\delta \m^\sharp\|_{\mcC^0_{\{x=0\},1}}$ is finite,
similarly for $(\partial_x-\partial_\tau) \delta \m^ \sharp$
when $\kappa \ge -(n-7)/2$. Finiteness of
$\|\delta\f^\sharp\|_{\mcC^0_{\{x=0\},1}}$ is straightforward
for $\kappa\ge  -(n-7)/2$ from \eq{T18}-\eq{TT20}. The estimate
on $\delta\Upsilon$ follows from \eq{T25} and \eq{xUpsilon}
provided again that $\kappa\ge (n-7)/2$.

For $\kappa\ge  -(n-5)/2$ the slowest decaying terms in $\f$,
$\Upsilon$, and in $(\partial_x-\partial_\tau) \m ^\sharp$  are
handled by the $\mcC^0_{\{x=0\},1}$-spaces equivalent of
\eq{1IX0.1},
\bean \lefteqn{
\|x^{\frac{(n-7)}2}\psi(\tau)\|_{\mcC^{0}_{\{x=0\},1}}
  }
  &&
\\
 &
  \le
  &
   \|x^{\frac{(n-7)}2}\psi(\tauz)\|_{\mcC^{0}_{\{x=0\},1}}
    +
     C(\tauz,\tau_1) \int_{\tauz}^\tau \|x^{\frac{(n-7)}2}\zeta (s)\|_{\mcC^{0}_{\{x=0\},1}} ds
     \;,
      \phantom{xxx}
\eeal{1IX0.2}
under the supplementary condition that $
   \|x^{\frac{(n-7)}2}\psi(\tauz)\|_{\mcC^{0}_{\{x=0\},1}}$ is
   finite.

For any $\sigma$ such that
\bel{7X0.2} \sigma<\kappa \ee
we have
$$
 \mcC^\kappa_{\{x=0\},1}\subset
 \mcB^\sigma_1
  \;.
$$
Hence the right-hand side of \eq{23VIII0.5} is finite for all
such $\sigma$'s, and so \eq{PPr2x} applies. It remains to show
that the integrand in the second line of \eq{PPr2x} can be
bounded by a multiple of the energy:
$$
  \| (\dm^\sharp,\delta
 \f^\sharp,{\dUpsilon} )\|^2_{\mcG^{\alpha-\mysigma}_{k}({\bf{\Large{H_{\tau}}}})}
 \le C
 E^\al_{k}[u(s)]
 \;.
$$
This is easily checked to hold under \eq{1IX0.6} or \eq{1IX0.7}
if we choose $\sigma$ so that
$$
 \sigma>-\frac {n-7}2
  \;.
$$
This, together with \eq{7X0.2}, explains \eq{1IX0.6}.

The property that the solutions of the Einstein-Maxwell
equations constructed by Loizelet are in
$\mcC^{\kappa}_{\{x=0\},1}$ on all hyperboloidal slices has
been verified in \eq{10I.1}. There $-\kappa=\delta\in (0,1/4)$
\qed

\bibliographystyle{amsplain}
\bibliography{%
../references/reffile,%
../references/newbiblio,%
../references/newbiblio2,%
../references/bibl,%
../references/howard,%
../references/bartnik,%
../references/myGR,%
../references/newbib,%
../references/Energy,%
../references/netbiblio,%
../references/PDE}

\end{document}